\documentclass[universe,review,accept,pdftex,moreauthors]{Definitions/mdpi} 

\newcommand{\msun}{$M_{\mathrm{\odot}} \,$}

\usepackage{acronym}
\firstpage{1} 
\pubvolume{1}
\issuenum{1}
\articlenumber{0}
\pubyear{2025}
\copyrightyear{2025}
\externaleditor{~~}
\datereceived{20 November 2024} 
\daterevised{3 February 2025} % Comment out if no revised date
\dateaccepted{18 February 2025} 
\datepublished{ } 
\hreflink{https://doi.org/} % If needed use \linebreak
\Title{Searching for New Physics in an Ultradense Environment:\linebreak A Review on Dark Matter Admixed Neutron Stars}

\TitleCitation{Searching for New Physics in an Ultradense Environment: A Review on Dark Matter Admixed Neutron Stars}

\Author{Francesco Grippa $^{1,2}$\orcidA{}, Gaetano Lambiase $^{1,2,}$*\orcidB{} and Tanmay Kumar Poddar $^{2}$\orcidC{}}

\AuthorNames{Francesco Grippa, Gaetano Lambiase and Tanmay Kumar Poddar}

\AuthorCitation{Grippa, F.; Lambiase, G.; Poddar, T.K.}
\address{%
$^{1}$ \quad Dipartimento di Fisica E.R. Caianiello, Università di Salerno,
Via Giovanni Paolo II 132 I-84084, \mbox{84084, Fisciano%MDPI: Please add the postal code. If the postal code is not available, Post Office Box number can be added instead.
, SA%MDPI: Please confirm if the province information is unnecesary and can be removed since its not the necessary information in affiliation.
, Italy}\\ %MDPI: We added the email addresses here according to those submitted online at susy.mdpi.com. Please confirm. So as the following ones with highlight.  
$^{2}$ \quad INFN %MDPI: Please provide the full name, if possible.
, Gruppo Collegato di Salerno, 84084, Fisciano, SA, Italy %MDPI: Please add the postal code and city information. If the postal code is not available, Post Office Box number can be added instead.
\\
}

\corres{Correspondence: lambiase@sa.infn.it}

\abstract{Neutron stars (NSs), among the densest objects in the universe, are exceptional laboratories for investigating the properties of dark matter (DM). Recent theoretical and observational developments have heightened interest in exploring the impact of DM on NS structure, giving rise to the concept of dark matter admixed neutron stars (DANSs). This review examines how NSs can accumulate DM over time, potentially altering their fundamental properties. We explore the leading models describing DM behavior within NSs, focusing on the effects of both bosonic and fermionic candidates on key features such as mass, radius, and tidal deformability. Additionally, we review how DM can modify the cooling and heating processes, trigger the formation of a black hole, and impact gravitational wave (GW) emissions from binary systems. By synthesizing recent research, this work highlights how DANSs might produce observable signatures, offering new opportunities to probe DM's properties through astrophysical phenomena.}

\keyword{\textls[-15]{neutron star; dark matter; gravitational waves; tidal deformability; \mbox{neutron decay}}}

\begin{document}
\nolinenumbers

%%%%%%%%%%%%%%%%%%%%%%%%%%%%%%%%%%%%%%%%%%
\section{Introduction}
Compact objects represent a unique class of astrophysical bodies characterized by incredibly high densities and strong gravitational fields. This group includes white dwarfs (WDs), neutron stars (NSs), and black holes (BHs), each formed through the evolutionary endpoints of stellar collapse and distinguished by the degeneracy pressure mechanisms or gravitational forces that prevent further collapse. WDs are sustained by electron degeneracy pressure, limiting their masses to the Chandrasekhar limit of approximately $1.4$ \msun \cite{Chandrasekhar:1939}, while NSs are held up by neutron degeneracy pressure and strong nuclear forces \cite{Lattimer:2004pg, Baym:2017whm}. BHs, by contrast, lack any such internal support mechanism and represent the ultimate state of stellar evolution, where collapse leads to singularities shielded by event horizons \cite{Shapiro:1983du}.

\textls[-15]{Among compact objects, NSs represent ideal environments to probe extreme states of matter, gravitational physics, and, potentially, DM interactions \cite{Bertone:2010zza}. Primarily composed of densely packed neutrons, NSs have masses ranging from about 1.4 to 2.5 \msun \cite{Lattimer:2004pg, Antoniadis:2013pzd, Ozel:2016oaf}}, compressed into a radius of roughly 10--12 km \cite{Lattimer:2006xb, Steiner:2010fz, Oertel:2016bki}. The central densities of NSs can exceed several times the nuclear saturation density ($\rho_\mathrm{nucl} \sim 2.7 \times 10^{14} \, \mathrm{g \, cm^{-3}}$), and can reach values as high as $5 \times 10^{15} \, \mathrm{g \, cm^{-3}}$ (Figure 4 in \cite{Lattimer:2010uk}) close to the core, orders of magnitude greater than the densities found in ordinary atomic nuclei \cite{Lattimer:2015nhk,Baym:2017whm}. Such extreme densities lead to intense gravitational fields, with surface gravity

\begin{equation}
    g_s = \frac{M}{R^2 \sqrt{1-2M/R}} \sim 2 \times 10^{12} \, \mathrm{m \; s^{-2}} \, ,
\end{equation}
for an average NS with mass $M = 1.4 \, M_\odot$ and radius $R = 10$ km \cite{Bejger:2004gz, Haensel:2007yy} (more than $10^{11}$ times that of Earth). This incredibly strong gravitational field allows for the probing of regimes of physics that are inaccessible to laboratory experiments on Earth.

The intense gravitational collapse leading to the formation of an NS heats its interior to a temperature of $\sim$$10 \, \mathrm{MeV} \simeq 10^{11}$ K \cite{1980SvA....24..303Y, Yakovlev:2004iq, Weber:2006ep}. However, NSs cool rapidly over time via neutrino emission, stabilizing to surface temperatures in the range of $10^6$--$10^8$ K after a few million years \cite{Pons:1998mm, Ho:2006uk, Potekhin:2015qsa}. Consequently, $T$ drops significantly, becoming about 3--4 orders of magnitude lower. That is why NSs are considered cold compact objects, for which the temperature, when expressed in MeV, is approximately zero. Despite this cooling, thermal and dynamic processes remain active and can provide clues about the underlying microphysics of dense matter.

The internal structure of NSs is typically modeled in layers. The outermost region is the crust, which extends to a depth of about 1--2 km and consists mainly of nuclei arranged in a crystal lattice immersed in a degenerate electron gas. As one moves deeper, towards the inner crust, neutrons begin to ``drip'' out of nuclei due to the increasing pressure. The inner core, which makes up most of the star's mass and volume, is composed of neutron-rich matter, and the exact composition remains an open question. It is theorized that it includes neutrons, protons, electrons, and possibly more exotic particles such as hyperons \cite{Logoteta:2021iuy,Oertel:2015fta,Chatterjee:2015pua}. Furthermore, phase transitions to deconfined quarks \cite{Bombaci:2016xuj, Alford:2006vz, Bombaci:2017tey} have been hypothesized to occur throughout the whole NS (forming bare quark stars \cite{Xu:2002ew, Page:2002bj, Weber:2012ta}) or near the core (hybrid stars \cite{Glendenning:1997wn, Burgio:2001mk, Alford:2002rj, Buballa:2003et, Alford:2004pf}). Both \emph{ud}- and \emph{uds}-quark matter---commonly referred to as strange quark matter (SQM) \cite{Bodmer:1971we, Witten:1984rs, Farhi:1984qu}---have been proposed \cite{Holdom:2017gdc} and studied in NSs \cite{Yuan:2022dxb}. A recent work by Bai et al. \cite{Bai:2024amm}, however, demonstrates the potential stability of 2-flavor quark matter, while excluding the possibility that (2+1)-flavor quark matter is the ground state in nature.

NS physics can be fully characterized by its Equation of State (EoS) \cite{Lattimer:2000nx, Sumiyoshi:2022uoj}, describing the relationship between pressure and density. Several models have been developed that are in agreement with observations. However, the exact EoS governing NSs at such high densities remains unknown due to the complex interactions between particles. The EoS is crucial for determining an NS's mass, radius, and stability against collapse into a BH. Observational constraints on the EoS have been significantly advanced by gravitational wave (GW) detections of binary neutron star (BNS) mergers, like GW170817 \cite{LIGOScientific:2018cki, LIGOScientific:2017ync,LIGOScientific:2017vwq, Radice:2017lry} and GW190425 \cite{LIGOScientific:2020aai}, which provided direct evidence of the deformability of NSs.

Recently, NSs have been identified as promising candidates for studying DM. Their high densities and temperatures, the strong gravitational field at their surfaces, and their ages (up to $10^{10} \, \mathrm{years}$) allow NSs to collect DM particles over time. This means NSs could serve as natural detectors, helping us learn more about DM, its properties, and its role in the universe \cite{Bertone:2007ae, deLavallaz:2010wp, Kouvaris:2010vv, Ciarcelluti:2010ji, Ellis:2018bkr}.

Accounting for approximately 27\% \cite{Planck:2018vyg} of the universe's total density, DM remains a mysterious component that exerts gravitational effects on visible matter, but neither emits nor absorbs light. This makes DM undetectable through traditional electromagnetic observations. First hypothesized in the 1930s by Fritz Zwicky, who noted the unusual motion of galaxies in the Coma Cluster \cite{Zwicky:1933gu}, DM's existence was further supported in the 1970s by observations of flat rotation curves in spiral galaxies \cite{Rubin:1970zza, Rubin:1978kmz, Mohan:2022kvb}. These curves suggest that galaxies contain far more mass than what is visible, implying the presence of an unseen gravitational source.

\textls[+15]{Early explanations focused on Massive Astrophysical Compact Halo Objects \mbox{(MACHOs)---such} as brown dwarfs, BHs, and faint stars \cite{Lasserre:2000xw}---or on modifications to gravitational laws, like the modified Newtonian dynamics (MOND) theory \cite{Famaey:2011kh}.} However, the inability of these hypotheses to match the results from microlensing experiments \cite{EROS-2:2006ryy} and observations within the Bullet Cluster \cite{Clowe:2006eq} led to the conclusion that a new type of matter (i.e., particles) was needed to explain cosmic structures. Standard Model (SM) particles like neutrinos were also discarded \cite{Boyarsky:2018tvu}, so the leading candidate became either weakly interacting massive particles (WIMPs) \cite{Pagels:1981ke, Jungman:1995df, Bertone:2004pz}, interacting via the weak nuclear force and gravity, or axions, lightweight particles originally proposed to solve the strong CP problem in quantum chromodynamics \cite{Weinberg:1972kfs, Peccei:1977hh}. Overall, several theories have been proposed throughout the decades to explain the nature of DM, spanning a mass range of more than 90 orders of magnitude \cite{Cirelli:2024ssz}; namely, from $\sim$$10^{-22}$ eV (fuzzy DM \cite{Hu:2000ke, Hui:2016ltb, Ferreira:2020fam}) to \mbox{$10^{2}$ \msun (primordial BHs \cite{Carr:2020gox, Carr:2020xqk}).
}

\begin{figure}[H]
    %\centering
 \hspace{-0.3cm}   \includegraphics[width=\columnwidth]{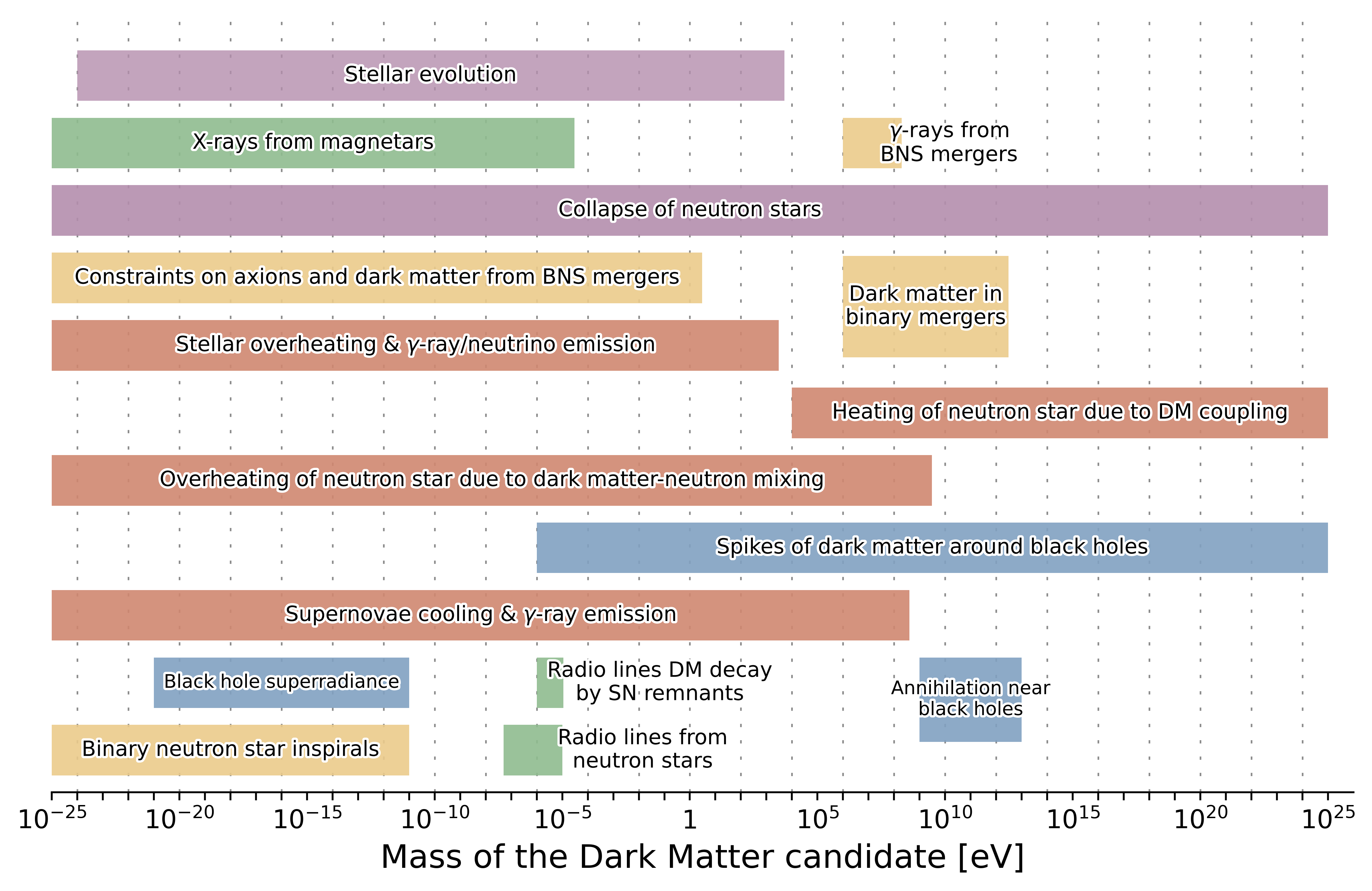}  
    \caption{Summary of astrophysical searches covering the wide range of DM masses probed by various methods. The schematic representation is adapted from \cite{Baryakhtar:2022hbu}.}
    \label{fig:dm_summary}
\end{figure}

Experimental efforts to detect DM have been extensive. Astrophysical searches primarily aim to capture signals of GWs and electromagnetic waves from DM interactions in extreme environments, namely, in isolated compact objects and in coalescences of binaries. The experimental techniques vary depending on the mass range under investigation: ultralight DM (<keV) may generate detectable radio or X-ray emissions from NSs, may influence the inspiral phase of GW emission from BNS mergers, and may %EE: Please check meaning retained
lead to coherent GW signals through BH superradiance. Light DM (keV--MeV) can be traced through X-ray/gamma-ray emissions from supernovae, temperature anomalies in NSs, or changes in stellar evolution influencing BH populations. Heavy DM (>GeV) may manifest through heating of NSs, DM spikes enhancing annihilation rates near BHs, and merger dynamics detected by future GW observatories. Such methods are described in \cite{Baryakhtar:2022hbu}, from which we obtain a schematic representation (Figure \ref{fig:dm_summary}) that summarizes how to probe DM within its wide mass range from astrophysical observations.

A significant role in the search for DM is played by direct detection experiments, such as those conducted by the XENON1T \cite{XENON:2018voc, XENON:2019rxp, XENON:2019gfn, XENON:2019zpr, XENON:2020gfr, XENON:2020kmp}, DAMA/LIBRA \cite{DAMA:2010gpn}, LUX--ZEPLIN \cite{LUX:2013afz}, CRESST \cite{CRESST:2015txj, CRESST:2019jnq}, CDMSLite \cite{SuperCDMS:2014cds, SuperCDMS:2017nns}, and CYGNUS \cite{Vahsen:2020pzb} collaborations, which aim to observe rare interactions between DM particles and atomic nuclei. Meanwhile, indirect detection approaches, such as those conducted by the  Fermi Large Area Telescope \cite{Fermi-LAT:2009ihh}, search for signals produced when DM particles potentially annihilate each other. Other promising methods include gravitational lensing studies \cite{Clowe:2006eq, Massey:2007wb}, where the gravitational effect of DM on light from distant galaxies helps map its distribution; and observations of cosmic microwave background (CMB) fluctuations \cite{Hooper:2018kfv}, which provide indirect evidence of DM’s influence on the early universe’s structure. Works like \cite{Battaglieri:2017aum, Bertone:2016nfn} provide comprehensive reviews of the history and current status of DM models and experimental searches.

DM can be accreted into NSs by gravitational capture \cite{Weinberg:1977ma, Marsh:2015xka, Eilers:2019gqs, Bell:2020jou, Busoni:2021zoe}, which is expected to be more efficient for massive particles such as WIMPs \cite{Kouvaris:2007ay, Feng:2010gw, Arcadi:2017kky}. Once accumulated, DM could influence the structure, the thermal evolution, and the long-term stability of the new compact object, known as a dark matter admixed neutron star (DANS). Indeed, the presence of DM can trigger different mechanisms. For instance, self-annihilating DM could inject energy into the star, potentially heating it and altering its cooling profile. This effect could be detectable in old NSs, where the heat generated by DM annihilation would cause them to appear hotter than expected \cite{Bertoni:2013bsa, Bell:2023ysh}. DM interactions can also be constrained by measuring the surface temperature of NSs (usually in systems where no other significant heating mechanism is expected) \cite{deLavallaz:2010wp, Kouvaris:2010vv, Kouvaris:2007ay, Giannotti:2015kwo}: Observations of Cassiopeia A \cite{Heinke:2010cr}, a young NS with a well-documented cooling history, could offer valuable data in this regard \cite{Shternin:2010qi, Giangrandi:2024qdb}. Moreover, ultralight DM particles might be emitted from the star, influencing both its spin-down rate and orbital period. These effects can be further constrained by observing GW radiation \cite{Hook:2017psm, KumarPoddar:2019ceq, KumarPoddar:2019jxe, Seymour:2019tir, Dror:2021wrl, Lambiase:2024dqe}. Additionally, the possible decay of neutrons into DM particles within the NS could even alter the mass--radius (M--R) relationship \cite{Cline:2018ami, Husain:2022brl, Husain:2022bxl}.

DM could also leave signatures in the GW signals emitted by BNS mergers. During the inspiral and merger phases of such collisions, the presence of DM could alter the tidal deformability ($\Lambda$) and the gravitational waveform, potentially leading to detectable deviations from the predictions of purely baryonic models \cite{Ellis:2017jgp, Kopp:2018jom, Bhattacharya:2023stq}. This idea gained traction following the detection of GWs from GW170817. Remarkably, the GW190814 \cite{LIGOScientific:2020zkf} event involved a $\sim$$23$ \msun BH and a compact object with a mass that was challenging to classify: either a high-mass NS or a low-mass BH \cite{Most:2020bba}. The uncertainty about the nature of the lighter companion led to speculation that it might be an NS with a significant fraction of DM \cite{Das:2021yny}. Future observations by both current detectors and incoming interferometers (such as the Einstein Telescope \cite{Hild:2010id, Branchesi:2023mws} and the Cosmic Explorer \cite{LIGOScientific:2016wof, Evans:2023euw}) could provide crucial insights into the existence of DM within NSs. Finally, ongoing X-ray observations by the NICER mission \cite{Gendreau:NICER}, which aims to measure NS masses and radii with great precision, offer another method for detecting DM effects by looking for any deviations from the mass--radius relationship predicted by purely baryonic frameworks.

So far, the numerous direct and indirect searches for DM have not yet provided any ``smoking gun'' proof. However, based on the scenarios discussed above, theoretical works have highlighted potential signatures which, if observed, would serve as compelling evidence for the presence of DM in NSs. Bezares et al. \cite{Bezares:2019jcb} have identified a characteristic $m = 1$ GW mode in the post-merger waveform of DANS mergers, attributed to the one-arm instability \cite{Paschalidis:2015mla} of the remnant. Additionally, they found a faster exponential decay of the $l = |m| = 2$ mode compared to BNS coalescences. Despite some degeneracy with ordinary NSs, these features may be distinguishable by interferometers. In a different context, Sun et al. \cite{Sun:2023cqr} have recently proposed a new diagnostic criterion: DM accumulation close to the NS core could lead to DANSs with smaller normal-matter radii but larger tidal deformabilities than ordinary NSs, in which this negative correlation does not arise. As such, if observations reveal two (or more) NSs with very similar mass and one exhibits a smaller radius but a greater $\Lambda$, this would strongly support the existence of DM in NSs. Even though more sophisticated models might modify or exclude this negative correlation, still a significant deviation from the expected M--R and/or $\Lambda$--M relations \cite{Ellis:2018bkr} could serve as a valuable clue.

In this work, we present a general overview of DANSs, emphasizing the impact of DM on their properties and examining how these effects align with current experimental bounds. In detail, the review is organized as follows. In Section \ref{sec:dark_matter_capture_and_accumulation}, we describe the accumulation mechanism of DM within NSs, the heating effects from self-annihilating DM (Section~\ref{sec:self_annihilating_dark_matter}), and the possible formation of BHs that is triggered by the prsence of DM (Section~\ref{sec:black_hole_formation}). Section \ref{sec:dark_matter_admixed_neutron_star_models} covers the models that have been studied when describing DANSs. In particular, in Section~\ref{sec:formalism} we present the numerical method typically used to solve the stellar structure of DANSs and to compute properties like radius, mass, and tidal deformability; theoretical models accounting for bosonic DM and fermionic DM are presented in Sections~\ref{sec:bosonic_dark_matter} and \ref{sec:fermionic_dark_matter}, respectively; while the effects of DM  on the mass--radius and tidal--mass relations are explored in Section~\ref{sec:mass_radius_and_tidal_mass_relations}. For fermionic DM, we mostly focus on scalar and vector DANSs, namely, DANS where the DM particles interact with each other by exchanging dark scalar or vector fields. The results are then compared to the current observational constraints from ground-based interferometers and NICER. An interesting approach for the computation of the speed of sound inside DANSs is described in Section~\ref{sec:speed_of_sound_in_dans}, though more detailed models are needed to analyze the influence of DM on the velocity of linear perturbation inside compact objects. Section \ref{sec:dark_matter_signals_in_gravitational_wave_emission} is dedicated to the potential DM imprint on signals arising from binary NS mergers. The possible decay of SM particles into DM is examined in Section \ref{sec:decay_of_standard_model_particles_into_DM}. Finally, the main conclusions are summarized in Section \ref{sec:conclusions_main}\textcolor{blue}.

In the following, we consider geometrized units $G = c = 1$, unless otherwise stated.

%%%%%%%%%%%%%%%%%%%%%%%%%%%%%%%%%%%%%%%%%%
\section{Dark Matter Capture and Accumulation}
\label{sec:dark_matter_capture_and_accumulation}

One of the primary mechanisms for DM accumulation within NSs is gravitational capture. As stars orbit around the center of a galaxy, %EE: Please check meaning retained
they encounter significant fluxes of DM particles. Interactions between DM and SM particles within stars can lead to energy loss, potentially trapping DM in the star’s gravitational field. This capture process would be particularly efficient due to the extreme density of NSs. Assuming a typical NS of $M \sim 1.5$ \msun and $R \sim 12$ km, the corresponding escape speed at its surface is $v_{\mathrm{esc}} = \sqrt{2M/R} \approx 0.6$, as measured by an observer at rest on the surface of the NS.

The estimation of the cross-section for collision between a DM particle $\chi$ with the NS is explicitly given in \cite{Goldman:1989nd}~(the original derivation was aimed at computing the mass flux of WIMPs that may collide with the NS, but an equal approach holds for any DM candidate). For a particle at infinity with velocity $v_\chi$ and impact parameter $b$, the closest approach of $\chi$ to NS is $R$ and the scattering cross-section is $\sigma = \pi b^2$. Recalling the motion of test particles in the Schwarzschild metric \cite{Weinberg:1972kfs}, and assuming $v_\chi^2 << 1$ in the derivation of the impact parameter $b = R \frac{v_{\mathrm{esc}}}{v_\chi} [1 - 2M/R]^{-1/2}$, allows us to compute the DM mass capture rate:

\begin{equation}
    \label{eq:DM_capture_rate}
    \dot{M} \equiv \frac{dM_\chi}{dt} = \rho_\chi v_\chi \pi b^2 P_{\mathrm{v}} P_{\sigma} \, ,
\end{equation}
where the overdot denotes the time derivative, $\rho_\chi$ is the local DM density, and $P_{\mathrm{v}}$ and $P_\sigma$ are the probabilities that a scattered DM particle loses enough energy to be captured and the probability that $\chi$ is scattered, respectively.

The capture rate in Equation~\eqref{eq:DM_capture_rate} depends primarily on the local DM density as well as its velocity, which can vary significantly across different cosmic environments/regions. Close to the Solar System, the local DM density is calculated as $\rho_\chi \approx 0.4 \, \mathrm{GeV/cm^3} \approx 6.7 \times 10^{-25} \, \mathrm{g/cm^3}$ and the non-relativistic DM velocity, $v_\chi \approx 250 \, \mathrm{km/s}$ \cite{Eilers:2019gqs}. Therefore, the resulting DM flux is approximately $\dot{M}_{\mathrm{ss}} \approx 53 \, \mathrm{g/s} \approx 3 \times 10^{25} \, \, \mathrm{GeV/s}$, as estimated in \cite{Goldman:1989nd}.

The probability of DM scattering can be expressed as $P_\sigma = 1 - e^{-\tau}$, where $\tau = \sigma_{\mathrm{n}\chi} / \sigma_{\mathrm{cap}}$, with $\sigma_{\mathrm{n}\chi}$ the nucleon-DM cross-section while $\sigma_{\mathrm{cap}}$ is the capture cross-section. From \cite{Bramante:2023djs}, the latter is

\begin{equation}
\label{eq:sigma_0}
    \sigma_{\mathrm{cap}} =
\begin{cases}
    \sigma_0 \left( \frac{m_\mathrm{n}}{m_\chi} \right), & \mathrm{if \;} m_{\mathrm{evap}} < m_\chi < m_\mathrm{n} \\
    \sigma_0, & \mathrm{if \;} m_\mathrm{n} \leq m_\chi \leq \mathrm{PeV} \\
    \sigma_0 \left( \frac{m_\chi}{\mathrm{PeV}} \right), & \mathrm{if \;} m_\chi > \mathrm{PeV} \, ,
\end{cases}
\end{equation}
where in Equation~\eqref{eq:sigma_0} $m_\mathrm{n}$ is the neutron mass, $\sigma_0 = \pi \left ( \frac{m_\mathrm{n}}{M} \right ) R^2$ the NS geometric cross-section accounting for the effective area in which DM can be captured, and $m_{\mathrm{evap}} \approx 20 \frac{T}{10^{3} \mathrm{K}}$~eV ($T$ is the temperature inside the NS) represents a threshold mass below which DM is ineffectively captured. In other words, if $m_\chi < m_{\mathrm{evap}}$, the NS's thermal energy would cause the captured DM to gain enough kinetic energy to escape from the stellar potential well \cite{Garani:2018kkd,Garani:2021feo}.

Not surprisingly, Equation~\eqref{eq:DM_capture_rate} implicitly depends on the DM mass $m_\chi$ through \mbox{Equation~\eqref{eq:sigma_0}}. The neutrons that scatter with DM particles are typically non-relativistic, with momenta smaller than $\sim$$\mathrm{GeV}$ (i.e., their mass). If $m_\chi < m_{\mathrm{n}}$ a few neutrons succeed in scattering to a state above the Fermi momentum because their recoil is insufficient to overcome the energy barrier set by the Fermi surface. Such a ``Pauli-blocking'' effect causes the capture cross-section to be $\sigma_{\mathrm{cap}} \propto m_\chi^{-1}$. For larger DM masses, the neutron recoil is larger, and the expression of $\sigma_0$ as a function of $m_\chi$ changes.

For the sake of completeness, a slightly different derivation of the capture rate of DM particles into compact objects is presented in \cite{Bell:2018pkk,Busoni:2021zoe}.

Once DM is captured, it undergoes a series of scattering events %EE: Please check meaning retained
with SM particles, gradually losing energy until it settles into a stable orbit within the gravitational pull of the star. These scatterings transfer kinetic energy to the NS, contributing to a gradual increase in its temperature. However, this heating effect is not the only factor influencing the star's thermal dynamics. Rather, the energy influx from DM is counter-balanced by the rate at which the NS emits photons, leading to a delicate equilibrium between heating and cooling. For NSs that are older than approximately one million years, the dominant mechanism for cooling is photon emission. At equilibrium, the two contributions (the heating due to in-falling DM and the cooling coming from photon emission) are equal.

A significant amount of research has explored the possibility of detecting DM by examining the cooling profiles of old NSs ($\tau \gtrsim 10^6$--$10^7$ years) and their surface temperatures. The time evolution of the red-shifted temperature of an NS is given by \cite{Hamaguchi:2019oev}

\begin{equation}
    \label{eq:full_cooling}
    C \frac{dT^\infty}{dt} = - L_\nu^\infty - L_\gamma^\infty + L_\mathrm{H}^\infty \, ,
\end{equation}
where $C$ is the total heat capacity of the NS; $L_\nu^\infty$ and $L_\gamma^\infty$ are the red-shifted luminosities of the neutrino and photon emissions, respectively; and $L_\mathrm{H}^\infty$ accounts for any mechanisms that contribute to raising the NS's temperature. In general, heat can be injected into the NS not only through the scattering and annihilation of DM particles, but also via internal processes such as conversion of magnetic, rotational, and chemical energies \cite{Gonzalez:2010ta}. For clarity, it is useful to distinguish these internal mechanisms from the contribution due to DM interactions, expressing $L_\mathrm{H}^\infty = L_\mathrm{int}^\infty + L_\mathrm{DM}^\infty$.

In Equation~\eqref{eq:full_cooling}, redshift correction occurs because the NS thermal relaxation is considered complete in $t \lesssim 10^2$ years. Beyond this, the red-shifted temperature, defined as $T^\infty = T(r) \left[ 1 - 2M/R\right]^{1/2}$, becomes constant in the core. Note that the most general definition $T^\infty = T(r) e^{\phi(r)}$ (with $e^{2\phi(r)} = g_{tt} (r)$ being the time component of the metric at position $r$) reduces to $T^\infty = T(r) \, \left [ 1 - 2M/R\right ]^{1/2}$ for a spherically symmetric spacetime described by the Schwarzschild metric, as we have implicitly assumed. The neutrino emission is enhanced by Urca processes \cite{Gusakov:2002hh, Page:2004fy} as well as pair-breaking and -formation processes \cite{Flowers:1976ux, Yakovlev:1998wr, Kaminker:1999ez}. During the first $t \lesssim 10^5$ years following their formation in supernova explosions, NSs primarily cool through neutrino emission from their interior \cite{Yakovlev:2000jp}. Hence, the neutrino luminosity is highly suppressed in older NSs. In contrast, the photon luminosity is expected to dominate at $t \gtrsim 10^5$ years and is intrinsically linked to the surface temperature $T_\mathrm{s}$ by

\begin{equation}
\label{eq:luminosity}
    L = g \dot{M} = 4 \pi \sigma R^2 T_\mathrm{s}^4 \, ,
\end{equation}
where $\sigma = \frac{\pi^2 k_\mathrm{B}^4}{60 \hbar^3 c^2} =  5.670400 \times 10^{-8} \; \mathrm{W \, m^{-2} \, K^{-4}}$ is the Stefan--Boltzmann constant (with $k_\mathrm{B} = 1.380 \; \mathrm{J \, K^{-1}}$ being the Boltzmann constant) and $g = G M / R$ accounts for gravitational effects and plays the role of an efficiency factor representing the fraction of the rest-mass energy converted into radiation during the accretion process ~\cite{Ikhsanov:2003dt}. The second equality, in which $\dot{M}$ is the rate at which mass is added to the star (Equation~\eqref{eq:DM_capture_rate}), holds by assuming that the NS behaves as a black-body radiator, whose luminosity $L = A \sigma T_\mathrm{s}^4$ follows the Stefan--Boltzmann law \cite{Yin:2011aa, book:Carroll_Ostlie_2017} for an approximately spherical NS of surface area $A = 4 \pi R^2$. The maximum photon luminosity $L_{\mathrm{max}}$ is obtained when $P_{\sigma_\chi} \approx 1$ and $P_{\sigma_{\mathrm{v}}} \approx 1$. Under these conditions, $L_{\mathrm{max}}$ can be computed close to the Solar System ($\sim$$10$~pc) as $L_{\mathrm{max}} \approx 7 \times 10^{24} \, \mathrm{GeV/s}$. Remarkably, for an ordinary NS with $R \sim 10$ km, this value corresponds to a superficial, effective temperature of $\sim$$2000$ K that may be detectable by cutting-edge infrared telescopes \cite{Chatterjee:2022dhp} such as the James Webb Space Telescope \cite{Gardner:2006ky, Kalirai:2018qfg, 2023PASP..135d8001R}.

To investigate the cooling profile, previous works \cite{Yakovlev:1999sk, Yakovlev:2000jp, Yakovlev:2004iq} have relied on the ``minimal cooling'' assumption, meaning that nucleons and charged leptons remain in $\beta$-equilibrium throughout the NS's evolution, with no internal mechanisms contributing to its heating ($L_\mathrm{int}^\infty = 0$). The resulting $T_s$ of isolated, old NSs was estimated to be $<$$1000$ K. However, when DM annihilation is included ($L_\mathrm{DM}^\infty \neq 0$), the predicted surface temperatures rise to $T_s \simeq (2 - 3)\times 10^3$ K. This suggests that the measurements of $T_s$ of old NSs could provide critical insights and constraints for DM models.

Recently, the $L_\mathrm{int}^\infty = 0$ assumption has been found to be invalid \cite{Yanagi:2019zne}, especially for fast-rotating NSs. Due to the gradual loss of rotational energy, the resulting reduction in centrifugal force causes the NS to slightly contract over time, altering the chemical equilibrium conditions among nucleons and leptons. This change cannot be sustained by weak processes, as they become more and more suppressed after $t \gtrsim 10^6$ years.

Such an imbalance provides a supplementary heating mechanism, known as \emph{rotochemical heating}. When included \cite{Fernandez:2005cg, Villain:2005ns, Petrovich:2009yh, Gonzalez-Jimenez:2014iia, Yanagi:2019vrr} ($L_\mathrm{int}^\infty = L_\mathrm{rc}^\infty \neq 0$), it can raise $T_s$ up to $10^6$~K, as confirmed by observations \cite{Kargaltsev:2003eb, Rangelov:2016syg, Pavlov:2017eeu, Gonzalez-Caniulef:2019wzi}. This makes the interpretation of NS cooling profiles and surface temperature measurements more challenging, as the intrinsic heating of the NS is now degenerate with the DM contribution. Interestingly, Hamaguchi et al. \cite{Hamaguchi:2019oev} have demonstrated that this degeneracy can be partially broken: the DM heating effect can still be observed if the initial period of ordinary NSs is relatively large, though we need to improve the current knowledge on nucleon pairing gaps as well as to evaluate the initial period of the pulsars more accurately. Furthermore, the lower limit \mbox{$T_s \gtrsim (2$--$3)\times 10^3$~K} should still hold because the rotochemical mechanism may only increase the surface temperature. Hence, observing an NS with much lower $T_s$ would ''exclude the DM heating, and severely constrain DM models''.

\subsection{Self-Annihilating Dark Matter}
\label{sec:self_annihilating_dark_matter}

In the case of self-annihilating DM (like WIMPs) into SM states, the accumulation of DM inside the NS can lead to significant heating. Generally, the more DM particles are concentrated in the core, the more efficient the annihilation mechanism, though the dependence on the theoretical model describing such an effect is not trivial \cite{Bramante:2023djs, Bertoni:2013bsa, Garani:2020wge}. If the DM particles have a non-negligible annihilation cross-section, they will annihilate each other, releasing energy into the star's core and altering its thermal evolution.

The most general equation that governs the evolution of the number $N_\chi(t)$ of DM particles accumulated inside an NS (at time $t$) is as follows \cite{Gaisser:1986ha, Zentner:2009is,Chen:2018ohx}:

\begin{equation}
    \label{eq:rate_annihilation_general}
        \frac{dN_\chi(t)}{dt} = C_{\mathrm{n}\chi} + C_{\chi\chi} N_\chi(t)- C_{\mathrm{ann}} N_\chi(t)^2 \, ,
\end{equation}
where $C_{\mathrm{n}\chi}$ is the DM capture rate via scattering on nucleons, $C_{\chi\chi}$ accounts for DM self-interactions (the complete expressions of $C_{\mathrm{n}\chi}$ and $C_{\chi\chi}$ are derived in Equation (3.7) in \cite{Guver:2012ba} and in Equation (9) in \cite{Zentner:2009is}, respectively), and $C_{\mathrm{ann}} = \langle \sigma v_\chi \rangle / V_\chi$ is the annihilation rate coefficient, with $\langle \sigma v_\chi \rangle$ representing the thermally averaged DM annihilation cross-section and ${V_\chi}$ being the volume of the DM core in which annihilation occurs.

Equation~\eqref{eq:rate_annihilation_general} can be used in cases where the DM self-annihilation (last term on the right-hand side) is negligible \cite{DelPopolo:2020hel}. Conversely, works as \cite{Kouvaris:2007ay, Kouvaris:2010vv, Acevedo:2019agu} have taken into account the self-annihilation but have ignored the DM self-capture (middle term on the right-hand side). In this latter, common framework, Equation~\eqref{eq:rate_annihilation_general} reads \cite{Bramante:2023djs}

\begin{equation}
    \label{eq:rate_annihilation}
    \frac{dN_\chi(t)}{dt} \simeq C_{\mathrm{cap}} - C_{\mathrm{ann}} N_\chi(t)^2 \, ,
\end{equation}
where $C_\mathrm{cap} = C_{\mathrm{n}\chi} + C_{\chi \chi} \simeq C_{\mathrm{n}\chi}$  describes the total DM accretion via both processes and may be computed through Equation~\eqref{eq:DM_capture_rate}. 

The simplified Equation~\eqref{eq:rate_annihilation} has an analytical solution:

\begin{equation}
    \label{eq:rate_annihilation_SOLUTION}
    N_\chi (t) = \sqrt{\frac{C_{\mathrm{cap}} V_\chi }  {\langle \sigma v_\chi \rangle }} \, \mathrm{tanh} \left ( \frac{t}{\tau_{\mathrm{eq}}} \right ) \,,
\end{equation}
with $\tau_{\mathrm{eq}} = \sqrt{ V_\chi / (C_{\mathrm{cap}} \langle \sigma v_\chi \rangle) }$ being the characteristic time-scale for the equilibrium between annihilation and capture to become settled (i.e., $dN_\chi(t) / dt \rightarrow 0$ at $t \geq \tau_{\mathrm{eq}}$). The energy released through annihilation can be computed \cite{Kouvaris:2007ay} as $E = C_\mathrm{ann} N_\chi^2 m_\chi = \frac{\langle \sigma v_\chi \rangle}{V_\chi} \mathrm{tanh}^2 \left ( \frac{t}{\tau_\mathrm{eq}} \right ) m_\chi$.

Assuming that the gravitational capture is the main mechanism (i.e., the first term on the right-hand side in Equation~\eqref{eq:rate_annihilation} dominates), the most general estimation of DM capture \cite{Bhattacharya:2023stq} is
\vspace{-6pt}
\begin{align}
    \label{eq:DM_capture_rate_numerical}
    \nonumber C = 1.4 \times 10^{20} \; \mathrm{s^{-1}} &\times \left ( \frac{\rho_\chi}{0.4 \; \mathrm{GeV \, cm^{-3}}} \right ) \left ( \frac{10^5 \; \mathrm{GeV}}{m_\chi} \right ) \left ( \frac{\sigma_{n\chi}}{10^{-45} \; \mathrm{cm^2}}  \right ) \times \\ &\times \left ( 1 - \frac{1-e^{-A^2}}{A^2} \right ) \left ( \frac{v_{\mathrm{esc}}}{1.9 \times 10^5 \; \mathrm{km \, s^{-1}}} \right )^2 \left ( \frac{220 \, \mathrm{km \, s^{-1}}}{v_\chi} \right )^2 \, ,
\end{align}
where, in addition to the above definitions, $A^2 = 6m_\chi m_n v_\mathrm{esc}^2 / v_\chi^2 \left ( m_\chi - m_n \right )^2$ accounts for inefficient momentum transfers at large $m_\chi$. The capture rate in Equation~\eqref{eq:DM_capture_rate_numerical} corresponds to a number of accumulated DM particles \cite{McDermott:2011jp} of $N_\chi \sim 10^{47}$ under standard conditions (whereas the average nuclear baryon number in an NS is $N_B \sim 10^{57}$ \cite{Berryman:2022zic}).

To understand whether equilibrium is reached within the lifetime of an NS, one needs to consider the minimum annihilation cross-section required. Following \cite{Garani:2020wge}, it is possible to adopt a partial-wave expansion $\langle \sigma v_\chi \rangle = a + b v_\chi^2 $ to express the equilibrium conditions for \emph{s-wave} and \emph{p-wave} dominations as

\begin{small}
\begin{align}
    \label{eq:s_and_p_waves}
    a &> 7.4 \times 10^{-54} \mathrm{cm^3 / s} \left ( \frac{\mathrm{Gyr}}{\tau} \right )^2  \left ( \frac{C_\mathrm{max}}{C_\chi} \right )  \left ( \frac{\mathrm{GeV}}{m_\chi} \frac{T}{10^3 \mathrm{K}} \right )^{\frac{1}{2}}, \\
    b &> 2.9 \times 10^{-44} \mathrm{cm^3 / s} \left ( \frac{\mathrm{Gyr}}{\tau} \right )^2  \left ( \frac{C_\mathrm{max}}{C_\chi} \right )  \left ( \frac{\mathrm{GeV}}{m_\chi} \frac{T}{10^3 \mathrm{K}} \right )^{\frac{1}{2}}, 
\end{align}
\end{small}
where $\tau$ is the average lifetime of an NS (with mass $\sim$$1.5$ \msun and radius $\sim$$12$ km) and $C_\mathrm{max}$ the maximum capture rate.

An intriguing method to investigate DM self-annihilation in NSs is by focusing on the annihilation products. In particular, if the primary products of DM annihilation are feebly interacting mediators between DM and SM particles, these particles could survive long enough to escape the star before decaying into detectable SM states. These mediators, generated through DM capture by NSs or other celestial bodies, could produce observable fluxes of radiation or neutrinos.

\subsection{Black Hole Formation}
\label{sec:black_hole_formation}

The capture mechanism described in Section \ref{sec:dark_matter_capture_and_accumulation} allows the DM fraction to increase more and more further into the NS. If DM could self-annihilate into SM particles upon reaching sufficient densities within the NS (like WIMPs), then the total amount of DM that settles close to the core would stop growing at some point. However, certain DM candidates, such as asymmetric DM do not annihilate due to the assumed asymmetry between the number density of particles and antiparticles. Instead, they continue to accumulate over time, creating a central core of DM that contributes to the gravitational field of the star \cite{Bramante:2014zca}. Asymmetric DM is such that DM particles and their antiparticles have unequal abundances. This disproportion means that asymmetric DM particles, unlike symmetric candidates, do not fully annihilate with their counterparts, allowing them to accumulate in celestial bodies over time. This kind of DM is widely studied in astrophysical contexts because it mirrors the asymmetry observed in the visible universe, which is primarily composed of matter (electrons, protons, and nucleons) and not anti-matter. This would possibly provide an explanation for why DM and ordinary matter exist in comparable proportions within the universe \cite{Kaplan:2009ag, Petraki:2013wwa, Zurek:2013wia}.

Once captured, DM particles may undergo one of the following:

\begin{itemize}
    \item A two-stage process if the DM--nucleon %EE: Please check meaning retained
cross-section is not too large \cite{Acevedo:2019gre, Acevedo:2020gro, Kouvaris:2010jy}. In this scenario, a first thermalization where DM orbits are larger than the size of the hosting NS is followed by a second thermalization once enough energy is lost and the orbit stabilizes to a final thermal radius within the star.

    \item Alternatively, in case of a large $\sigma_{n \chi}$, all captured DM particles will slow down rapidly in the outer shells of the star and only drift, on much longer timescales, towards the core under the influence of a viscous drag force  \cite{Gould:1989gw, Starkman:1990nj, Mack:2007xj, Bramante:2019fhi}.
\end{itemize}

In both scenarios, the thermal radius, as well as the thermalization time, can be estimated as \cite{Goldman:1989nd, Kouvaris:2010vv, Bertoni:2013bsa}
\begin{align}
    \label{eq:thermal_radius}
    r_\mathrm{th} &= 2.2 \, \mathrm{m} \, \left ( \frac{T}{10^5} \right)^{\frac{1}{2}} \left ( \frac{\mathrm{GeV}}{m_\chi} \right)^{\frac{1}{2}}  \, , \\
    \label{eq:thermalization_time}
    t_\mathrm{th} &= 0.2 \, \mathrm{yr^{-1}} \left ( \frac{m_\chi}{\mathrm{TeV}} \right )^2 \left ( \frac{\sigma}{10^{-43} \, \mathrm{cm^2}} \right )^{-1} \left ( \frac{T}{10^5} \, \mathrm{K} \right )^{-1},
\end{align}
by assuming that the typical DM particle becomes non-relativistic after having been captured and having lost most of its energy.

\begin{figure}[H]
    %\centering
    \includegraphics[width=0.5\columnwidth]{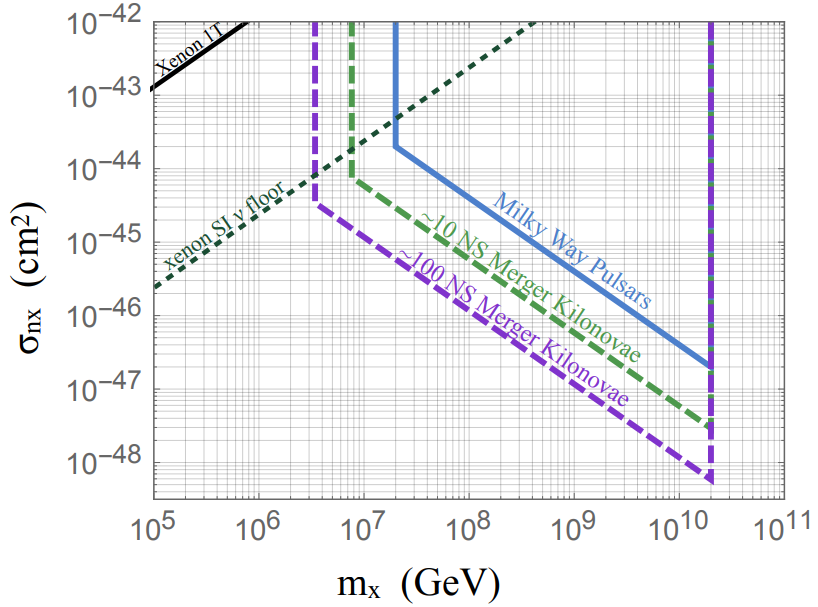}
    \hspace{0.2cm}
    \includegraphics[width = 0.47\columnwidth]{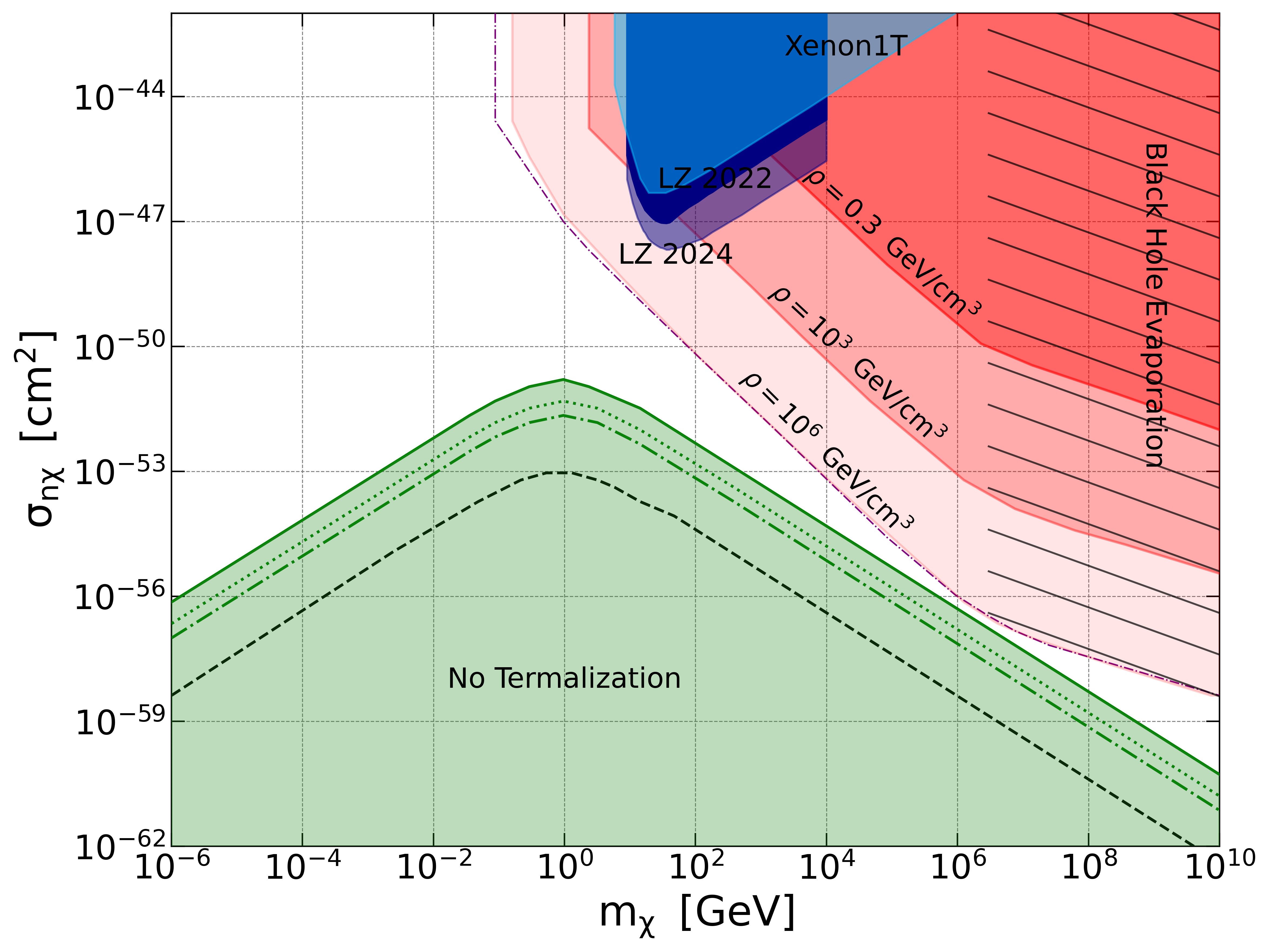}  
    \caption{\textbf{Left panel}: Bounds on the (heavy) fermionic DM--nucleon cross-section obtained from the observation of old pulsars in the Milky Way that have not been converted to BHs. Ref. \cite{Bramante:2017ulk} (from which the plot is taken) also reported constraints from terrestrial direct experiments such as Xenon1T \cite{XENON:2019rxp, XENON:2019gfn, XENON:2019zpr, XENON:2020gfr} and xenon neutrino floor \cite{Ruppin:2014bra}, as well as potential observations of binary NS mergers accompanied by kilonovae (with localization precision up to 1 kpc within Milky Way-like spiral galaxies). \textbf{Right panel}: Bounds on DM-neutron cross-section obtained with the gravitational collapse condition applied to an old NS with a thermalized core of bosonic DM when a BEC is not formed. The representative NS is assumed to have an average lifetime of $10^{10}$ years and a mean temperature of $10^5$ K, according to Garani et al.'s \cite{Garani:2018kkd} model, from which the plot is adapted. In the red areas (corresponding to different values of $\rho_\chi$), the accumulated DM triggers the collapse to a BH. In the green areas, DM fails to thermalize with neutrons. The hatched regions indicate where the BH evaporates before it can destroy the NS, thereby weakening the constraints. The Xenon1T limit on DM--neutron cross-sections is represented by the dark blue-shaded region. The updated upper limits from the Lux-Zeplin 2022 \cite{LZ:2022lsv} and 2024 \cite{LZCollaboration:2024lux} campaigns are reported in light blue. Note that these experiments explored the WIMP range up to $\sim$$10^4$ GeV (see also Figure \ref{fig:sigma_DM} below). Remarkably, in \cite{Garani:2018kkd} similar analyses have been performed for DM scattering with protons and muons as well as when considering bosonic DM that forms a BEC or fermionic DM.}
    \label{fig:dm_bh_constraints}
\end{figure}

At some point a critical threshold may be reached, meaning that enough DM has been accumulated into the thermalized region within the NS to trigger the collapse to a BH. This condition can be exploited to limit the DM parameter space as in Fig. \ref{fig:dm_bh_constraints}. Such a process is strongly dependent on the properties of DM particles, especially their mass, interaction cross-section, possible self-interactions, and the degeneracy pressure they experience. Assuming the presence of a WIMP sphere, Kouvaris et al. \cite{Kouvaris:2010jy} estimated that the onset of the gravitational collapse occurs when the total number $N$ of accumulated WIMPs is $N \sim 5 \times 10^{43} \, \left ( m_\chi / \mathrm{TeV} \right )^{-3}$. For such a value, the potential energy of (semi-relativistic) WIMPs exceeds the Fermi momentum, and therefore Pauli blocking cannot prevent the collapse anymore. The properties of the DM candidate may also influence the likelihood and/or the speed of the gravitational collapse. For example, if the captured DM consists of bosons that form a Bose--Einstein condensate within the star, it would exert minimal pressure and the collapse would be further accelerated \cite{Kouvaris:2011fi, McDermott:2011jp, Kouvaris:2012dz, Jamison:2013yya, Garani:2022quc}.

According to \cite{Acevedo:2020gro}, in order for DM to trigger the conversion of the NS, the following conditions must hold:

\begin{enumerate}

    \item The sound crossing time $t_\mathrm{s} = r_\mathrm{th} / c_\mathrm{s}$ (where $c_\mathrm{s}$ is the sound speed in a DANS modeled as a perfect fluid) should be greater than the DM freefall time $t_\mathrm{ff}$, which is computed by considering the free fall of a test mass onto an object of mass $\frac{4}{3} \pi \rho_\chi r_\mathrm{th}^3$. The \mbox{condition is}
    \begin{equation}
        \frac{3}{\sqrt{4 \pi \rho_\mathrm{NS}}} = t_\mathrm{s} \geq t_\mathrm{ff} = \sqrt{\frac{3 \pi}{32 \rho_\chi}} \, ,
    \end{equation}
    
    which is also known as \emph{Jeans stability} (see \cite{Herrera:1997plx, Chavanis:2011zi} and references therein for similar analysis in other contexts) and is approximately realized when the density of the DM sphere is greater than the solar or terrestrial core density \cite{Acevedo:2020gro}.

    \item The mass of captured DM should overcome a threshold mass, given as
    
    \begin{equation}
        M_\mathrm{coll} = \sqrt{\frac{3 T_\mathrm{NS}^3}{\pi m_\chi^3 \rho_\mathrm{NS}}} \approx 2.2 \times 10^{54} \, \mathrm{GeV} \, \left ( \frac{10^7 \mathrm{GeV}}{m_\chi} \right )^{\frac{3}{2}} \, ,
    \end{equation}

    computed for an average NS with core density $\rho_\mathrm{NS} \sim 10^{15} \, \mathrm{g/cm^3}$ and absolute temperature $\sim$$10^8$ K.
    
    For $M_\mathrm{DM} < M_\mathrm{coll}$, namely, prior to collapse, the DM particles are in thermal equilibrium within a dense sphere of radius $r_\mathrm{th}$ at temperature  $T_\mathrm{NS}$ of the hosting NS's core ($r_\mathrm{th}$ and $T_\mathrm{NS}$ are related through the virial theorem $2 \langle K \rangle = - \langle U \rangle$, where in turn, $\langle K \rangle = 3/2 \cdot T_\mathrm{th}$ and $\langle U \rangle = \frac{4}{3} \pi r_\mathrm{th}^2 \rho_\mathrm{NS} m_\chi$). Once the condition $M_\mathrm{DM} > M_\mathrm{coll}$ is reached, the DM core can no longer be stabilized and the collapse to a BH is \mbox{irreversibly triggered.}

    In addition, the central DM sphere must be sufficiently massive such that any additional contributions from self-interactions or degeneracy pressure are not enough to stabilize it and prevent collapse. More precisely, the critical mass beyond which collapse occurs should be defined as $M_\mathrm{cr} = \mathrm{max} [M_\mathrm{coll}, \, M_i]$, where $M_i$ \cite{Bramante:2023djs} represents the maximum mass that can be stabilized by interactions and is given by
    
    \begin{equation}
        M_i \equiv
        \begin{cases} 
            M_f \sim M_\mathrm{pl}^3 / m_\chi^2 \sim \, M_\odot \left ( \frac{\mathrm{GeV^2}}{m_\chi^2} \right ) \, \;\; &\mathrm{for \;\; fermionic \;\; DM} \, , \\
            M_b \sim \frac{2M_\mathrm{pl}^2}{m_\chi} \left (1 + \frac{\lambda}{32\pi} \frac{M_\mathrm{pl}^2}{m_\chi^2} \right )^{1/2} \;\; &\mathrm{for \;\; bosonic \;\; DM} \, ,
        \end{cases}   
    \end{equation}
    with $M_\mathrm{pl} = 1.2209 \times 10^{19}$ GeV being the Planck mass and $\lambda$ the coupling constant of the possible boson repulsive self-interaction ($\mathcal{L} \supset -(\lambda/4!)\chi^4$).
\end{enumerate}
Once the BH is formed close to the center, its time evolution depends on competing effects coming from accretion (of both nuclear and DM) and Hawking radiation. Specifically, a small enough BH could even evaporate completely; in contrast, a large BH is likely to accumulate surrounding material and possibly convert the whole NS. The rate at which the mass of the BH, $M_\mathrm{BH}$, changes has been derived in \cite{Garani:2018kkd, Acevedo:2020gro}:

\begin{equation}
\small
\label{eq:rate_BH_mass}
    \frac{dM_\mathrm{BH}}{dt} = \dot{M}^{\mathrm{(NS \; acc)}} + \dot{M}^{(\mathrm{DM \; acc)}} - \dot{M}^{(\mathrm{Haw \; rad})} = \frac{4 \pi \rho_\mathrm{NS} M_\mathrm{BH}^2 }{c_s^3} + e_\chi m_\chi C_\mathrm{cap} - \frac{f(M_\mathrm{BH})}{M_\mathrm{BH}^2} \, ,
\end{equation}
where $e_\chi \in [0, 1]$ is an efficiency factor accounting for additional DM falling into the newborn BH, $C_\mathrm{cap}$ is the DM capture rate defined in Equation~\eqref{eq:rate_annihilation}, and $f(M_\mathrm{BH})$ is the Page factor \cite{Page:1976df} that relates the SM degrees of freedom %EE: Please check meaning retained
radiated from the BH to the gray-body corrections \cite{MacGibbon:1990zk, MacGibbon:1991tj}.

On the right-hand side of Equation~\eqref{eq:rate_BH_mass}, the baryonic matter (BM) accretion is assumed to be primarily described by the Bondi process
\cite{1944MNRAS.104..273B, 1954MNRAS.114..437M, 1952MNRAS.112..195B}, a spherical accretion through which a compact object draws in and accretes gas from its surroundings. This process has been widely investigated in \cite{Kouvaris:2013kra, Giffin:2021kgb}, along with possible small deviations from the Bondi behavior. In general, spherical accretion can be influenced by the angular momentum of the material being drawn in. If this initial angular momentum is large, it might lead to an orbital path instead of direct in-fall. However, for an NS with a mass around $\sim$$M_\odot$, the BM's angular momentum is expected to be minimal, meaning it is unlikely to interfere with spherical Bondi accretion onto BHs that originate from DM collapse. East et al. \cite{East:2019dxt} demonstrated that even a near-maximally spinning NS may accumulate baryons with a rate matching Bondi accretion. The middle term on the right-hand side of Equation~\eqref{eq:rate_BH_mass} accounts for any additional in-fall %EE: Please check meaning retained
of DM. In case the DM is self-thermalized, this mechanism can be described as a nearly Bondi process as well; if not, the individual trajectories of DM particles should be considered \cite{Doran:2005vm, Bramante:2013nma}. The last term represents the Hawking evaporation rate \cite{Hawking:1974rv, Ray:2023auh, Basumatary:2024uwo}.

Overall, if $\frac{dM_\mathrm{BH}}{dt} > 0$, the BH grows in the NS; otherwise, $\frac{dM_\mathrm{BH}}{dt} < 0$ yields a partial or total evaporation. Remarkably, the BH's fate is almost entirely determined by its initial mass \cite{McDermott:2011jp, Kouvaris:2011fi}. From Equation~\eqref{eq:rate_BH_mass}, Garani et al. \cite{Garani:2018kkd} obtained lower bounds for the mass of different DM candidates that would cause the evaporation of such a BH as

\begin{equation}
    m_\chi \gtrsim
    \begin{cases}
    3 \times 10^6 \mathrm{\; GeV \; for \; bosons \; that \; do \; not \; form \; a \; BEC;} \\
    16 \mathrm{\; GeV \; for \; bosons \; that \; do \; form \; a \; BEC;} \\
    10^{10} \mathrm{\; GeV \; for \; fermions.}
    \end{cases}
\end{equation}
Experimental searches for DM signatures in NSs have increasingly focused on identifying anomalies in NS populations and characteristics, particularly in regions of high DM density, such as the galactic center or within dwarf spheroidal galaxies. Several experimental areas are worth focusing on to search for BH formation due to DM accretion in NSs. The existence of old pulsars (in our galaxy, many pulsars with ages on the order of several billion years have been observed \cite{Fermi-LAT:2023zzt}; most of these are located within 1 kpc, where the DM density is approximately $\rho_\chi = 0.3 \, \mathrm{GeV/cm^3}$) in regions with known DM density can serve as a constraint on the amount of DM captured by NSs \cite{Kouvaris:2007ay, Kouvaris:2010vv, Bramante:2015dfa} (see left panel in \mbox{Figure \ref{fig:dm_bh_constraints}}). These considerations about old NSs also significantly limit the DM parameter space and the DM--nucleon cross-section (right panel in \mbox{Figure \ref{fig:dm_bh_constraints}}). These exclusion curves have been discussed in \cite{McDermott:2011jp, Bertoni:2013bsa,Garani:2018kkd}, who also provided a similar analysis for muons. Additionally, the observed absence of millisecond pulsars in the Galactic Center of the Milky Way \cite{Perna:2003ck, Dexter:2013xga, Suresh:2022vmf} has been hypothesized to result from NS implosions triggered by DM accumulation \cite{deLavallaz:2010wp, Fuller:2014rza, Bramante:2015cua}. A possible source of r-process elements may be supernovae that follow NSs' collapse \cite{Bramante:2016mzo, Bramante:2017ulk}.

%%%%%%%%%%%%%%%%%%%%%%%%%%%%%%%%%%%%%%%%%%
\section{Dark Matter Admixed Neutron Star Models}
\label{sec:dark_matter_admixed_neutron_star_models}

\begin{figure}[H]
    \centering
    \includegraphics[width=0.6\columnwidth]{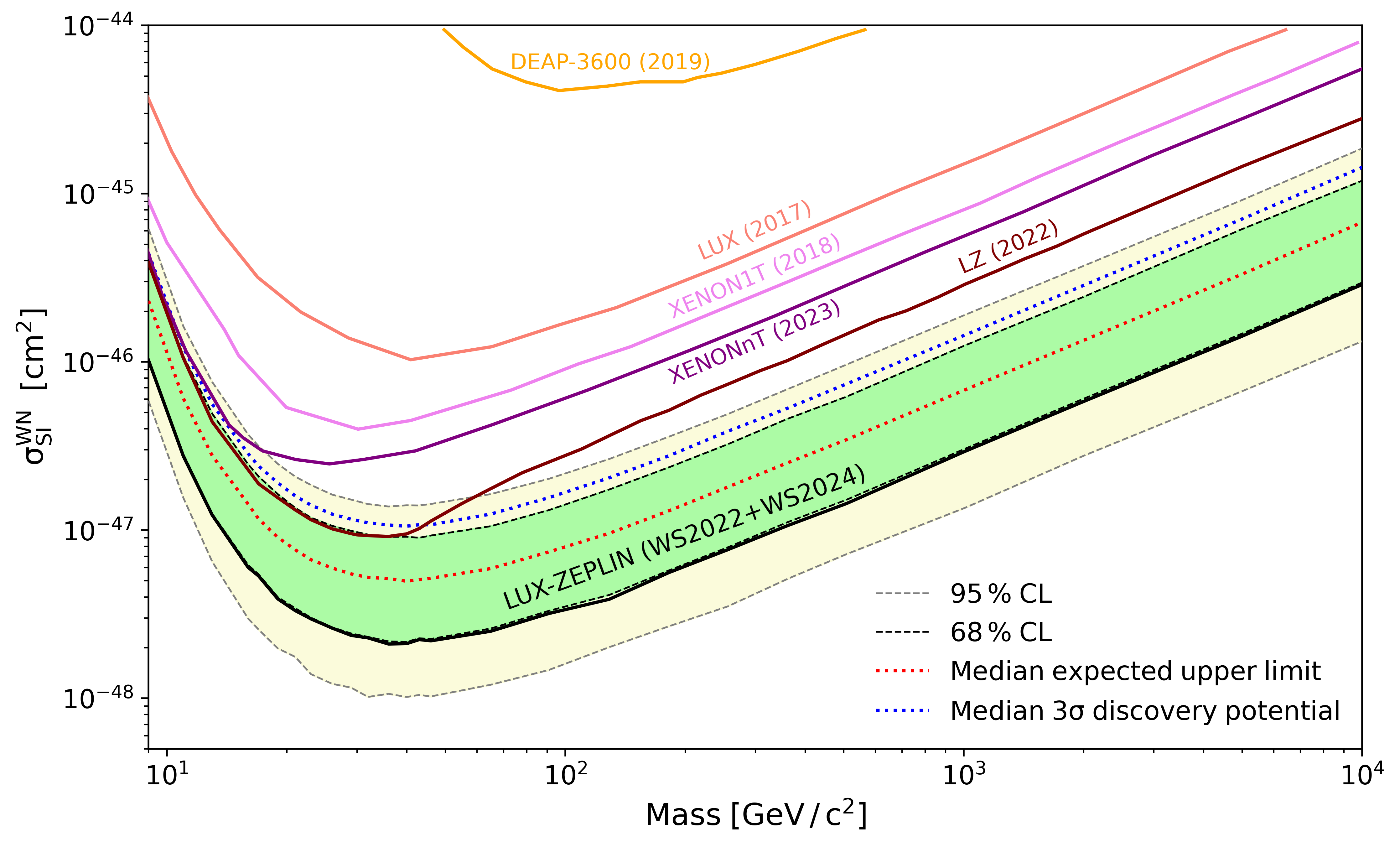}

    \vspace{0.4cm}
    
    \includegraphics[width=0.49\columnwidth]{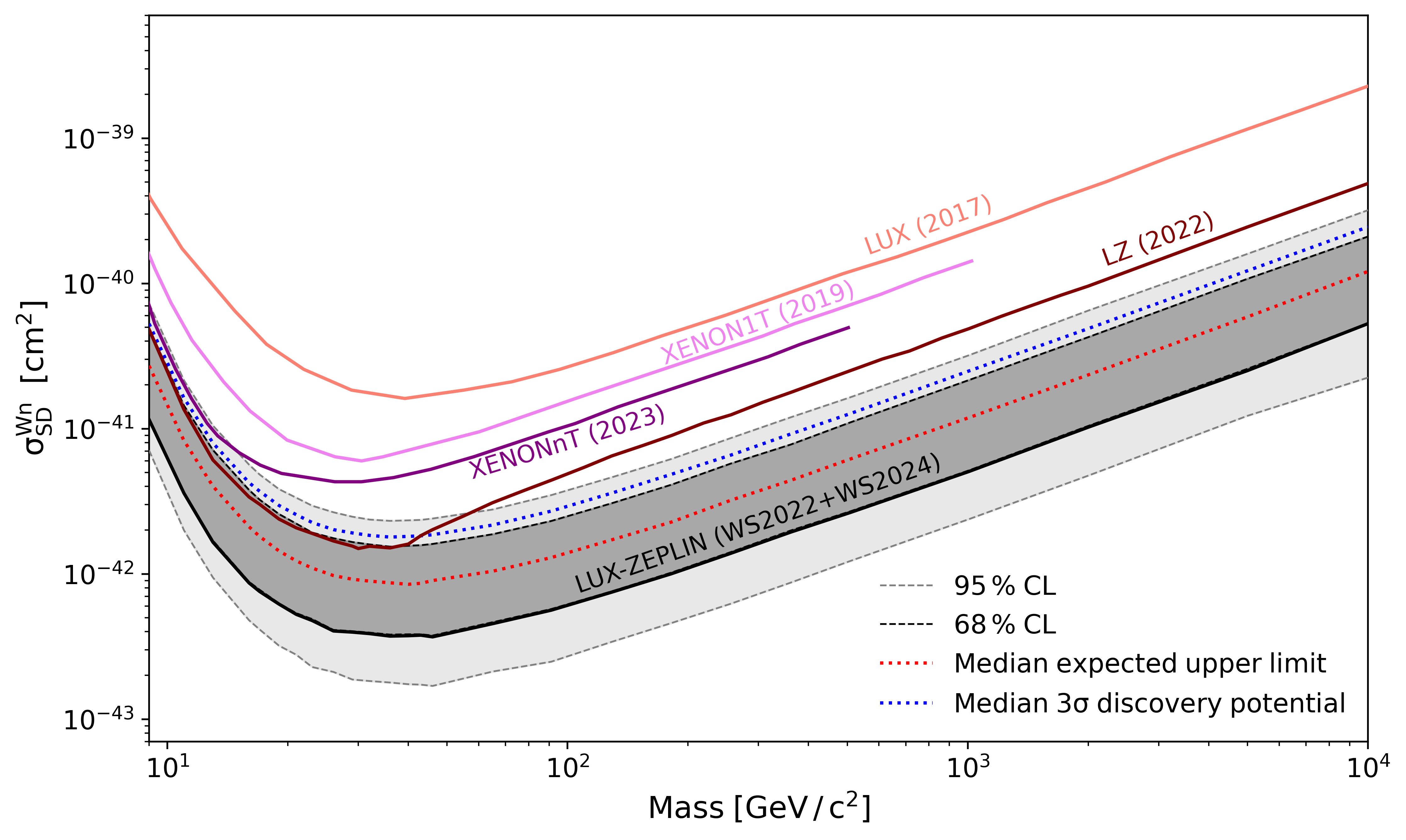}
    \includegraphics[width=0.49\columnwidth]{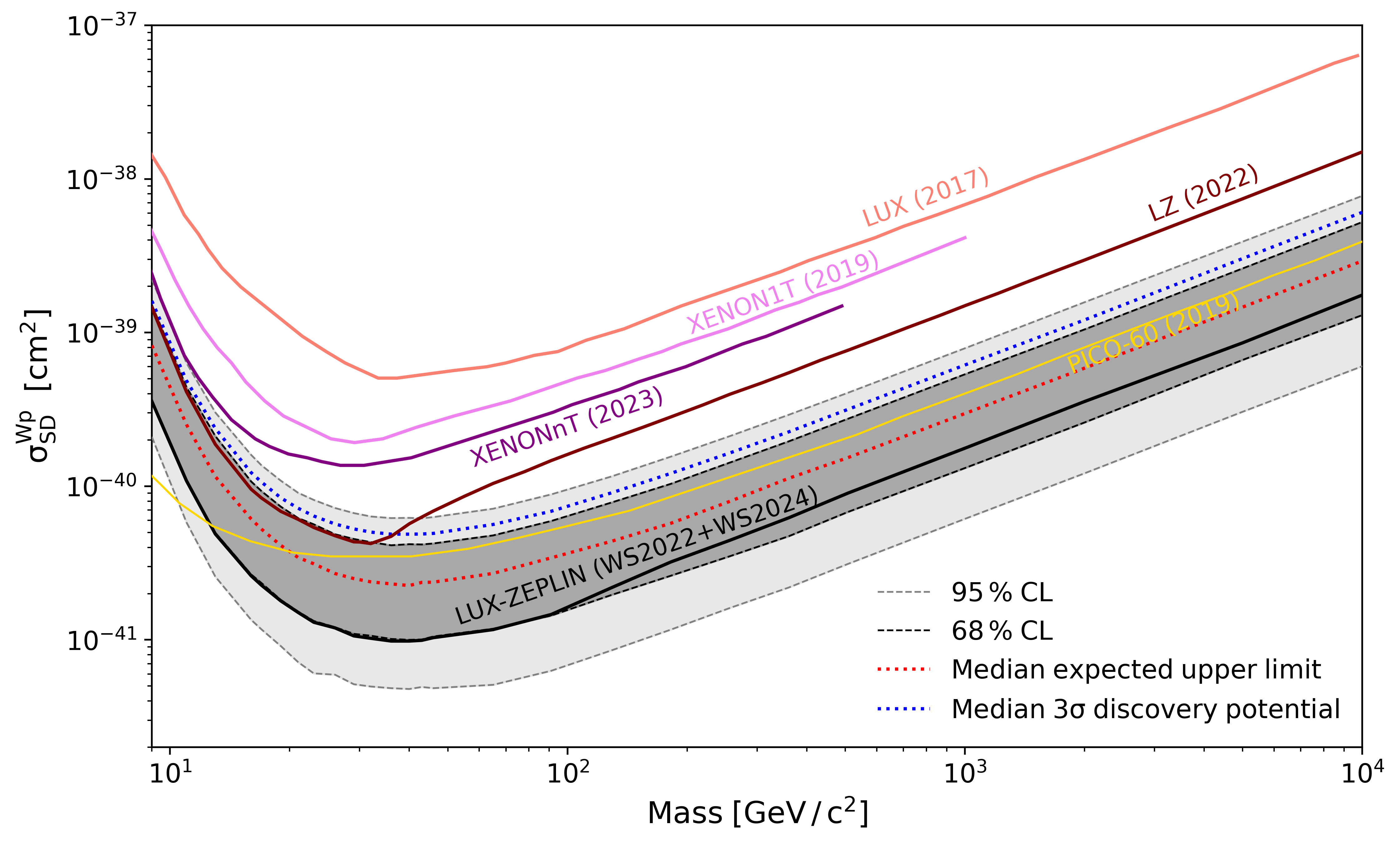}
    
    \caption{Upper limits (68\% and 95\% CL) on the DM--nucleon cross-section for spin-independent (SI) interactions (\textbf{top panel}) as well as spin-dependent (SD) DM--neutron (\textbf{bottom left}) and DM--proton (\textbf{bottom right}) interactions from the latest public data release of the LUX--ZEPLIN collaboration \cite{LZCollaboration:2024lux} (from which the plots are adapted). The results are obtained by combining the first 60-live-day exposure (WS2022) with a new 220-live-day exposure campaign (WS2024). Experimental limits derived from WS2022 only \cite{LZ:2022lsv}, LUX \cite{LUX:2016ggv}, Xenon1T \cite{XENON:2018voc}, XENONnT \cite{XENON:2023cxc}, DEAP3600 \cite{DEAP:2019yzn}, and PICO--60 \cite{PICO:2019vsc} are also shown.}
    \label{fig:sigma_DM}
\end{figure}

DANSs are astrophysical environments where a fraction of DM has been accreted into the stellar volume through the physical mechanisms described in Section \ref{sec:dark_matter_capture_and_accumulation}. To model the influence of DM on NSs, one must first consider how the two types of matter---the baryonic and the dark sectors---interact. While a number of interactions are possible, including direct coupling through forces beyond gravity, null experimental evidence has placed strict limits on non-gravitational interactions between these components. In particular, DM direct detection experiments and the Bullet Cluster \cite{Clowe:2006eq, Randall:2008ppe, MarrodanUndagoitia:2015veg} have constrained the DM-BM cross-section to be many orders of magnitude lower than the typical nuclear one $\sigma_\mathrm{DM-BM} \sim 10^{-45} \, \mathrm{cm^2} << \sigma_\mathrm{nucl} \sim 10^{-24} \, \mathrm{cm^2}$. Figure \ref{fig:sigma_DM} shows the latest experimental limits on the nucleon cross-section for spin-independent DM--nucleon interactions (top panel) and spin-dependent interactions (bottom panel) derived from the publicly available data of the LUX--ZEPLIN collaboration \cite{LZCollaboration:2024lux}. Interestingly, Kouvaris et al. \cite{Kouvaris:2010jy} pointed out that for massive DM candidates, the spin-dependent cross-section can be further constrained from the existence of NSs in globular clusters (see Figure~2 in \cite{Kouvaris:2010jy}, while Figure~1 represents a similar analysis when WDs are assumed to capture DM), that is, by looking at WIMPs' accretion into NSs or in a progenitor supramassive star that later collapses to an NS.

These experimental constraints significantly reduce the likelihood of strong coupling, guiding theoretical efforts toward more conservative approaches. In other words, from the perspective of astrophysical dense environments as NSs, this generally means that any interaction between DM and the ordinary matter of the NS is negligibly small \cite{Nelson:2018xtr, Deliyergiyev:2019vti}. As a consequence, the most common and physically motivated framework to solve DANSs' structure is the two-fluid model; namely, a two-component version of the Tolman--Oppenheimer--Volkoff (TOV) equations \cite{PhysRev.55.374, Tolman:1939jz}, describing the hydrodynamical equilibrium of a spherically symmetric, non-rotating star under general relativity (GR). In DANS models, the two components are the baryonic (neutron-rich) matter and the DM. They are treated as separate fluids, interacting only gravitationally. The two-fluid approach is particularly advantageous for several reasons. First, it respects the minimal interaction scenario, supported by current cosmological and astrophysical observations as well as direct detection experiments. Second, this model is computationally accessible, allowing for simulations of NS structures while incorporating distinct equations of state for both baryonic matter and DM that separately maintain their own thermodynamic properties (i.e., pressure and energy density). Additionally, the two-fluid model enables the exploration of a wide variety of DM candidates (Sections \ref{sec:bosonic_dark_matter} and \ref{sec:fermionic_dark_matter} below). Each of these modifies the stellar structure differently, but the two-fluid framework can be adapted to accommodate these different scenarios.

Overall, the theoretical models analyzing the influence of DM on NS properties must account for both the gravitational coupling between the two components as well as any potential self-interactions within the dark sector. Research in this area has employed different approaches to describe either the baryonic or the dark fluid. While variations in the EoS for BM yield only minor differences due to the stringent observational constraints they must satisfy (see \cite{Burgio:2021vgk} for a comprehensive review on nuclear EoSs), the uncertain nature of DM leads to the consideration of a variety of scenarios, in which DM is either gravitationally captured by the NS or is embedded from the time of the star's formation. Regarding the nature of DM, it can consist of either bosons \cite{Chavanis:2011cz, Giangrandi:2022wht, RafieiKarkevandi:2021hcc, Karkevandi:2021ygv}  or fermions \cite{Narain:2006kx, Goldman:2013qla}. In the former case, the DM candidates form a Bose--Einstein condensate close to the star's center; in the latter, particles may be free or interact through the exchange of scalar and vector dark mediators. Detailed models include work by Leung et al. \cite{Leung:2011zz} and Ivanytskyi et al. \cite{Ivanytskyi:2019wxd}, who modeled DM as a free Fermi gas, while Ellis et al. \cite{Ellis:2018bkr} explored both self-interacting bosonic DM, which can form a Bose--Einstein condensate, and asymmetric fermionic DM. Further studies by Karkevandi et al. \cite{Karkevandi:2024vov, Shakeri:2022dwg} and Liu et al. \cite{Liu:2023ecz, Liu:2024rix} have examined the effects of both bosonic and fermionic DM. Additional research has expanded the range of DM models: Jockel et al. \cite{Jockel:2023rrm} studied fermionic Proca stars; Diedrichs et al. \cite{Diedrichs:2023trk} focused on the properties of ultralight DM; mirror DM was investigated in \cite{Ciancarella:2020msu, Berezhiani:2020zck, Kain:2021hpk, Berezhiani:2021src}; strange stars with mirror DM have been studied in \cite{Yang:2021bpe, Yang:2024sxi, Yang:2024ycl}; finally, Rezaei \cite{Rezaei:2023iif} analyzed fuzzy DM. For the sake of completeness, we report that alternative ideas have been proposed and investigated by  Shahrbaf et al. \cite{Shahrbaf:2022upc, Shahrbaf:2024gdm}, who assumed (bosonic) DM consists of stable sexa-quark states; \emph{scalarization} of DM inside NSs within modified theories of gravity has been considered in \cite{Chen:2015zmx, Morisaki:2017nit, Sotani:2017pfj, Davoudiasl:2024grq}; and superfluid models have been applied to either the DM component \cite{Reddy:2021rln} or the nuclear matter \cite{Acevedo:2019agu, Fujiwara:2023hlj}.

It is also worth mentioning that recently some authors made use of a single-fluid framework. Works like \cite{Panotopoulos:2017idn,Das:2018frc,Das:2020ptd,Das:2021wku,Das:2021yny, Sen:2021wev, Lenzi:2022ypb, Dutra:2022mxl,  Flores:2024hts, Hong:2024sey, Thakur:2024btu} do not consider gravitational interactions, rather they exploit a Higgs portal mechanism to couple DM and BM. DANSs with feeble interactions between DM and hadronic matter have been studied in \cite{Guha:2021njn, Guha:2024pnn, Sen:2021wev}. Alternatively, Refs. \cite{Motta:2018bil, Husain:2022bxl, Shirke:2023ktu, Shirke:2024ymc, Bastero-Gil:2024kjo} assumed that DANSs may form from neutron decay into dark matter (see Section \ref{sec:decay_of_standard_model_particles_into_DM}). These several different approaches highlight the extensive exploration of DM's potential impacts on NSs' characteristics. 

\subsection{Formalism}
\label{sec:formalism}

\textls[-15]{The general relativistic metric describing a static and spherically symmetric \mbox{spacetime is}}

\begin{equation}
\label{eq:metric}
    ds^2 = -e^{\nu(r)}dt^2 + \frac{dr^2}{1-2m(r)/r} + r^2(d\theta^2 + sin^2\theta d\phi^2)\, ,
\end{equation}
where $m(r)$ is the enclosed mass within radius $r$ and $\nu(r)$ is the metric function that decouples for a static metric. The baryonic and dark sectors are usually modeled as perfect fluids, whose stress--energy tensors are

\begin{equation}
    T_i^{\mu\nu} = (\epsilon_i + P_i) u^\mu u^\nu + P_i \eta^{\mu\nu} \, ,    
\end{equation}
where $\epsilon_i$ and $P_i$ ($i = \mathrm{BM}, \mathrm{DM}$) are the energy densities and the pressures of the two fluids separately. From such definitions, the total stress--energy tensor of the DANS is computed as their sum $T_{\mathrm{tot}}^{\mu\nu} = T_\mathrm{BM}^{\mu\nu} + T_{\mathrm{DM}}^{\mu\nu}$. Since the two fluids only interact gravitationally, the conservation law holds separately for each fluid, i.e., $\nabla_{\mu} T_{i}^{\mu\nu} = 0$. Then, one can obtain the coupled TOV equations by solving the Einstein field equations for the metric in Equation~\eqref{eq:metric}. They describe the hydrostatic equilibrium of the DANS and they take \mbox{the form}

\begin{align}
\label{eq:TOV1}
    \frac{dP_i}{dr} &= - (P_i + \epsilon_i) \frac{4 \pi r^3 P_{\mathrm{tot}} + m(r)}{r(r-2m(r))} \, , \\
\label{eq:TOV2}
    \frac{dm(r)}{dr} &= 4 \pi \epsilon_{\mathrm{tot}} r^2 \, ,
\end{align}
where $P_{\mathrm{tot}} = P_\mathrm{BM} + P_{\mathrm{DM}}$, $\epsilon_{\mathrm{tot}} = \epsilon_\mathrm{BM} + \epsilon_{\mathrm{DM}}$ are the total pressure and energy density, respectively.

Equations~\eqref{eq:TOV1} and \eqref{eq:TOV2} form a system of equations that is closed by the equation of state $P_i \equiv P_i(\epsilon_i)$, usually expressed in the barotropic approximation \cite{Bauswein:2012ya}. To numerically solve such a system, one first needs to specify the central density $(\epsilon^\mathrm{c})$ of the star that acts as a boundary condition. Equations~\eqref{eq:TOV1} and \eqref{eq:TOV2} are then integrated outward, starting from the core of the star (where $r=0$) and reaching the surface (defined as the locus of points  where $P_{\mathrm{tot}}(r=R) = 0$; this condition implicitly defines the star's physical radius and ensures that the solution is consistent with the hydrostatic equilibrium of the DANS). 

In addition, it is important to note that solving the equations for two-fluid systems requires specifying the fraction $f_\mathrm{DM}$. While a few studies \cite{Das:2020ecp} express it in terms of the central densities of the two stars, i.e., $f_{\mathrm{DM}} = \epsilon_{\mathrm{DM}}^\mathrm{c} / \epsilon_{\mathrm{BM}}^\mathrm{c}$, most of the works \cite{Karkevandi:2021ygv, Panotopoulos:2017idn, Ivanytskyi:2019wxd} define it as the ratio between the DM mass and the total mass of the star, i.e., $ f_{\mathrm{DM}} = M_{\mathrm{DM}} / M_{\mathrm{star}} = M_{\mathrm{DM}} / ( M_{\mathrm{BM}} + M_{\mathrm{DM}} )$. Fixing this fraction allows for the computation of a one-dimensional mass--radius relation, facilitating comparisons with the M--R curves of ordinary NSs and validating or refuting models. However, this approach can be significantly model-dependent, as it assumes a specific mechanism for DM accumulation. The capture details are sensitive to the nature of DM self-interactions, interactions between DM and standard particles, and other uncertain astrophysical factors.
A more consistent approach would treat the DM fraction as a variable parameter, leading to a two-dimensional space of solutions rather than a single curve \cite{Hippert:2022snq}. While this enhances the exploration of the stellar properties impacted by the DM fraction, it complicates direct comparisons with NS models. This complexity is one reason many authors choose to fix $f_\mathrm{DM}$ in their analyses.

Another key observable that has emerged from the study of NSs, especially in the GW era, is the tidal deformability, a measure of how easily an NS's shape is distorted by an external gravitational field. This property is particularly important in the context of BNS mergers, where tidal interactions between the stars leave an imprint on the GW signal \cite{Radice:2017lry,Raithel:2018ncd,Zhao:2018nyf}. The dimensionless tidal deformability parameter \cite{Hinderer:2007mb, Ellis:2018bkr} is 

\begin{equation}
\label{eq:tidal}
    \Lambda = \frac{2}{3} k_2 \left ( \frac{M}{R}\right)^{-5} \, ,
\end{equation}
where $k_2$ is the $\ell = 2$ tidal Love number \cite{Flanagan:2007ix, Damour:2009vw}, with a typical range from $0.05$ to $0.15$ \cite{Hinderer:2009ca, Postnikov:2010yn}. From \cite{Hinderer:2007mb}, the Love number is computed  as
\begin{eqnarray}
    k_2 &= \frac{8C^5}{5} (1-2C)^2 [2 + 2C (y_R - 1) - y_R] \Big\{2C(6-3y_R + 3C(5y_R-8)) + \nonumber \\ & 4C^3  \big[13-11y_R + C(3y_R-2)+2C^2(1+y_R)\big] \times \nonumber \\ &\times 3(1-2C)^2 \big[2-y_R+2C(y_R-1)\big]\mathrm{\ln}(1-2C) \Big\}^{-1} \,,
\label{eq:k_2}
\end{eqnarray}
where $C = M/R$ is the compactness parameter and $y_R \equiv y (r=R)$ is the solution of the two-fluid differential equation
\begin{equation}
\label{eq:y_R}
    r \frac{dy(r)}{dr} + y(r)^2 + y(r)F(r) + r^2Q(r) = 0\, ,
\end{equation}
where in turn,
\begin{align}
\label{eq:F(r)}
    F(r) &= \frac{r-4 \pi r^3(\rho_{tot} - P_{tot})}{r-2m(r)} \, , \\
\label{eq:Q(r)}
    Q(r) &= \frac{4 \pi r [5 \rho_{tot} + 9P_{tot} + \sum_{i} \frac{\rho_i + P_i}{\partial P_i / \partial \rho_i}  - \frac{6}{4 \pi r^2}}{r - 2m(r)} - 4 \left [ \frac{m(r) + 4 \pi r^3 P_{tot}}{r^2(\-2m(r)/r)}\right]^2 \, .
\end{align}

\subsection{Bosonic DM}
\label{sec:bosonic_dark_matter}

To model bosonic DM, a complex scalar field with mass $m_\chi$ and chemical potential $\mu_\chi$ is considered. These particles are not subject to the Pauli exclusion principle and they can accumulate in close proximity. At sufficiently low temperatures, bosonic DM exists in the form of a Bose--Einstein condensate (BEC) \cite{Li:2012qf, Arbey:2003sj, Suarez:2013iw}. In the absence of interactions, a BEC has zero pressure and is mechanically unstable against gravitational compression. To stabilize the BEC, a repulsive interaction mediated by a real vector field $\omega_\mu$ (that couples with $\chi$) must be introduced. The minimal Lagrangian accounting for this interaction has been discussed in detail by Giangrandi et al. \cite{Giangrandi:2022wht}. This model is equivalent to a massive $U(1)$ gauge theory (like scalar electrodynamics with a massive photon). Thus, the Lagrangian includes the mass and kinetic terms of the fields $\chi$ and $\omega_\mu$ that are coupled through the covariant derivative $D_\mu = \partial_\mu - i g \omega_\mu$ ($g$ is the coupling constant):

\begin{equation}
    \label{eq:bosonic_lagrangian}
    \mathcal{L}_\mathrm{B} = (D_\mu \chi)^* D^\mu \chi - m_\chi^2 \chi^*\chi - \frac{\Omega_{\mu\nu}\Omega^{\mu\nu}}{4} + \frac{m_\omega^2\omega_\mu \omega^\mu}{2} \, .
\end{equation}
In Equation~\eqref{eq:bosonic_lagrangian}, $m_\chi$ and $m_\omega$ are the masses of the scalar and the vector field, respectively, and $\Omega_{\mu\nu} = \partial_\mu \omega_\nu - \partial_\nu \omega_\mu$ is the vector field tensor. 

It is worth noting that other studies have examined variations to this model. Nelson et al. \cite{Nelson:2018xtr} and Rutherford et al. \cite{Rutherford:2022xeb} investigated asymmetric bosonic DM, considering not only repulsive self-interactions but also an additional term $g_\mathrm{B} \omega_\mu J^\mu_\mathrm{B}$, where $g_\mathrm{B}$ represents the interaction strength of $\omega_\mu$ with the SM baryon number current $J^\mu_\mathrm{B}$ and the whole term accounts for small attractive interactions. Alternatively, Karkevandi et al. \cite{Karkevandi:2021ygv} considered a self-interaction potential $V(\chi) = \lambda |\phi|^4$, where $\lambda$ is the dimensionless coupling constant of the scalar field forming the BEC. This modification leads to a different EoS of the form 

\begin{equation}
    P = \frac{m_\chi^4}{9\lambda} \left (\sqrt{ 1 + \frac{3\lambda}{m_\chi^4}} -1 \right) \, .
\end{equation}
Overall, while the qualitative impact of bosonic DM on DANSs remains similar, different models may be adopted due to the unknown nature of DM.

The Lagrangian in Equation~\eqref{eq:bosonic_lagrangian} gives rise a Noether current that corresponds to the invariance of the action under global $U(1)$ transformations $j^\mu = i (\chi^* \partial \chi - \chi \partial \chi^*) + 2g\omega^\mu \chi^* \chi$. Note that the last term in the right-hand side does not contribute to the EoS. Indeed, for the computation of $\epsilon$ and $\rho$, the relevant quantity is the density of conserved charge associated with the DM particles. This density is obtained by averaging the zeroth component of $j^\mu$. When performing this average, $2g\omega^\mu \chi^* \chi$ provides a vanishing contribution.

By assuming cold NSs ($T \sim 0$), the bosonic field totally converts into a BEC, and thermal fluctuations are suppressed. This allows for the use of a relativistic mean-field (RMF) approach \cite{Shen:1998gq, Diener:2008bj} to derive the EoS. In this framework, the vector field $\omega_\mu$ can be replaced by its constant mean value $\langle \omega^\mu \rangle$. After some algebra, the expressions of pressure and energy density are obtained from the Euler--Lagrange equations as

\begin{align}
    \label{eq:e_bosonic}
    \epsilon_\chi = \frac{m_\mathrm{I}^4}{4} \left ( \frac{\mu_\chi^3}{\sqrt{2m_\chi^2 - \mu_\chi^2}} - m_\chi^2 \right ) \, , \\
    \label{eq:pressure_bosonic}
    P_\chi = \frac{m_\mathrm{I}^4}{4} \left ( m_\chi^2 - \mu_\chi \sqrt{2m_\chi^2 - \mu_\chi^2} \right ) \, ,
\end{align}
with $m_\mathrm{I}$ being a fictional mass that controls the strength of the coupling ($m_\mathrm{I} \rightarrow \infty$ when $g \rightarrow 0$ and vice-versa). It can be shown \cite{Giangrandi:2022wht} that no values of $m_\mathrm{I}$ lead to infinite density, %EE: Please check meaning retained
but, to remain consistent with experimental observations from the Bullet Cluster, a conservative bound of $m_\mathrm{I} \, \mathrm{[MeV]} > 18.24 \sqrt[4]{m_\chi \mathrm{[MeV]}}$ is needed to keep the parameter space as broad \mbox{as possible.}

\subsection{Fermionic DM}
\label{sec:fermionic_dark_matter}

Fermionic DM, that must obey the Pauli exclusion principle, is described by a Dirac field $\Psi$. Some studies have modeled DM as a relativistic Fermi gas of non-interacting particles with mass $m_\chi$. Starting from the Lagrangian density of a Dirac field, the corresponding EoS for a cold DANS reads \cite{Shapiro:1983du, Glendenning:1997wn}
\vspace{-3pt}
\begin{align}
    \label{eq:eps_free_fermion}
    \epsilon &= \frac{m_\Psi^4}{2} \, \eta(x) = \, \frac{m_\Psi^4}{2} \frac{1}{8 \pi^2} \left [ x \sqrt{1+x^2} (1 + 2x^2) - \mathrm{ln} (x + \sqrt{1 + x^2}) \right ] \, ,\\
    \label{eq:pressure_free_fermion}
    P &= \frac{m_\Psi^4}{2} \, \Xi(x) = \frac{m_\Psi^4}{2} \, \frac{1}{8 \pi^2} \left [ x \sqrt{1+x^2} \left (\frac{2}{3} x^2 - 1 \right) + \mathrm{ln} (x + \sqrt{1 + x^2}) \right ] \, ,
\end{align}
where $x = k_\mathrm{F} / m_\Psi$ is the dimensionless Fermi momentum.

While this model provides reliable results when coupled with updated nuclear EoSs for BM (APR in \cite{Leung:2011zz}, IST in \cite{Ivanytskyi:2019wxd}, SLy in \cite{Kain:2021hpk}, NL3$\omega\rho$ in \cite{Miao:2022rqj}), there is room for improvement by taking into account interactions with \emph{dark} scalar and/or vector fields. Works such as \cite{Collier:2022cpr, Ellis:2018bkr} included a Yukawa potential of the form $V(r) = \frac{g e^{-m_v r}}{4 \pi r}$ ($g$ represents the DM--mediator coupling constant) that models the self-repulsion mediated by a vector field $V_\mu$ with mass $m_v$. The resulting energy density and pressure acquire an equal additional term $\frac{g^2 m_\Psi^6 x^6}{2 (3\pi^2)^2 m_v^2}$. An even more comprehensive model has been studied in \cite{Das:2020ecp, Routaray:2023spb} (who adopted an RMF Lagrangian for the nuclear sector) and \cite{Grippa:2024sfu} (in combination with Brussels--Montreal energy density functional BSk22 EoS for BM), where interactions with scalar fields were also considered. The full Lagrangian, incorporating both attractive and repulsive interactions, takes the form 
\vspace{-3pt}
\begin{align}
    \nonumber \mathcal{L}_\mathrm{F} &= \bar{\Psi} (x) \left \{ \gamma_{\mu} \left [ i \partial^{\mu}- g_v V^{\mu}(x) \right] - \left[m_\Psi - g_s \phi(x) \right] \right \} \Psi (x) + \\ \label{eq:fermion_lagrangian} &+ \frac{1}{2} \big[ \partial_{\mu} \phi (x) \partial^{\mu} \phi (x) - m_s^2 \phi^2(x) \big ] - \frac{1}{4} V_{\mu\nu}V^{\mu\nu} + \frac{1}{2} m_v^2 V_\mu (x) V^\mu (x),
\end{align}
where $g_v$ and $m_v$ are, respectively, the coupling constants and the mass of the vector field $V_\mu$, $g_s$ and $m_s$ those of the scalar field $\phi$ and $V_{\mu\nu} = \partial_{\mu}V_\nu(x) - \partial_{\nu}V_\mu(x)$.

The exchanges of mesons described by the two terms $g_s\phi\bar{\Psi}\Psi$  and $g_v \bar{\Psi} \gamma_\mu \Psi  V^\mu$ give rise to an effective Yukawa potential,

\begin{equation}
    \label{eq:y_pot_fermion_DM}
    V(r) = \frac{g_{v}^2}{4\pi} \frac{e^{-m_v r}}{r} - \frac{g_s^2}{4\pi} \frac{e^{-m_s r}}{r} \, ,
\end{equation}
whose behavior is governed by the couplings $g_i$ and the masses $m_i$ ($i= s, v$) of the mediators. There is no unique choice for these values, as the DM Lagrangian leads to an EoS that depends only on the ratios $c_i = g_i / m_i$. Typically $c_i \sim \mathcal{O}(1-15) \, \mathrm{GeV^{-1}}$, with $c_s < c_v$ so that the potential is attractive at large distances and repulsive at short distances \cite{Xiang:2013xwa, Das:2020ecp}.

Within the RMF approximation, one can replace the mediator field operators with their ground state expectation values $\left \langle \phi(x) \right \rangle = \phi_0$, $\left \langle V_\mu(x) \right \rangle = V_0$. These values are computed from the Euler--Lagrange equations and depend on the effective mass of the DM fermion candidate, which reduces due to the scalar coupling as $m_\Psi^* = m_\Psi - g_s \phi_0$.

Finally, the energy density and the pressure corresponding to the Lagrangian in Equation~\eqref{eq:fermion_lagrangian} are given by

\begin{align}
\label{eq:energy_density_fermionic}
    \epsilon &= \frac{{m_\Psi^*}^4}{\pi^2} \eta(x) + \frac{g_v^2}{2m_v^2} \rho_\Psi^2 + \frac{g_s^2}{2m_s^2} (m_\Psi - m_\Psi^*)^2 \, , \\
\label{eq:pressure_fermionic}
    P &=  \frac{{m_\Psi^*}^4}{3\pi^2} \Xi (x) + \frac{g_v^2}{2m_v^2} \rho_\Psi^2 - \frac{g_s^2}{2m_s^2} (m_\Psi - m_\Psi^*)^2 \, ,
\end{align}
where the functions $\eta(x)$ and $\Xi(x)$ correspond to the ``free'' terms defined in Equations~\eqref{eq:eps_free_fermion} and \eqref{eq:pressure_free_fermion}. Equations~\eqref{eq:energy_density_fermionic} and \eqref{eq:pressure_fermionic} represent the equation of state for fermionic DM that interacts both with scalar and vector mediators.

As follows from the EoSs in Eqs.~\eqref{eq:e_bosonic}-\eqref{eq:pressure_bosonic} and Eqs.~\eqref{eq:energy_density_fermionic}-\eqref{eq:pressure_fermionic}, the phenomenology of DANSs is largely controlled by a set of poorly known microscopic parameters, whose estimation requires a dedicated statistical framework. In recent years, Bayesian inference has become a widely adopted approach to constrain NS models using astrophysical observations, and has been extensively applied to infer the properties of dense nuclear matter from mass, radius, and tidal-deformability measurements \cite{Huang:2023grj, Grundler:2025mcz, Imam:2025lut, Cartaxo:2025jpi}.

The same framework is even more relevant for bosonic/fermionic DM, whose parameters -- such as the DM candidate mass and the strength of its interactions -- are far less constrained than their baryonic counterparts, and can span wide ranges without being excluded by current observations.

In this context, theoretical information on the microscopic parameters is combined with astrophysical data within the Bayesian inference through Bayes’ theorem

\begin{equation}
    \label{eq:bayes_theorem}
    p \left( \vec{\theta}; \vec{x} \right ) = \frac{\pi \left ( \vec{\theta} \right ) \times \mathcal{L} \left (\vec{x}; \vec{\theta} \right)}{p \left (\vec{x} \right )} \; ,
\end{equation}

where $\vec{\theta}$ and $\vec{x}$ denote the set of model parameters and observational data, respectively. In Eq.~\eqref{eq:bayes_theorem}, $\pi(\vec{\theta})$ and $\mathcal{L}(\vec{x};\vec{\theta})$ represent the prior distribution and the likelihood function, while $p(\vec{x})$, known as evidence, serves only as a normalization.

Therefore, the combination of available data, such as GW signals from the LIGO-Virgo collaboration and NICER constraints on the mass and the radius of NSs, allows to construct posterior distributions for the DM couplings and masses. This strategy has already proven effective in recent studies of bosonic and fermionic DANSs \cite{Grippa:2024sfu, Liu:2025cwy, Arvikar:2025dwl, Santos:2025xep}, and is expected to play an increasingly important role as observational constraints improve.

\subsection{Mass--Radius and Tidal--Mass Relations}
\label{sec:mass_radius_and_tidal_mass_relations}

In this subsection, we summarize the main effects of DM on the mass--radius and tidal--mass relations of DANSs. Indeed, the existence of a DM component can significantly alter the NS’s properties. This happens because the accreted DM can either form a dense core within the NS or a halo surrounding the compact object \cite{Kain:2021hpk}. Unless much extended halos are formed, DM accumulation typically reduces a DANS’s mass. In an NS, all the BM contributes to the internal pressure, whereas in a DANS the DM fraction typically has a negligible effect on this support. As a result, an NS composed solely of BM could exert more pressure and sustain a higher mass.

\begin{figure}[H]
    %\centering
    \includegraphics[width=0.48\textwidth]{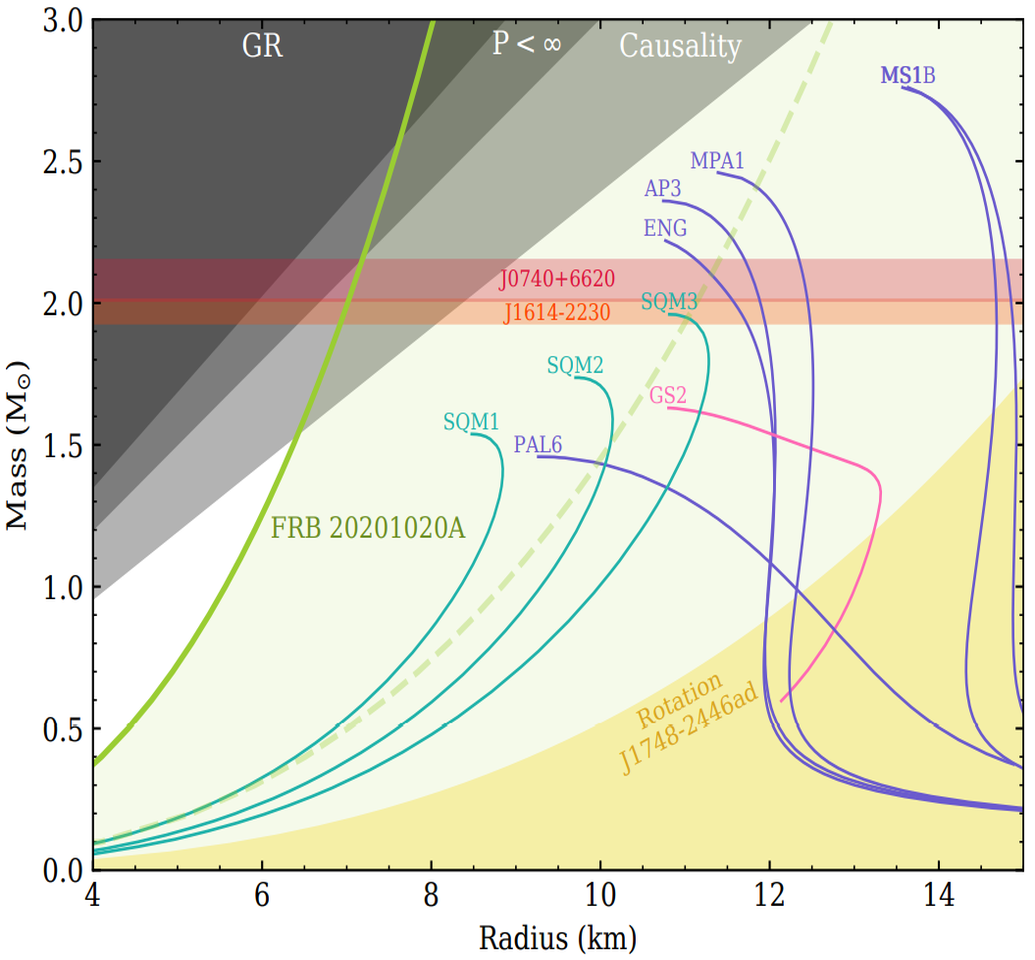}
    \hspace{0.3cm}
    \includegraphics[width=0.48\textwidth]{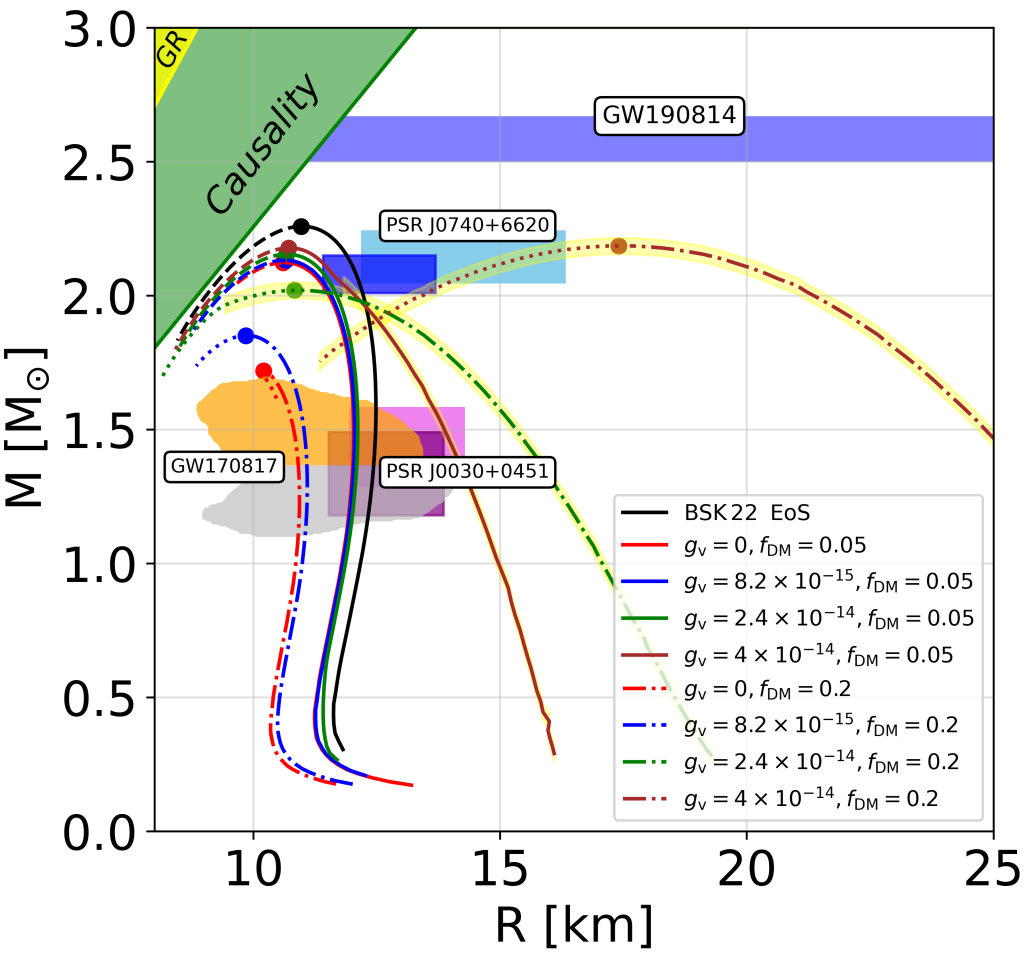}
    \caption{\textbf{Left}: Mass--radius relations for different equations of state of BM (solid lines). The gray-shaded regions are excluded due to GR ($R > 2GM /c^2$), finite pressure ($R > 9GM /(4c^2)$), and causality ($R > 8GM /(3c^2)$) bounds \cite{Bucciantini:Lecture_notes}, respectively. The rotation of the fastest spinning pulsar, J1748-2446ad \cite{Hessels:2006ze}, excludes the yellow-shaded area, whereas the red and orange strips correspond to the masses of the two heaviest pulsars: J0740+6620 \cite{NANOGrav:2019jur, Fonseca:2021wxt} and J1614-2230 \cite{Demorest:2010bx}. The figure is taken from \cite{Pastor-Marazuela:2022pnp}. \textbf{Right}: Mass--radius diagram when DM accumulates within an NS. The plot is adapted from \cite{Grippa:2024sfu}, where fermionic, interacting DM described by the Lagrangian in Equation~\eqref{eq:fermion_lagrangian} is considered. The black curve corresponds to the BSk22 nuclear EoS, namely, without DM. The other solid lines represent the mass--radius relations with different vector coupling $g_v$ and/or DM fractions $f_\mathrm{DM}$. Configurations highlighted in yellow represent DANSs with a DM halo, otherwise a DM core forms. The blue strip is drawn for GW190814 \cite{LIGOScientific:2020zkf}; the orange and gray areas correspond to the estimates of the masses of the NSs involved in GW170817 \cite{LIGOScientific:2017vwq, LIGOScientific:2018cki}; the magenta and purple regions come from the 2019 NICER data of PSR J0030+0451 from Miller et al. \cite{Miller:2019cac} and Riley et al. \cite{Riley:2019yda}; and the dark and light blue areas represent the 2021 NICER data of PSR J0740+6620 from the same groups \cite{Miller:2021qha, Riley:2021pdl}.}
    \label{fig:mass_radius}
\end{figure}

\begin{figure}[H]
    \centering
    \includegraphics[width=0.5\textwidth]{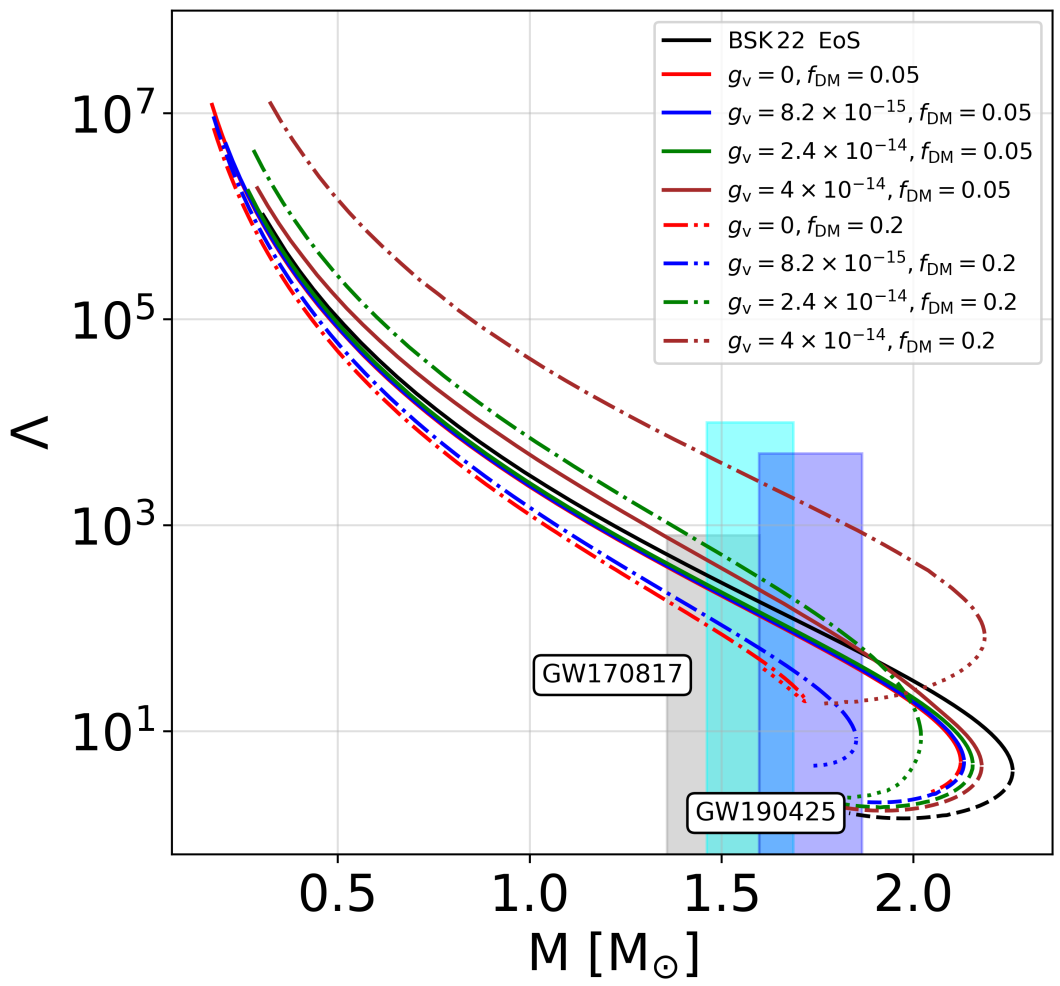}
    \caption{$\Lambda$--M relations for a DANS with fermionic DM described by the Lagrangian in Equation~\eqref{eq:fermion_lagrangian}. As in Figure \ref{fig:mass_radius}, the black curve is obtained when no DM accumulates, whereas changing $g_v$ and/or $f_\mathrm{DM}$ provides significant modifications. The gray region represents the 90\% confidence upper bound from GW170817 \cite{LIGOScientific:2018hze}, and the light and dark blue regions correspond to the same bounds for the primary and secondary compact objects of GW190425 \cite{LIGOScientific:2020aai, Yang:2022ees} (assuming low-spin priors). Plot is adapted from \cite{Grippa:2024sfu}.}
    \label{fig:tidal_mass}
\end{figure}

In detail, we compare the M--R (Figure \ref{fig:mass_radius}) and $\Lambda$--M (Figure \ref{fig:tidal_mass}) diagrams for nuclear equations of state (left panels) with DANS models (right panels). Fermionic, interacting DM is assumed to accumulate in the NSs, as studied in \cite{Grippa:2024sfu}.

In the right panels of Figure \ref{fig:mass_radius}, fixed masses of dark particles are considered. In particular a nucleon-like DM candidate (this is particularly interesting because such a DM particle may be related to the mirror particle of ordinary BM \cite{Ciarcelluti:2010ji,Khlopov:1989fj}) with \mbox{$m_\Psi = 1$ GeV} and ultralight mediators with $m_s = 10^{-14}$ eV and $m_v = 10^{-6}$ eV. These choices result in $c_s \sim 5 \, \mathrm{GeV^{-1}}$ and $c_v \sim 10 \, \mathrm{GeV^{-1}}$, so that the Yukawa potential in Equation~\eqref{eq:y_pot_fermion_DM} is attractive at large distances and repulsive at small distances. Depending on the strength of the coupling $g_v$ and the amount of captured DM $f_\mathrm{DM}$, DM can form a core (unstressed points), a halo (shadowed yellow points), or both along an entire M--R curve. A transition between a DM core and halo occurs as $f_\mathrm{DM}$ increases.

Increasing the coupling $g_v$ and the DM fraction $f_\mathrm{DM}$ enhances the amount of DM settling in the NS's core and the star's self-attraction intensifies. The same happens with increasing scalar coupling $g_s$; similar effects of $g_s$ and $g_v$ on DANSs are anticipated because the scalar and the vector contributions in Equations~\eqref{eq:eps_free_fermion} and \eqref{eq:pressure_free_fermion} share the same form and combine additively. Since it is assumed that DM does not interact with BM, it does not significantly contribute to the degeneracy pressure that counters the contraction. The weak interactions among DM fermions yield a less dense packing compared to BM within NSs. The resulting degeneracy pressure adds a negligible contribution to the BM. Consequently, the outward pressure can support a lower mass, leading to DANSs that are less massive and more compact (i.e., with smaller radii). Note that the radius $R_\mathrm{DANS}$ of the DANS corresponds to the outermost radius. If a DM core forms, then $R_\mathrm{DANS} = R_\mathrm{BM}$ because the DM remains entirely within the star. Otherwise, if a DM halo surrounds the star, $R_\mathrm{DANS} = R_\mathrm{DM}$ because DM extends beyond the ``baryonic region''. In both scenarios, however, the visible radius still \mbox{remains $R_\mathrm{BM}$.}

According to the EoS in Equations~\eqref{eq:energy_density_fermionic} and \eqref{eq:pressure_fermionic}, the interaction with the dark vector field contributes additively to the pressure. As such, if a DM core is formed, the maximum mass slightly increases as $g_v$ becomes greater, though always being below the TOV limit in the absence of DM (black curve). This happens because now more pressure can counteract the gravitational force pushing inward. However, if $g_v$ and/or $f_\mathrm{DM}$ are increased further, DM no longer settle in the core; rather, a halo surrounding the DANS forms. Interestingly, as for bosonic DM \cite{Karkevandi:2021ygv}, a core--halo transition may occur along the same M--R curve. For given $m_\Psi$, $g_i$, and $f_\mathrm{DM}$, the outermost radius may interchange between $R_\mathrm{BM}$ and $R_\mathrm{DM}$. Configurations exhibiting a DM halo possess much larger radii, although the maximum mass is compatible. While DM halos do not directly impact the optical radius measured by NICER, their inclusion alters the overall mass and radius profile of the NS, which could affect the compatibility of the model with multi-messenger observational constraints.

Interestingly, a small variation in the coupling is sufficient to produce significant changes in the M--R relations. This feature aligns with findings in \cite{Collier:2022cpr}. The vector (and scalar) coupling constants are characterized by a narrow interval within which they produce relevant effects. If the couplings are too large, the pressure might become negative, making the EoS nonphysical. Conversely, if the couplings are too small, the interactions would become ineffective, virtually resulting in a free Fermi gas, described by Equations~\eqref{eq:eps_free_fermion} and \eqref{eq:pressure_free_fermion}. Moreover, the monotonic increase in the stellar mass with greater coupling(s) emerges because both $\epsilon$ and $P$ are functions of the ratio $c_i$. 

While the masses of the DM candidates have been considered fixed so far, the effects of a variable $m_\chi$ have been demonstrated by Routaray et al. \cite{Routaray:2023spb}. Their findings indicate that lighter DM particles ($\sim$$300$ MeV) are more likely to form extended DM halos, whereas heavier DM candidates tend to be confined to the core of DANSs. Furthermore, they stressed that with intermediate DM mass ($\sim$$500$ MeV) cores form only with high $f_\mathrm{DM}$.

Regarding bosonic DM, as highlighted by Ellis et al. \cite{Ellis:2018bkr}, once a positive pressure is established for the DM component to prevent the formation of a BH at the center of the NS, both bosonic and fermionic DM produce similar effects on DANSs. This was found in \cite{Giangrandi:2022wht}, where bosonic DM is coupled with nuclear EoSs as IST, DD2, and $\mathrm{DD2}\Lambda$ (the latter accounting for the presence of hyperons that soften the EoS). In all three cases, when considering a DM candidate with mass $1$ GeV and relatively small DM fractions $f_\mathrm{DM}$ (1\%, 3\%, and 5\%), a DM core is formed within the NS that leads to a decrease in both the mass and the radius. As $f_\mathrm{DM}$ increases, the core becomes heavier, and $M_\mathrm{DANS}$ and $R_\mathrm{DANS}$ decrease further. Thus, the presence of a DM core effectively softens the baryonic EoS. However, if too much DM is accumulated near the core and/or light DM candidates are considered ($\sim$$100$ MeV), a core no longer exists; rather, a halo forms due to the weaker gravitational interaction.

Such a trend is confirmed even when considering different EoSs for BM. For instance, Ref. \cite{Liu:2023ecz} derives the BM EoS from a Brueckner--Hartree--Fock approach, covering all $f_\mathrm{DM} \in [0, 1]$. Their findings confirm that massive DM particles, low fractions, and weak couplings favor core formation and, in turn, result in smaller and denser DANSs. Additionally, they found a couple of interesting outcomes. First, stable high-$f_\mathrm{DM}$ stars can be produced under specific conditions: the hosting NSs must be very light (with mass $\leq 0.5$ \msun if $f_\mathrm{DM} \sim 0.9$---see Figure 3 in \cite{Liu:2023ecz}) and DM should accumulate through exotic mechanisms such as accretion of normal matter onto a dark star \cite{Ciarcelluti:2010ji, Ellis:2018bkr, Gleason:2022eeg}. Second, twin DANSs may exist, namely, a low-$f_\mathrm{DM}$ star with a DM core and a high-$f_\mathrm{DM}$ star with a DM halo both with $M_\mathrm{DANS} \sim 0.5$ \msun and $R_\mathrm{DANS} \sim 12$ km. These twins would be indistinguishable based solely on mass and optical radius measurements, yet their internal structure and related properties would be fundamentally different (due to the way DM is distributed). %EE: Please check meaning retained

Similar trends are observed in \cite{Nelson:2018xtr, Karkevandi:2021ygv, RafieiKarkevandi:2021hcc} and other studies. A few differences may include the following: the ``threshold'' DM mass above which only DM cores %EE: Please check meaning retained
form (with a realistic radius) and the possible existence of intermediate regions where the behavior of the maximum mass is not constant due to the core--halo transition. For instance, Karkevandi et al. \cite{Karkevandi:2021ygv} found that, within their model, if $ 105 \, \mathrm{MeV} < m_\chi < 200 \, \mathrm{MeV}$, then the maximum mass decreases for low fractions because a core is formed, but increases for larger $f_\mathrm{DM}$ as a halo forms, potentially exceeding 2 \msun. Conversely, if $m_\chi \geq 200$ MeV, only a core forms and, even with very large fractions, the maximum mass never reaches 2 \msun. 

We emphasize that for both fermionic and bosonic DM the potential shift in the M--R curve induced by the accumulation of DM could significantly aid in detecting its presence within NSs. In this regard, Ciarcelluti et al. \cite{Ciarcelluti:2010ji} demonstrated that the stellar properties of the sources 4U 1608-52 \cite{Guver:2008gc}, 4U 1820-30 \cite{Guver:2010td}, and EXO 1745-248 \cite{Ozel:2010fw} are more effectively explained by incorporating a heavy DM core (at least 25\%) rather than relying solely on pure nucleonic EoSs. However, without additional supporting evidence, the characteristic M--R curves alone cannot be considered definitive proof for the existence of DANSs. Several analyses have found that, within certain ranges of DM parameters, DM models can be degenerate with pure nuclear EoSs. This could occur for several reasons. First, the DM fraction may be so small that no detectable effects arise. Second, two-fluid DM models and nuclear EoSs can both be compatible with observations. Scordino et al. \cite{Scordino:2024ehe} estimated that for a typical $1.4 M_\odot$ NS with a DM fraction of $f_\mathrm{DM} = 6\%$, the observable stellar radius is reduced by approximately 0.6 km (around 5\%). These changes fall within the current observational uncertainties of instruments such as NICER, which measures NSs radii. Also in the model presented here (Figure \ref{fig:mass_radius}), the black curve, representing a pure NS described by the BSk22 EoS, and all the 5\% DM models with $g_v \sim 10^{-14}$, are simultaneously consistent with experimental constraints posed from GW170817, PSR J0740+6620, and PSR J0030+0451. Third, the effects of DM on observable properties are not easy to isolate because they may mimic those caused by exotic baryonic states or variations in the nuclear EoS. As described, the softer EoS caused by DM accumulation leads to more compact DANSs with smaller radii, but the same effect might also arise from the inclusion of hyperons \cite{Zdunik:2003vg, Weissenborn:2011kb, Vidana:2022tlx} or deconfined quarks \cite{Schertler:2000xq, Baym:2017whm} in pure nuclear models.

On the other hand, these analyses are not only valuable for strengthening the case for the presence of DM but also for ruling out some combinations of baryonic and DM models. For instance, Ellis et al. \cite{Ellis:2018bkr} demonstrated that well-established nuclear EoSs, such as ALF2, APR4, ENG, H4, and SLy4, are incompatible with the mass measurement of PSR J0348+0432 if the accreted DM fraction exceeds 5\%.

The (dimensionless) tidal deformability $\Lambda$ is computed by means of Equation~\eqref{eq:tidal}, where $M \equiv M_\mathrm{DANS}$ and $R \equiv R_\mathrm{DANS}$, i.e., the outermost radius. As for the mass--radius relations, in Figure \ref{fig:tidal_mass} we report how a nuclear EoS (as BSk22) is modified if DM accumulates within the star according to the afore-mentioned model of fermionic, interacting DM. For a given total gravitational mass, the presence of DM condensed in a core results in a smaller tidal deformability \cite{Ivanytskyi:2019wxd,Liu:2024rix} compared to a pure NS. When a DM core forms, both the mass and radius of the NS decrease. However, in the scaling relation $\Lambda \propto (M_\mathrm{DANS}/R_\mathrm{DANS})^{-5}$ (Equation~\eqref{eq:tidal}), the reduction in radius has a dominant effect over the reduction in mass, resulting in an overall decrease in tidal deformability. From the perspective of a distant observer, these changes would be interpreted as an effective softening of the EoS. Indeed, with a softer EoS less pressure can be exerted to balance the gravitational self-interaction, so the star becomes more compact with a smaller radius. This increased compactness results in the star being more resistant to deformation, thus lowering its $\Lambda$. However, as pointed out in \cite{Giangrandi:2022wht}, this effect may be not easy to identify. As for the M--R curves,  the modifications in the $\Lambda$--M relations alone are not sufficient to establish the presence of DM within NSs. Indeed, both the DM capture and the presence of an additional degree of freedom (such as hyperons) can contribute to the softening of the EoS. As a consequence, distinguishing these scenarios based on the measurement of $\Lambda$ and/or the observed radius mass would be challenging. In other words, despite completely different internal structures, the same mass and radius could correspond to two distinct situations: a DANS with a certain fraction of DM and a pure NS containing hyperons, SQM, or other new particles. In contrast, the existence of a DM halo leads to a substantial rise in the tidal deformability because of the significant increase in  $R_\mathrm{DANS}$.

As for the dependence on the coupling, $\Lambda$ increases monotonically with larger $g_v$, eventually favoring the formation of a DM halo. Das et al. \cite{Das:2020ecp} have emphasized this evidence is important in differentiating a single-fluid model from a two-fluid approach when describing DANSs. Indeed, from the analysis in \cite{Das:2018frc} (where DM is coupled to BM through a Higgs portal interaction, forming one fluid), the stellar properties, including $\Lambda$, always decrease in the presence of DM. In contrast, in the two-fluid scenario, $\Lambda$ can rise beyond the observational limits when a halo extends over $\sim$$20$ km. Due to such a great radius, the resulting curves do not usually match experimental observations, making configurations with a halo disfavored by these DM models.

\subsection{Speed of Sound in Dark Matter Admixed Neutron Stars}
\label{sec:speed_of_sound_in_dans}
The stiffness of matter in compact objects is described by the sound speed $c_s$, which plays a critical role in preventing a static NS from collapsing into a BH. Unlike the mass or the radius, the sound speed cannot be directly measured by experiments, nor it is easily inferred. However, the speed of sound is still a key parameter for characterizing the NS properties and, in particular, for quantifying how resistant the matter is to compression in extreme conditions. In this regard, Ecker et al. \cite{Ecker:2022xxj} have demonstrated that in light stars (whose $c_s$ increases gradually towards the center), the outer layers are soft while the core is rather stiff; whereas heavy NSs have a stiffer outer shell and a softer core (where the sound speed reaches a local minimum).

By definition, the sound speed represents the velocity at which linear perturbations propagate in a uniform fluid \cite{Rezzolla:2013dea} and can be derived from the EoS  as $c_s^2 = \left ( \frac{\partial P}{\partial \epsilon} \right)_s$, that is, assuming thermodynamic equilibrium and computing the derivative at constant specific entropy $s$. This definition implies that a stiffer EoS leads to a larger $c_s$, whereas smaller sound speeds are obtained with softer EoSs. Since NSs are not uniform spheres of constant density, $c_s$ varies as one moves from the surface to the core. Due to the uncertainty about the nuclear EoS governing the interiors of NSs, the exact trend in $c_s$ remains a crucial open question for understanding dense matter properties and GW emission from \mbox{compact objects.}

Minimal physical constraints on the sound speed are posed by causality \cite{Bludman:1968zz} and thermodynamic stability, requiring $0 \leq c_s \leq 1$. Beyond that, theoretical constraints can be derived. At low densities $n \lesssim n_0 = 0.16 \, \mathrm{fm^{-3}}$ ($n_0$ is the nuclear saturation number density), the condition $c_s << c$ has been well established \cite{Hebeler:2013nza, Altiparmak:2022bke}. Conversely, at high densities ($n >> n_0$) exotic states of matter (like free quarks and gluons) may exist, making it hard to predict the profile of $c_s$. In this regime, an important open question is whether the so-called sub-conformal limit $c_s < c/\sqrt{3}$ should hold for the NS's EoS \cite{Bedaque:2014sqa, Hoyos:2016cob, Landry:2020vaw, Roy:2022nwy}. Such a limit reflects the restoration of the conformal symmetry of quantum chromodynamics (QCD) at asymptotically large densities \cite{Kurkela:2009gj, Fraga:2013qra}. Hence, in conformal theories, the sound speed is expected to approach the value $c/\sqrt{3}$, realized in ultrarelativistic fluids from below \cite{Tews:2018kmu, Greif:2018njt}. Some studies argue for violations of this limit \cite{Kojo:2014rca, Alsing:2017bbc, BitaghsirFadafan:2018uzs, McLerran:2018hbz}, which is further supported by its tension with measurements of the most massive NSs \cite{Demorest:2010bx, Antoniadis:2013pzd, Fonseca:2021wxt} as well as theoretical predictions on the maximum (gravitational) mass \cite{Bauswein:2017vtn, Margalit:2017dij, Rezzolla:2017aly, Ruiz:2017due, Shibata:2017xdx, Shibata:2019ctb, Nathanail:2021tay} and from QCD \cite{Altiparmak:2022bke} (and references therein).

DM accumulation inside NSs alters the total pressure and energy density, thus affecting $c_s$. While the literature lacks comprehensive analyses of DM's influence on DANSs' speed of sound, an important contribution comes from  Giangrandi et al. \cite{Giangrandi:2022wht}, who derive an effective sound speed for a two-fluid system of (bosonic) DM and BM:

\begin{equation}
\label{eq:effective_sound_speed}
    c_{s\mathrm{, \, eff}}^2 = \eta c_{s\mathrm{, \, BM}}^2 + (1 - \eta) c_{s\mathrm{, \, DM}}^2 \, ,
\end{equation}
where $c_{s\mathrm{, \, BM}}^2$ and $c_{s\mathrm{, \, DM}}^2$ are the speed of sound for the baryonic and the DM fluid, respectively. %EE: Please check meaning retained
In Equation~\eqref{eq:effective_sound_speed}, $\eta$ is a parameter related to the thermodynamic properties of the \mbox{two components}:

\begin{equation}
    \label{eq:eta_sound_speed}
    \eta = \frac{\partial \epsilon_\mathrm{BM}}{\partial \mu_\mathrm{BM}} \left ( \frac{\partial n_\mathrm{BM}}{\partial \mu_\mathrm{BM}} + \xi^2 \frac{\partial n_\mathrm{DM}}{\partial \mu_\mathrm{DM}} \right )^{-1} \, ,
\end{equation}
where $\xi = \mu_\mathrm{DM}/\mu_\mathrm{BM}$ is the ratio of chemical potentials. This formulation ensures that $\eta \in [0, 1]$ and, in turn, $c_{s\mathrm{, \, eff}}^2$, lies between the ratios of the %EE: Please check meaning retained
pure components. In particular, its lower value is obtained for a pure DM fluid ($\eta = 0$); whereas the upper edge, reached with $\eta = 1$, corresponds to an ``ordinary'' NS of only BM.

\begin{figure}[H]
    %\centering
    \includegraphics[width=0.47\columnwidth]{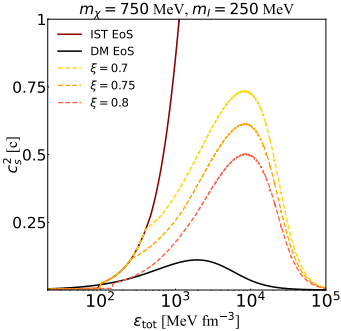}
    \hspace{0.4cm}\includegraphics[width=0.47\columnwidth]{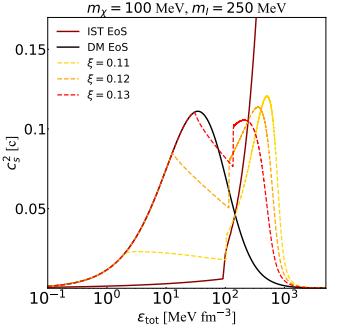}
    \caption{The (squared) effective sound speed $c_{s\mathrm{, \, eff}}^2$ versus total energy density $\epsilon_\mathrm{tot}$ for DANSs with a DM core (\textbf{left}) or a DM halo (\textbf{right}) for two different values of the DM candidate mass $m_\chi$ and the fictional mass $m_I$ defined in Equations~\eqref{eq:pressure_bosonic} and \eqref{eq:e_bosonic} according to Giangrandi's model \cite{Giangrandi:2022wht} (from which the plots are taken). In the scenario of DM core formation, $c_{s\mathrm{, \, eff}}^2$ is in between the speed of sound of a pure fluid of BM (red curve, derived from the nuclear IST EoS) and DM (black curve). The effective sound speed decreases with a larger DM fraction (represented by the parameter $\xi$ defined in Equation~\eqref{eq:eta_sound_speed}), and the discrepancy increases with greater energy density, meaning that near the core, where DM is accumulated, the DANS's EoS is significantly softer. In contrast, the presence of a DM halo makes $c_{s\mathrm{, \, eff}}^2$ close to the sound speed of a BM fluid at large densities, while in the crust (i.e., near the surface) the DM begins to dominate, as expected.}
    \label{fig:sound_speed_Giangrandi}
\end{figure}

Figure \ref{fig:sound_speed_Giangrandi} \cite{Giangrandi:2022wht} clarifies the influence of DM on the NS's effective sound speed. In detail, if a DM core forms (left panel in Figure \ref{fig:sound_speed_Giangrandi}), then $c_{s\mathrm{, \, eff}}^2$ is significantly smaller than $c_{s\mathrm{, \, BM}}^2$ and the DANS's EoS becomes softer overall. As expected, the difference is more pronounced moving towards the core, where DM is accumulated. In contrast, if a DM halo forms (right panel in Figure \ref{fig:sound_speed_Giangrandi}), no significant distinctions arise between $c_{s\mathrm{, \, eff}}^2$ and $c_{s\mathrm{, \, BM}}^2$ as long as the density is large enough. However, near the crust and up to the surface (i.e., at low energy densities), the DM component begins to dominate, consistently having a halo configuration.

A similar analysis by Thakur et al. \cite{Thakur:2024btu} examined fermionic DM accreted through neutron decay (see Section \ref{sec:decay_of_standard_model_particles_into_DM}). Their findings, along with the previous discussion, show that the presence of DM %EE: Please check meaning retained
softens the EoS at intermediate densities, reducing the sound speed compared to a pure BM NS and meaning that DM tends to settle close to the NS's core.

Before ending this section, we emphasize that significant knowledge about extremely dense matter in compact objects can be derived from the sound speed because its own definition is directly related to the EoS. Analyzing the trend in $c_s$ allows for comparison of theoretical models that incorporate DM and those that exclude it with observations in order to constrain the DM parameter space and to support/rule out some EoS frameworks.

However, as for the M--R and $\Lambda$--M relations discussed in Section \ref{sec:mass_radius_and_tidal_mass_relations}, some degeneracy between purely baryonic stars and DANSs may arise. Again, this would manifest because baryonic models, with suitable parameter adjustments, can replicate variations in $c_s$ that resemble those predicted in DANSs. Hence, a comprehensive approach that synthesizes both observational (a robust dataset containing information on mass, radius, $\Lambda$, $c_s$ and thermal evolution) and theoretical insights is needed to resolve this ambiguity and identify the distinct signatures of DM in NSs.

%%%%%%%%%%%%%%%%%%%%%%%%%%%%%%%%%%%%%%%%%%
\section{Dark Matter Signal in Gravitational Wave Emission}
\label{sec:dark_matter_signals_in_gravitational_wave_emission}

As DM accumulates in NSs, it can alter their structure and potentially leave a distinct signature in the GW emission from binary systems observed by ground-based detectors such as Virgo \cite{VIRGO:2014yos}, LIGO \cite{LIGOScientific:2014pky}, and KAGRA \cite{Somiya:2011np}. Analytical studies \cite{Ellis:2017jgp, Kopp:2018jom, Bhattacharya:2023stq, Croon:2017zcu, Choi:2018axi, Alexander:2018qzg} have computed the contribution of DM to the GW spectrum. If DANSs involved in coalescence contain significant amounts of DM, interaction between their dark components via long-range forces may occur. To remain as model-independent as possible, these interactions can be parametrized by a Yukawa-type potential \cite{Alexander:2018qzg}, expressed as $V_Y(r) = \alpha \frac{m^2 \eta}{r} e^{-r/l}$, where $l = m_\chi^{-1}$ is the length scale of the interaction, $m = m_1 + m_2$ is the binary's total mass, $\eta = m_1 m_2/m$ the reduced mass of the inspiraling compact objects, $r$ the orbital separation, and $\alpha = \pm (q_1 q_2)/(m_1 m_2)$ represents the strength of the potential relative to gravity (with the positive sign in case of an attractive force, whereas the negative sign means repulsion). During the (early) inspiral, the DANSs can be modeled as point masses/charges and, as long as the distance between the companions greatly exceeds $l$, the system’s GW emission follows the predictions of GR. However, once their relative distance dropped below the range of that force, the Yukawa interaction would manifest and modify the potential energy of the binary system. This yields a couple of important observable effects. First, detectors would observe a modification in the chirp mass $\mathcal{M} = \eta^{3/5} (m_1 + m_2)^{2/5}$. The additional \emph{dark} force influences the rate of orbital decay by affecting the time evolution of the system's frequency and the time of the merger. This, in turn, would cause deviations in the GW waveforms that are degenerate with a shift in the DANS's masses. This degeneracy is further explored in \cite{Croon:2017zcu}, where it is demonstrated that dark photons with masses $\mathcal{O} (10) \; \mathrm{km} \lesssim (m_\chi)^{-1} \lesssim \mathcal{O} (1000) \; \mathrm{km}$ emitted by inspiraling NSs would manifest as a time-dependent evolution of the chirp mass, quantified as $\sim$$(1-\alpha')^{2/5}$, with $\alpha' \sim 1/3$ \cite{Sagunski:2017nzb} lying within the LIGO-Virgo sensitivity band. Second, a dipole radiation may emerge and influence the period evolution of companion stars with significantly asymmetric charge-to-mass ratio. This occurs once the orbital angular frequency of the binary exceeds $l^{-1}$, affecting the inspiral \cite{Alexander:2018qzg, Hook:2017psm, Kopp:2018jom,  Poddar:2023pfj}.

For such effects to be detectable and distinguishable from ordinary NS mergers, significant DM accumulation is required. This cannot be achieved through gravitational capture alone (Section \ref{sec:dark_matter_capture_and_accumulation}), which typically results in DM cores no larger than $\leq 10^{-10} M_\odot$. Indeed, assuming a DM candidate with $m_\chi \lesssim 1$ GeV and ambient density $\rho_\chi \sim 1 \; \mathrm{GeV/cm^3}$, the total number of DM particles (computed through Equation ~\eqref{eq:DM_capture_rate_numerical} or Equation~(6) in \cite{McDermott:2011jp}) accreted solely via gravitational capture is much smaller than the baryon number $N_\mathrm{B}$ in NSs, i.e., $N_\chi \sim 10^{47} << 10^{57} \sim N_\mathrm{B}$. To obtain a much higher value that is consistent with the DM fractions assumed in Section~\ref{sec:mass_radius_and_tidal_mass_relations}, either DM is accreted in much denser regions or alternative accumulation mechanisms must be taken into account.

By assuming that any DM particle that passes through an NS over its lifetime is captured, Kopp et al. \cite{Kopp:2018jom} established upper limits on the number $N_\chi^\mathrm{capt}$ of DM particles accreted in an NS and on the coupling $\alpha \lesssim 10^{-15}$ that leads to an almost negligible Yukawa force. To obtain much larger values of $N_\chi^\mathrm{capt}$ and  $\alpha$ and, in turn, to observe meaningful effects, the assumption that DM can be accreted only through direct capture should be relaxed, and more exotic accumulation mechanisms need to be considered. One proposal involves DM production through bremsstrahlung in a newly formed hot NS \cite{Nelson:2018xtr, Ellis:2018bkr}. Another possibility is the generation of DM via hypothetical decay of SM particles. Section \ref{sec:decay_of_standard_model_particles_into_DM} is dedicated to \mbox{this topic.}

Additional phenomena have been investigated in scenarios where DANS merge. Ellis et al. \cite{Ellis:2017jgp} found that the power spectral density of emitted GWs could exhibit additional peaks beyond the primary mode $f_2$, typically seen in the range \mbox{2--4 kHz \cite{Bauswein:2011tp, Clark:2015zxa, Prakash:2021wpz, Prakash:2023afe}}; though similar extra peaks have sometimes been found also in NS merger simulations, where secondary oscillation modes can emerge due to non-linear interaction between the fundamental quadrupolar mode $f_2$ and quasi-radial oscillations of the remnant \cite{Stergioulas:2011gd, Takami:2014zpa, Bauswein:2015vxa}.  Hook et al. \cite{Hook:2017psm} highlighted that a force mediated by light DM particles, such as axions, could impact the inspiral phase and be detected. Bezares et al. \cite{Bezares:2019jcb} observed a couple of interesting characteristics: %EE: Please check meaning retained
In DANS mergers, the three-body interaction between the two DM cores and the fermionic matter may trigger a pronounced one-arm instability of the remnant, thus breaking the traditional quadrupolar symmetry and forming an $|m| = 1$ over-density. The gravitational waveform would then exhibit an additional $|m| = 1$ mode, occurring at half the frequency of the dominant mode $|m| = 2$, which, in turn, would decay more rapidly over time compared to standard BNS coalescences. Remarkably, while these latter can sometimes mimic similar features (especially in asymmetric binaries with high spin and/or eccentricity \cite{Paschalidis:2015mla, East:2015vix, Lehner:2016wjg, Lehner:2016lxy, Radice:2016gym}), some features might break this degeneracy. In DANS mergers, the one-arm instability develops even in equal-mass, non-spinning systems, a behavior almost absent in standard NS binaries. Both kinds of coalescence may show the same exponential decay of the dominant $|m| = 2$ mode, but when DM is accreted its attenuation is fast enough to be detectable.

Mergers between NSs and their mirror counterparts (containing mirror DM \cite{Hodges:1993yb, Berezhiani:2000gw, Berezhiani:2003xm, Ignatiev:2003js, Foot:2014mia}) have been analyzed in \cite{Hippert:2022snq}. The fate of the remnant depends on $f/v$, namely, the ratio of the vacuum expectation values of standard matter and mirror-sector Higgs fields, respectively. For $f/v \geq 5$, a stable (mirror) DANS may form from the coalescence; whereas smaller ratios yield prompt or retarded collapse of the remnant to a BH. In general, a numerical relativity approach is needed to establish the features of such a merger. Numerical simulations of DANS mergers have been successfully carried out in \cite{Bezares:2019jcb, Emma:2022xjs, Bauswein:2020kor}, along with the construction of quasi-equilibrium configurations \cite{Ruter:2023uzc} (where it is also shown that bosonic DM halos can induce a ``mass-shedding'' effect in binary systems). Furthermore, dark photons \cite{Diamond:2021ekg} and axions \cite{Diamond:2023cto} produced in these collisions might convert to observable $\gamma$-rays.

Overall, these advancements open a new avenue for studying the dynamics of DANS coalescence and obtaining waveforms to compare with observations. By incorporating DM into models, researchers are now equipped to investigate its potential signatures in GW signals and to test DM theories with current interferometers like Advanced Virgo \cite{VIRGO:2014yos}, Advanced LIGO \cite{LIGOScientific:2014pky}, and KAGRA \cite{Somiya:2011np} and the next-generation detectors such as the Einstein Telescope \cite{Hild:2010id, Branchesi:2023mws}, the Cosmic Explorer \cite{LIGOScientific:2016wof, Evans:2023euw}, and LISA \cite{LISA:2017pwj, LISA:2022kgy, LISA:2022yao}. Future detectors may also reveal the influence of extended DM halos on NSs' tidal deformability and second Love number \cite{Dengler:2021qcq}.

%%%%%%%%%%%%%%%%%%%%%%%%%%%%%%%%%%%%%%%%%%
\section{Neutron Decay to Dark Matter}
\label{sec:decay_of_standard_model_particles_into_DM}
DM interactions can be constrained by investigating neutron decay within NSs. The longstanding discrepancy in neutron decay lifetime, observed using the beam and bottle methods \cite{Berezhiani:2005hv,Serebrov:2007gw,Fornal:2018eol}, provides an avenue to explore these interactions. In the bottle method, neutrons are trapped, making it sensitive to the total neutron width, yielding a slightly shorter lifetime of $\tau_n^\mathrm{bottle}=879.6\pm 0.6~\mathrm{s}$ \cite{Pichlmaier:2010zz,Serebrov:2004zf,Steyerl:2012zz,Arzumanov:2015tea}. In contrast, the beam method measures the number of protons produced through $\beta$-decay in a fixed volume from a neutron beam, resulting in a longer lifetime of $\tau_n^\mathrm{beam}=888.0\pm 2.0~\mathrm{s}$ \cite{Byrne:1990ig,Yue:2013qrc}. This discrepancy can be reconciled by introducing a new decay channel for the neutron into a light dark baryon $(\chi)$, as $n\rightarrow \chi+\cdots$, where the additional particles $(\cdots)$ conserve energy and momentum. Consequently, the branching fraction for this decay mode becomes
\begin{equation}
    f_{n\rightarrow \chi}=1-\frac{\tau^\mathrm{bottle}_n}{\tau^\mathrm{beam}_n}=(0.9\pm 0.2)\times 10^{-2} \, .
\end{equation}
However, more precise measurements of the neutron axial-vector coupling constant $g_A$ place stringent limits on the non-standard branching fraction of neutron decay, restricting it to less than $2.7\times 10^{-3}$ at the $95\%$ C.L \cite{Czarnecki:2018okw}.

A study in \cite{Fornal:2018eol} explores how the $\sim$$8~\mathrm{s}$ discrepancy in neutron lifetime measurements, with a $4\sigma$ tension, could be resolved by introducing an additional neutron decay channel involving a fermionic DM particle $(\chi)$. This particle’s mass falls within the narrow range $m_p-m_e<m_\chi<m_n$, ensuring consistency with proton stability. While the simplest decay mode, $n\rightarrow \chi\gamma$, is excluded by the absence of a detected monochromatic photon \cite{Tang:2018eln}, alternative invisible decay channels, such as $n\rightarrow \chi \phi$, face stringent constraints from NS observations. The presence of new states within NSs could significantly reduce their maximum mass. Observations of NSs with masses exceeding $0.7~M_\odot$ impose strong constraints, requiring $\chi$ particles with a baryon number %EE: Please check meaning retained
to have masses greater than $1.2~\mathrm{GeV}$ or to exhibit strongly repulsive self-interactions \cite{McKeen:2018xwc,Cline:2018ami}. Neutron--mirror neutron oscillations have also been investigated \cite{Berezhiani:2017jkn, Berezhiani:2018eds, Berezhiani:2018udo, Ayres:2021zbh}, while an alternative approach, distinct from DM, has been proposed to address the neutron lifetime measurement puzzle, as discussed in \cite{Fuwa:2024cdf}.

%%%%%%%%%%%%%%%%%%%%%%%%%%%%%%%%%%%%%%%%%%
\section{Conclusions}
\label{sec:conclusions_main}

In this review article, we have described the role of DM in NSs, focusing on how its accumulation can influence the structure, evolution, and observable properties of these compact objects. NSs, with their extreme densities and gravitational fields, provide unique environments where DM could accumulate over time. The mechanisms of DM capture and accretion reveal that these compact objects can gather significant amounts of DM, which could lead to heating effects due to self-annihilating interactions. The cooling profile of NSs may be significantly affected by DM scattering and/or annihilation processes. In the latest theoretical framework that incorporates internal mechanisms such as rotochemical heating, the intrinsic heating of NSs turns out to be degenerate with the DM contribution. This degeneracy may potentially be resolved in NSs with long periods \cite{Hamaguchi:2019oev} once more sophisticated methods to evaluate the nucleon pairing and to determine the initial period are developed. Furthermore, if enough DM settles in an NS's core, it could even trigger the formation of a BH, which in turn could evaporate due to Hawking radiation or consume the whole NS in a relatively short time.

Modeling DANSs often needs a two-fluid framework, where the baryonic and the dark sectors are treated as separate fluids, interacting only through gravity. This approach allows the exploration of different DM candidates, including bosonic and fermionic DM, and the verification of how the formation of DM cores or halos alters observable properties like mass--radius relations and tidal deformability. These properties are closely tied to observable signatures, particularly by GWs produced in BNS mergers. Hence, these models are suitable to be effectively constrained by current interferometers such as LIGO, Virgo, and KAGRA, as well as future third-generation detectors like the Einstein Telescope and Cosmic Explorer. A brief summary is as follows:
 
\begin{itemize}
    \item The formation of a DM core yields a softer EoS and results in a more compact DANS, whose corresponding mass and radius are smaller than those of a star made entirely of BM. In contrast, if DM forms a halo, the EoS turns out to be stiffer, resulting in a larger radius and tidal deformability and slightly greater gravitational mass. In the latter scenario, since $R_\mathrm{DANS}$ represents the outermost radius, then $R_\mathrm{DANS} = R_\mathrm{DM}$ and it can extend significantly beyond the BM. As a consequence, configurations with extended halos are unlikely to match observational constraints and the corresponding EoSs are usually disfavored. 

    \item Weak interaction couplings, low fractions of DM, and large masses of DM candidates favor the formation of a core.

    \item Despite the degeneracy among DM and pure nuclear models, the analyses of M--R and $\Lambda$--M relations can help to rule out certain combinations of DM and baryonic EoSs, while also possibly supporting the evidence for the existence of DANSs.

    \item The tidal deformability $\Lambda$ (computed through Equation~\eqref{eq:tidal}) is typically reduced in the presence of a DM core, as the star becomes more compact and it is harder for it to be deformed. In contrast, if a DM halo forms, $\Lambda$ increases significantly due to the larger effective radius. This is particularly relevant for GW signals from NS mergers, as the tidal interactions between merging stars leave a distinct imprint on the waveforms. 
\end{itemize}

DANSs may possibly be distinguished in merging binaries. The exertion of a dark force between the dark components of the stars would manifest both in a shift of the masses of the stars that would alter the chirp mass of the system and in a dipole radiation that would reduce the binding energy and influence the binary period evolution. 

A fundamental quantity characterizing NSs is $c_s$. While it has been widely analyzed for baryonic NSs, the literature lacks a similar comprehensive study of DM's effect on the sound speed in a DANS. However, an effective sound speed has been derived in \cite{Giangrandi:2022wht}. We review how DM cores yield a softer EoS and smaller $c_s$, whereas the distribution of DM in a halo favors a larger speed of sound.

An alternative/additional mechanism of DM accumulation in NSs could be the decay of SM particles into DM. The observed $\sim$$8$ s discrepancy between beam and bottle neutron lifetime measurements suggests a possible decay channel involving a dark particle $\chi$ with a branching fraction constrained to be $f_{n\rightarrow \chi} \lesssim 2.7 \times 10^{-3}$ (90\% CL). In particular, a fermionic DM candidate with mass $m_\chi \gtrsim 1.2 \, \mathrm{GeV}$ or exhibiting strong self-repulsion could resolve that tension.

In conclusion, our review of DANSs has shown that the inclusion of DM in NS models significantly expands the range of possible configurations and observable signatures. As future GW detectors improve and more data become available, matching theoretical predictions with these observations will be critical for confirming the role of DM in \mbox{compact objects.}

%\clearpage

%%%%%%%%%%%%%%%%%%%%%%%%%%%%%%%%%%%%%%%%%%

%%%%%%%%%%%%%%%%%%%%%%%%%%%%%%%%%%%%%%%%%%
\vspace{12pt}

\authorcontributions{F. Grippa, T. K. Poddar, G. Lambiase conceived the idea of the project. F. Grippa wrote the first version of the manuscript with input from all authors; T. K. Poddar and G. Lambiase supervised the study. All authors have read and agreed to the published version of this review article.}

\funding{This research received no external funding.}

\dataavailability{No new data were created.}

%\dataavailability{We encourage all authors of articles published in MDPI journals to share their research data. In this section, please provide details regarding where data supporting reported results can be found, including links to publicly archived datasets analyzed or generated during the study. Where no new data were created, or where data is unavailable due to privacy or ethical restrictions, a statement is still required. Suggested Data Availability Statements are available in section ``MDPI Research Data Policies'' at \url{https://www.mdpi.com/ethics}.} 

\acknowledgments{G.L. and T.K.P. thank COST Action COSMIC WISPers CA21106, supported by COST (European Cooperation in Science and Technology).}

\conflictsofinterest{The authors declare no conflicts of interest.}

%\conflictsofinterest{Declare conflicts of interest or state ``The authors declare no conflicts of interest.'' Authors must identify and declare any personal circumstances or interest that may be perceived as inappropriately influencing the representation or interpretation of reported research results. Any role of the funders in the design of the study; in the collection, analyses or interpretation of data; in the writing of the manuscript; or in the decision to publish the results must be declared in this section. If there is no role, please state ``The funders had no role in the design of the study; in the collection, analyses, or interpretation of data; in the writing of the manuscript; or in the decision to publish the results''.} 

\abbreviations{Abbreviations}{
The following abbreviations are used in this manuscript:\\

\noindent 
\begin{tabular}{@{}ll}
NS & Neutron star \\
DANS & Dark matter admixed neutron star \\
WD & White dwarf \\
BH & Black hole \\
EoS & Equation of state \\
DM & Dark matter \\
BM & Baryonic matter \\
GW & Gravitational wave \\
BNS & Binary neutron star \\
$\Lambda$ & Tidal deformability \\
SM & Standard Model \\
WIMP & Weakly interacting massive particle \\
CMB & Cosmic microwave background \\
TOV & Tolman--Oppenheimer--Volkoff \\
GR & General relativity \\
M--R & Mass--radius \\
$\Lambda$--M & Tidal deformability--mass \\
BEC & Bose--Einstein condensate \\
RMF & Relativistic mean field \\
SQM & Strange quark matter \\
QCD & Quantum chromodynamics \\
SI & Spin independent \\
SD & Spin dependent \\
CL & Confidence level 
\end{tabular}
}

%\clearpage

\begin{adjustwidth}{-\extralength}{0cm}

\reftitle{References}

%\begin{thebibliography}{999}
% Reference 7
%\bibitem[Author7(year)]{ref-thesis}
%Author 1, A.B. Title of Thesis. Level of Thesis, Degree-Granting University, Location of University, Date of Completion.
% Reference 8
%\bibitem[Author8(year)]{ref-url}
%Title of Site. Available online: URL (accessed on Day Month Year).
%\end{thebibliography}

%\bibliographystyle{apsrev4-1}
%\bibliography{bibliography}

\begin{thebibliography}{999}

\bibitem[Chandrasekhar(1939)]{Chandrasekhar:1939}
Chandrasekhar, S.
\newblock {\em An Introduction to the Study of Stellar Structure};
  Astrophysical Monographs, The University of Chicago Press: Chicago, IL, USA, 1939.

\bibitem[Lattimer and Prakash(2004)]{Lattimer:2004pg}
Lattimer, J.M.; Prakash, M.
\newblock {The physics of neutron stars}.
\newblock {\em Science} {\bf 2004}, {\em 304},~536--542.
\newblock {\url{https://doi.org/10.1126/science.1090720}}.

\bibitem[Baym et~al.(2018)Baym, Hatsuda, Kojo, Powell, Song, and
  Takatsuka]{Baym:2017whm}
Baym, G.; Hatsuda, T.; Kojo, T.; Powell, P.D.; Song, Y.; Takatsuka, T.
\newblock {From hadrons to quarks in neutron stars: A review}.
\newblock {\em Rept. Prog. Phys.} {\bf 2018}, {\em 81},~056902.
%  %\href{http://arxiv.org/abs/1707.04966}{{\normalfont
%  %[arXiv:astro-ph.HE/1707.04966]}}.
\newblock {\url{https://doi.org/10.1088/1361-6633/aaae14}}.

\bibitem[Shapiro and Teukolsky(1983)]{Shapiro:1983du}
Shapiro, S.L.; Teukolsky, S.A.
\newblock {\em {Black Holes, White Dwarfs, and Neutron Stars: The Physics of
  Compact Objects}}; John Wiley \& Sons: Hoboken, NJ, USA, 1983.
\newblock {\url{https://doi.org/10.1002/9783527617661}}.

\bibitem[Silk et~al.(2010)]{Bertone:2010zza}
Silk, J.; Moore B.; Bullock J.; Kaplinghat M.; Strigari L.; Mellier Y.; Merritt D.; Bekenstein J.; Gelmini G.; Gondolo P.  et~al.
\newblock {\em {Particle Dark Matter: Observations, Models and Searches}};
  Cambridge University Press: Cambridge, UK, 2010.
\newblock {\url{https://doi.org/10.1017/CBO9780511770739}}.

\bibitem[Antoniadis et~al.(2013)]{Antoniadis:2013pzd}
Antoniadis, J.; Freire, P.C.; Wex, N.; Tauris, T.M.; Lynch, R.S.; Van Kerkwijk, M.H.; Kramer, M.; Bassa, C.; Dhillon, V.S.; Driebe, T.;  et~al.
\newblock {A Massive Pulsar in a Compact Relativistic Binary}.
\newblock {\em Science} {\bf 2013}, {\em 340},~6131.
%  %\href{http://arxiv.org/abs/1304.6875}{{\normalfont
%  %[arXiv:astro-ph.HE/1304.6875]}}.
\newblock {\url{https://doi.org/10.1126/science.1233232}}.

\bibitem[\"Ozel and Freire(2016)]{Ozel:2016oaf}
\"Ozel, F.; Freire, P.
\newblock {Masses, Radii, and the Equation of State of Neutron Stars}.
\newblock {\em Ann. Rev. Astron. Astrophys.} {\bf 2016}, {\em 54},~401--440.
%  %\href{http://arxiv.org/abs/1603.02698}{{\normalfont
%  %[arXiv:astro-ph.HE/1603.02698]}}.
\newblock {\url{https://doi.org/10.1146/annurev-astro-081915-023322}}.

\bibitem[Lattimer and Prakash(2007)]{Lattimer:2006xb}
Lattimer, J.M.; Prakash, M.
\newblock {Neutron Star Observations: Prognosis for Equation of State
  Constraints}.
\newblock {\em Phys. Rept.} {\bf 2007}, {\em 442},~109--165.
%  %\href{http://arxiv.org/abs/astro-ph/0612440}{{\normalfont
%  %[astro-ph/0612440]}}.
\newblock {\url{https://doi.org/10.1016/j.physrep.2007.02.003}}.

\bibitem[Steiner et~al.(2010)Steiner, Lattimer, and Brown]{Steiner:2010fz}
Steiner, A.W.; Lattimer, J.M.; Brown, E.F.
\newblock {The Equation of State from Observed Masses and Radii of Neutron
  Stars}.
\newblock {\em Astrophys. J.} {\bf 2010}, {\em 722},~33--54.
%  %\href{http://arxiv.org/abs/1005.0811}{{\normalfont
%  %[arXiv:astro-ph.HE/1005.0811]}}.
\newblock {\url{https://doi.org/10.1088/0004-637X/722/1/33}}.

\bibitem[Oertel et~al.(2017)Oertel, Hempel, Kl\"ahn, and Typel]{Oertel:2016bki}
Oertel, M.; Hempel, M.; Kl\"ahn, T.; Typel, S.
\newblock {Equations of state for supernovae and compact stars}.
\newblock {\em Rev. Mod. Phys.} {\bf 2017}, {\em 89},~015007.
  %\href{http://arxiv.org/abs/1610.03361}{{\normalfont
  %[arXiv:astro-ph.HE/1610.03361]}}.
\newblock {\url{https://doi.org/10.1103/RevModPhys.89.015007}}.

\bibitem[Lattimer and Prakash(2011)]{Lattimer:2010uk}
Lattimer, J.M.; Prakash, M. {What a Two Solar Mass Neutron Star Really Means}.
\newblock In {\em {From Nuclei to Stars: Festschrift in Honor of Gerald E
  Brown}}; Lee, S., Ed.; World Scientific: Singapore, 2011; pp. 275--304.
  %\href{http://arxiv.org/abs/1012.3208}{{\normalfont
  %[arXiv:astro-ph.SR/1012.3208]}}.
\newblock {\url{https://doi.org/10.1142/9789814329880_0012}}.

\bibitem[Lattimer and Prakash(2016)]{Lattimer:2015nhk}
Lattimer, J.M.; Prakash, M.
\newblock {The Equation of State of Hot, Dense Matter and Neutron Stars}.
\newblock {\em Phys. Rept.} {\bf 2016}, {\em 621},~127--164.
  %\href{http://arxiv.org/abs/1512.07820}{{\normalfont
  %[arXiv:astro-ph.SR/1512.07820]}}.
\newblock {\url{https://doi.org/10.1016/j.physrep.2015.12.005}}.

\bibitem[Bejger and Haensel(2004)]{Bejger:2004gz}
Bejger, M.; Haensel, P.
\newblock {Surface gravity of neutron stars and strange stars}.
\newblock {\em Astron. Astrophys.} {\bf 2004}, {\em 420},~987--991.
  %\href{http://arxiv.org/abs/astro-ph/0403550}{{\normalfont
%  %[astro-ph/0403550]}}.
\newblock {\url{https://doi.org/10.1051/0004-6361:20034538}}.

\bibitem[Haensel et~al.(2007)Haensel, Potekhin, and Yakovlev]{Haensel:2007yy}
Haensel, P.; Potekhin, A.Y.; Yakovlev, D.G.
\newblock {\em {Neutron Stars 1: Equation of State and Structure}}; 
  Springer: New York, NY, USA,  2007; Volume 326.
\newblock {\url{https://doi.org/10.1007/978-0-387-47301-7}}.

\bibitem[Yakovlev and Urpin(1980)]{1980SvA....24..303Y}
Yakovlev, D.G.; Urpin, V.A.
\newblock {Thermal and Electrical Conductivity in White Dwarfs and Neutron
  Stars}.
\newblock {\em Sov. Astron.} {\bf 1980}, {\em 24},~303.

\bibitem[Yakovlev and Pethick(2004)]{Yakovlev:2004iq}
Yakovlev, D.G.; Pethick, C.J.
\newblock {Neutron star cooling}.
\newblock {\em Ann. Rev. Astron. Astrophys.} {\bf 2004}, {\em 42},~169--210.
  %\href{http://arxiv.org/abs/astro-ph/0402143}{{\normalfont
%  %[astro-ph/0402143]}}.
\newblock {\url{https://doi.org/10.1146/annurev.astro.42.053102.134013}}.

\bibitem[Weber et~al.(2007)Weber, Negreiros, Rosenfield, and
  Stejner]{Weber:2006ep}
Weber, F.; Negreiros, R.; Rosenfield, P.; Stejner, M.
\newblock {Pulsars as Astrophysical Laboratories for Nuclear and Particle
  Physics}.
\newblock {\em Prog. Part. Nucl. Phys.} {\bf 2007}, {\em 59},~94--113.
  %\href{http://arxiv.org/abs/astro-ph/0612054}{{\normalfont
  %[astro-ph/0612054]}}.
\newblock {\url{https://doi.org/10.1016/j.ppnp.2006.12.008}}.

\bibitem[Pons et~al.(1999)Pons, Reddy, Prakash, Lattimer, and
  Miralles]{Pons:1998mm}
Pons, J.A.; Reddy, S.; Prakash, M.; Lattimer, J.M.; Miralles, J.A.
\newblock {Evolution of protoneutron stars}.
\newblock {\em Astrophys. J.} {\bf 1999}, {\em 513},~780.
  %\href{http://arxiv.org/abs/astro-ph/9807040}{{\normalfont
  %[astro-ph/9807040]}}.
\newblock {\url{https://doi.org/10.1086/306889}}.

\bibitem[Ho et~al.(2007)Ho, Kaplan, Chang, van Adelsberg, and
  Potekhin]{Ho:2006uk}
Ho, W.C.G.; Kaplan, D.L.; Chang, P.; van Adelsberg, M.; Potekhin, A.Y.
\newblock {Magnetic Hydrogen Atmosphere Models and the Neutron Star RX
  J1856.5-3754}.
\newblock {\em Mon. Not. Roy. Astron. Soc.} {\bf 2007}, {\em 375},~821--830.
  %\href{http://arxiv.org/abs/astro-ph/0612145}{{\normalfont
  %[astro-ph/0612145]}}.
\newblock {\url{https://doi.org/10.1111/j.1365-2966.2006.11376.x}}.

\bibitem[Potekhin et~al.(2015)Potekhin, Pons, and Page]{Potekhin:2015qsa}
Potekhin, A.Y.; Pons, J.A.; Page, D.
\newblock {Neutron stars---Cooling and transport}.
\newblock {\em Space Sci. Rev.} {\bf 2015}, {\em 191},~239--291.
  %\href{http://arxiv.org/abs/1507.06186}{{\normalfont
  %[arXiv:astro-ph.HE/1507.06186]}}.
\newblock {\url{https://doi.org/10.1007/s11214-015-0180-9}}.

\bibitem[Logoteta(2021)]{Logoteta:2021iuy}
Logoteta, D.
\newblock {Hyperons in Neutron Stars}.
\newblock {\em Universe} {\bf 2021}, {\em 7},~408.
\newblock {\url{https://doi.org/10.3390/universe7110408}}.

\bibitem[Oertel et~al.(2015)Oertel, Providencia, Gulminelli, and
  Raduta]{Oertel:2015fta}
Oertel, M.; Providencia, C.; Gulminelli, F.; Raduta, A.R.
\newblock {Hyperons in neutron star matter within relativistic mean-field
  models}.
\newblock {\em Phys. Part. Nucl.} {\bf 2015}, {\em 46},~830--834.
\newblock {\url{https://doi.org/10.1134/S1063779615050214}}.

\bibitem[Chatterjee and Vida\~na(2016)]{Chatterjee:2015pua}
Chatterjee, D.; Vida\~na, I.
\newblock {Do hyperons exist in the interior of neutron stars?}
\newblock {\em Eur. Phys. J. A} {\bf 2016}, {\em 52},~29.
  %\href{http://arxiv.org/abs/1510.06306}{{\normalfont
  %[arXiv:nucl-th/1510.06306]}}.
\newblock {\url{https://doi.org/10.1140/epja/i2016-16029-x}}.

\bibitem[Bombaci et~al.(2016)Bombaci, Logoteta, Vida\~na, and
  Provid\^encia]{Bombaci:2016xuj}
Bombaci, I.; Logoteta, D.; Vida\~na, I.; Provid\^encia, C.
\newblock {Quark matter nucleation in neutron stars and astrophysical
  implications}.
\newblock {\em Eur. Phys. J. A} {\bf 2016}, {\em 52},~58.
  %\href{http://arxiv.org/abs/1601.04559}{{\normalfont
  %[arXiv:astro-ph.HE/1601.04559]}}.
\newblock {\url{https://doi.org/10.1140/epja/i2016-16058-5}}.

\bibitem[Alford et~al.(2007)Alford, Blaschke, Drago, Klahn, Pagliara, and
  Schaffner-Bielich]{Alford:2006vz}
Alford, M.; Blaschke, D.; Drago, A.; Klahn, T.; Pagliara, G.;
  Schaffner-Bielich, J.
\newblock {Quark matter in compact stars?}
\newblock {\em Nature} {\bf 2007}, {\em 445},~E7--E8.
  %\href{http://arxiv.org/abs/astro-ph/0606524}{{\normalfont
  %[astro-ph/0606524]}}.
\newblock {\url{https://doi.org/10.1038/nature05582}}.

\bibitem[Bombaci and Logoteta(2017)]{Bombaci:2017tey}
Bombaci, I.; Logoteta, D.
\newblock {Quark deconfinement in neutron stars and astrophysical
  implications}.
\newblock {\em Int. J. Mod. Phys.} {\bf 2017}, {\em 1},~904--923.
\newblock {\url{https://doi.org/10.1142/S021827181730004X}}.

\bibitem[Xu(2003)]{Xu:2002ew}
Xu, R.
\newblock {Bare strange quark stars: Formation and emission}.
\newblock {\em Astrophys. Space Sci. Libr.} {\bf 2003}, {\em 298},~73.
  %\href{http://arxiv.org/abs/astro-ph/0211563}{{\normalfont
  %[astro-ph/0211563]}}.
\newblock {\url{https://doi.org/10.1007/978-94-017-0403-8_11}}.

\bibitem[Page and Usov(2002)]{Page:2002bj}
Page, D.; Usov, V.V.
\newblock {Thermal evolution and light curves of young bare strange stars}.
\newblock {\em Phys. Rev. Lett.} {\bf 2002}, {\em 89},~131101.
  %\href{http://arxiv.org/abs/astro-ph/0204275}{{\normalfont
  %[astro-ph/0204275]}}.
\newblock {\url{https://doi.org/10.1103/PhysRevLett.89.131101}}.

\bibitem[Weber et~al.(2013)Weber, Orsaria, Rodrigues, and Yang]{Weber:2012ta}
Weber, F.; Orsaria, M.; Rodrigues, H.; Yang, S.H.
\newblock {Structure of Quark Stars}.
\newblock {\em IAU Symp.} {\bf 2013}, {\em 291},~61--66.
  %\href{http://arxiv.org/abs/1210.1910}{{\normalfont
  %[arXiv:astro-ph.SR/1210.1910]}}.
\newblock {\url{https://doi.org/10.1017/S1743921312023174}}.

\bibitem[Glendenning(1997)]{Glendenning:1997wn}
Glendenning, N.K.
\newblock {\em {Compact Stars: Nuclear Physics, Particle Physics, and General
  Relativity}}; Springer Science \& Business Media: Berlin, Germany, 1997.

\bibitem[Burgio et~al.(2002)Burgio, Baldo, Sahu, Santra, and
  Schulze]{Burgio:2001mk}
Burgio, G.F.; Baldo, M.; Sahu, P.K.; Santra, A.B.; Schulze, H.J.
\newblock {Maximum mass of neutron stars with a quark core}.
\newblock {\em Phys. Lett. B} {\bf 2002}, {\em 526},~19--26.
  %\href{http://arxiv.org/abs/astro-ph/0111440}{{\normalfont
  %[astro-ph/0111440]}}.
\newblock {\url{https://doi.org/10.1016/S0370-2693(01)01479-4}}.

\bibitem[Alford and Reddy(2003)]{Alford:2002rj}
Alford, M.; Reddy, S.
\newblock {Compact stars with color superconducting quark matter}.
\newblock {\em Phys. Rev. D} {\bf 2003}, {\em 67},~074024.
  %\href{http://arxiv.org/abs/nucl-th/0211046}{{\normalfont [nucl-th/0211046]}}.
\newblock {\url{https://doi.org/10.1103/PhysRevD.67.074024}}.

\bibitem[Buballa et~al.(2004)Buballa, Neumann, Oertel, and
  Shovkovy]{Buballa:2003et}
Buballa, M.; Neumann, F.; Oertel, M.; Shovkovy, I.
\newblock {Quark mass effects on the stability of hybrid stars}.
\newblock {\em Phys. Lett. B} {\bf 2004}, {\em 595},~36--43.
  %\href{http://arxiv.org/abs/nucl-th/0312078}{{\normalfont [nucl-th/0312078]}}.
\newblock {\url{https://doi.org/10.1016/j.physletb.2004.05.064}}.

\bibitem[Alford et~al.(2005)Alford, Braby, Paris, and Reddy]{Alford:2004pf}
Alford, M.; Braby, M.; Paris, M.W.; Reddy, S.
\newblock {Hybrid stars that masquerade as neutron stars}.
\newblock {\em Astrophys. J.} {\bf 2005}, {\em 629},~969--978.
  %\href{http://arxiv.org/abs/nucl-th/0411016}{{\normalfont [nucl-th/0411016]}}.
\newblock {\url{https://doi.org/10.1086/430902}}.

\bibitem[Bodmer(1971)]{Bodmer:1971we}
Bodmer, A.R.
\newblock {Collapsed nuclei}.
\newblock {\em Phys. Rev. D} {\bf 1971}, {\em 4},~1601--1606.
\newblock {\url{https://doi.org/10.1103/PhysRevD.4.1601}}.

\bibitem[Witten(1984)]{Witten:1984rs}
Witten, E.
\newblock {Cosmic Separation of Phases}.
\newblock {\em Phys. Rev. D} {\bf 1984}, {\em 30},~272--285.
\newblock {\url{https://doi.org/10.1103/PhysRevD.30.272}}.

\bibitem[Farhi and Jaffe(1984)]{Farhi:1984qu}
Farhi, E.; Jaffe, R.L.
\newblock {Strange Matter}.
\newblock {\em Phys. Rev. D} {\bf 1984}, {\em 30},~2379.
\newblock {\url{https://doi.org/10.1103/PhysRevD.30.2379}}.

\bibitem[Holdom et~al.(2018)Holdom, Ren, and Zhang]{Holdom:2017gdc}
Holdom, B.; Ren, J.; Zhang, C.
\newblock {Quark matter may not be strange}.
\newblock {\em Phys. Rev. Lett.} {\bf 2018}, {\em 120},~222001.
  %\href{http://arxiv.org/abs/1707.06610}{{\normalfont
  %[arXiv:hep-ph/1707.06610]}}.
\newblock {\url{https://doi.org/10.1103/PhysRevLett.120.222001}}.

\bibitem[Yuan et~al.(2022)Yuan, Li, Miao, Zuo, and Bai]{Yuan:2022dxb}
Yuan, W.L.; Li, A.; Miao, Z.; Zuo, B.; Bai, Z.
\newblock {Interacting ud and uds quark matter at finite densities and quark
  stars}.
\newblock {\em Phys. Rev. D} {\bf 2022}, {\em 105},~123004.
  %\href{http://arxiv.org/abs/2203.04798}{{\normalfont
  %[arXiv:nucl-th/2203.04798]}}.
\newblock {\url{https://doi.org/10.1103/PhysRevD.105.123004}}.

\bibitem[Bai and Chen(2024)]{Bai:2024amm}
Bai, Y.; Chen, T.K.
\newblock {Approaching Stable Quark Matter}. \emph{arXiv} {\bf 2024}, arXiv:2410.19678.
\newblock  %\href{http://arxiv.org/abs/2410.19678}{{\normalfont
  %[arXiv:hep-ph/2410.19678]}}.

\bibitem[Lattimer and Prakash(2001)]{Lattimer:2000nx}
Lattimer, J.M.; Prakash, M.
\newblock {Neutron star structure and the equation of state}.
\newblock {\em Astrophys. J.} {\bf 2001}, {\em 550},~426.
  %\href{http://arxiv.org/abs/astro-ph/0002232}{{\normalfont
  %[astro-ph/0002232]}}.
\newblock {\url{https://doi.org/10.1086/319702}}.

\bibitem[Sumiyoshi et~al.(2023)Sumiyoshi, Kojo, and
  Furusawa]{Sumiyoshi:2022uoj}
Sumiyoshi, K.; Kojo, T.; Furusawa, S. {Equation of State in Neutron Stars and
  Supernovae}.
\newblock In {\em {Handbook of Nuclear Physics}}; Tanihata, I., Toki, H.,
  Kajino, T., Eds.; Springer Nature: Berlin, Germany, 2023; pp. 1--51.
  %\href{http://arxiv.org/abs/2207.00033}{{\normalfont
  %[arXiv:nucl-th/2207.00033]}}.
\newblock {\url{https://doi.org/10.1007/978-981-15-8818-1_104-1}}.

\bibitem[Abbott et~al.(2018)]{LIGOScientific:2018cki}
Abbott, B.P.; Abbott, R.; Abbott, T.D.; Acernese, F.; Ackley, K.; Adams, C.; Adams, T.; Addesso, P.; Adhikari, R.X.; Adya, V.B.;  et~al.
\newblock {GW170817: Measurements of neutron star radii and equation of state}.
\newblock {\em Phys. Rev. Lett.} {\bf 2018}, {\em 121},~161101.
  %\href{http://arxiv.org/abs/1805.11581}{{\normalfont
  %[arXiv:gr-qc/1805.11581]}}.
\newblock {\url{https://doi.org/10.1103/PhysRevLett.121.161101}}.

\bibitem[Abbott et~al.(2017{\natexlab{a}})]{LIGOScientific:2017ync}
Abbott, B.P.
\newblock {Multi-messenger Observations of a Binary Neutron Star Merger}.
\newblock {\em Astrophys. J. Lett.} {\bf 2017}, {\em 848},~L12.
  %\href{http://arxiv.org/abs/1710.05833}{{\normalfont
  %[arXiv:astro-ph.HE/1710.05833]}}.
\newblock {\url{https://doi.org/10.3847/2041-8213/aa91c9}}.

\bibitem[Abbott et~al.(2017{\natexlab{b}})]{LIGOScientific:2017vwq}
Abbott, B.P.; Abbott, R.; Abbott, T.; Acernese, F.; Ackley, K.; Adams, C.; Adams, T.; Addesso, P.; Adhikari, R.X.; Adya, V.B.;  et~al.
\newblock {GW170817: Observation of Gravitational Waves from a Binary Neutron
  Star Inspiral}.
\newblock {\em Phys. Rev. Lett.} {\bf 2017}, {\em 119},~161101.
  %\href{http://arxiv.org/abs/1710.05832}{{\normalfont
  %[arXiv:gr-qc/1710.05832]}}.
\newblock {\url{https://doi.org/10.1103/PhysRevLett.119.161101}}.

\bibitem[Radice et~al.(2018)Radice, Perego, Zappa, and
  Bernuzzi]{Radice:2017lry}
Radice, D.; Perego, A.; Zappa, F.; Bernuzzi, S.
\newblock {GW170817: Joint Constraint on the Neutron Star Equation of State
  from Multimessenger Observations}.
\newblock {\em Astrophys. J. Lett.} {\bf 2018}, {\em 852},~L29.
  %\href{http://arxiv.org/abs/1711.03647}{{\normalfont
  %[arXiv:astro-ph.HE/1711.03647]}}.
\newblock {\url{https://doi.org/10.3847/2041-8213/aaa402}}.

\bibitem[Abbott et~al.(2020)]{LIGOScientific:2020aai}
Abbott, B.P.; Abbott, R.; Abbott, T.D.; Abraham, S.; Acernese, F.; Ackley, K.; Adams, C.; Adhikari, R.X.; Adya, V.B.; Affeldt, C.;  et~al.
\newblock {GW190425: Observation of a Compact Binary Coalescence with Total
  Mass $\sim$$3.4 M_{\odot}$}.
\newblock {\em Astrophys. J. Lett.} {\bf 2020}, {\em 892},~L3.
  %\href{http://arxiv.org/abs/2001.01761}{{\normalfont
  %[arXiv:astro-ph.HE/2001.01761]}}.
\newblock {\url{https://doi.org/10.3847/2041-8213/ab75f5}}.

\bibitem[Bertone and Fairbairn(2008)]{Bertone:2007ae}
Bertone, G.; Fairbairn, M.
\newblock {Compact Stars as Dark Matter Probes}.
\newblock {\em Phys. Rev. D} {\bf 2008}, {\em 77},~043515.
  %\href{http://arxiv.org/abs/0709.1485}{{\normalfont
  %[arXiv:astro-ph/0709.1485]}}.
\newblock {\url{https://doi.org/10.1103/PhysRevD.77.043515}}.

\bibitem[de~Lavallaz and Fairbairn(2010)]{deLavallaz:2010wp}
de~Lavallaz, A.; Fairbairn, M.
\newblock {Neutron Stars as Dark Matter Probes}.
\newblock {\em Phys. Rev. D} {\bf 2010}, {\em 81},~123521.
  %\href{http://arxiv.org/abs/1004.0629}{{\normalfont
  %[arXiv:astro-ph.GA/1004.0629]}}.
\newblock {\url{https://doi.org/10.1103/PhysRevD.81.123521}}.

\bibitem[Kouvaris and Tinyakov(2010)]{Kouvaris:2010vv}
Kouvaris, C.; Tinyakov, P.
\newblock {Can Neutron stars constrain Dark Matter?}
\newblock {\em Phys. Rev. D} {\bf 2010}, {\em 82},~063531.
  %\href{http://arxiv.org/abs/1004.0586}{{\normalfont
  %[arXiv:astro-ph.GA/1004.0586]}}.
\newblock {\url{https://doi.org/10.1103/PhysRevD.82.063531}}.

\bibitem[Ciarcelluti and Sandin(2011)]{Ciarcelluti:2010ji}
Ciarcelluti, P.; Sandin, F.
\newblock {Have neutron stars a dark matter core?}
\newblock {\em Phys. Lett. B} {\bf 2011}, {\em 695},~19--21.
  %\href{http://arxiv.org/abs/1005.0857}{{\normalfont
  %[arXiv:astro-ph.HE/1005.0857]}}.
\newblock {\url{https://doi.org/10.1016/j.physletb.2010.11.021}}.

\bibitem[Ellis et~al.(2018)Ellis, H\"utsi, Kannike, Marzola, Raidal, and
  Vaskonen]{Ellis:2018bkr}
Ellis, J.; H\"utsi, G.; Kannike, K.; Marzola, L.; Raidal, M.; Vaskonen, V.
\newblock {Dark Matter Effects On Neutron Star Properties}.
\newblock {\em Phys. Rev. D} {\bf 2018}, {\em 97},~123007.
  %\href{http://arxiv.org/abs/1804.01418}{{\normalfont
  %[arXiv:astro-ph.CO/1804.01418]}}.
\newblock {\url{https://doi.org/10.1103/PhysRevD.97.123007}}.

\bibitem[Aghanim et~al.(2020)]{Planck:2018vyg}
Aghanim, N.; Akrami Y.; Ashdown M.; Aumont J.; Bacciagalupi C.; Ballardini M.; Banday A. J.; Barreiro R. B.; Bartolo N.; Basak S. et~al.
\newblock {Planck 2018 results. VI. Cosmological parameters}.
\newblock {\em Astron. Astrophys.} {\bf 2020}, {\em 641},~A6;
  %\href{http://arxiv.org/abs/1807.06209}{{\normalfont
  %[arXiv:astro-ph.CO/1807.06209]}}.
\newblock Erratum in \emph{Astron. Astrophys.} \textbf{2021}, \emph{652}, C4.  {\url{https://doi.org/10.1051/0004-6361/201833910}}.

\bibitem[Zwicky(1933)]{Zwicky:1933gu}
Zwicky, F.
\newblock {Die Rotverschiebung von extragalaktischen Nebeln}.
\newblock {\em Helv. Phys. Acta} {\bf 1933}, {\em 6},~110--127.
\newblock {\url{https://doi.org/10.1007/s10714-008-0707-4}}.

\bibitem[Rubin and Ford(1970)]{Rubin:1970zza}
Rubin, V.C.; Ford, Jr., W.K.
\newblock {Rotation of the Andromeda Nebula from a Spectroscopic Survey of
  Emission Regions}.
\newblock {\em Astrophys. J.} {\bf 1970}, {\em 159},~379--403.
\newblock {\url{https://doi.org/10.1086/150317}}.

\bibitem[Rubin et~al.(1978)Rubin, Ford, and Thonnard]{Rubin:1978kmz}
Rubin, V.C.; Ford, W.K., Jr.; Thonnard, N.
\newblock {Extended rotation curves of high-luminosity spiral galaxies. IV.
  Systematic dynamical properties, Sa through Sc}.
\newblock {\em Astrophys. J. Lett.} {\bf 1978}, {\em 225},~L107--L111.
\newblock {\url{https://doi.org/10.1086/182804}}.

\bibitem[Mohan and Goswami(2024)]{Mohan:2022kvb}
Mohan, G.; Goswami, U.D.
\newblock {Galactic rotation curves of spiral galaxies and dark matter in
  f(R,T) gravity theory}.
\newblock {\em Int. J. Geom. Meth. Mod. Phys.} {\bf 2024}, {\em 21},~2450082.
  %\href{http://arxiv.org/abs/2211.02948}{{\normalfont
  %[arXiv:gr-qc/2211.02948]}}.
\newblock {\url{https://doi.org/10.1142/S0219887824500828}}.

\bibitem[Lasserre(2000)]{Lasserre:2000xw}
Lasserre, T.
\newblock {Not enough stellar mass machos in the galactic halo}.
\newblock {\em Astron. Astrophys.} {\bf 2000}, {\em 355},~L39--L42.
  %\href{http://arxiv.org/abs/astro-ph/0002253}{{\normalfont
  %[astro-ph/0002253]}}.

\bibitem[Famaey and McGaugh(2012)]{Famaey:2011kh}
Famaey, B.; McGaugh, S.
\newblock {Modified Newtonian Dynamics (MOND): Observational Phenomenology and
  Relativistic Extensions}.
\newblock {\em Living Rev. Rel.} {\bf 2012}, {\em 15},~10.
  %\href{http://arxiv.org/abs/1112.3960}{{\normalfont
  %[arXiv:astro-ph.CO/1112.3960]}}.
\newblock {\url{https://doi.org/10.12942/lrr-2012-10}}.

\bibitem[Tisserand et~al.(2007)]{EROS-2:2006ryy}
Tisser, ; P.; Le Guillou, L.; Afonso, C.; Albert, J.N.; Andersen, J.; Ansari, R.; Aubourg, E; Bareyre, P.; Beaulieu, J.P.; Charlot, X.;  et~al.
\newblock {Limits on the Macho Content of the Galactic Halo from the EROS-2
  Survey of the Magellanic Clouds}.
\newblock {\em Astron. Astrophys.} {\bf 2007}, {\em 469},~387--404.
  %\href{http://arxiv.org/abs/astro-ph/0607207}{{\normalfont
  %[astro-ph/0607207]}}.
\newblock {\url{https://doi.org/10.1051/0004-6361:20066017}}.

\bibitem[Clowe et~al.(2006)Clowe, Bradac, Gonzalez, Markevitch, Randall, Jones,
  and Zaritsky]{Clowe:2006eq}
Clowe, D.; Bradac, M.; Gonzalez, A.H.; Markevitch, M.; Randall, S.W.; Jones,
  C.; Zaritsky, D.
\newblock {A direct empirical proof of the existence of dark matter}.
\newblock {\em Astrophys. J. Lett.} {\bf 2006}, {\em 648},~L109--L113.
  %\href{http://arxiv.org/abs/astro-ph/0608407}{{\normalfont
  %[astro-ph/0608407]}}.
\newblock {\url{https://doi.org/10.1086/508162}}.

\bibitem[Boyarsky et~al.(2019)Boyarsky, Drewes, Lasserre, Mertens, and
  Ruchayskiy]{Boyarsky:2018tvu}
Boyarsky, A.; Drewes, M.; Lasserre, T.; Mertens, S.; Ruchayskiy, O.
\newblock {Sterile neutrino Dark Matter}.
\newblock {\em Prog. Part. Nucl. Phys.} {\bf 2019}, {\em 104},~1--45.
  %\href{http://arxiv.org/abs/1807.07938}{{\normalfont
  %[arXiv:hep-ph/1807.07938]}}.
\newblock {\url{https://doi.org/10.1016/j.ppnp.2018.07.004}}.

\bibitem[Pagels and Primack(1982)]{Pagels:1981ke}
Pagels, H.; Primack, J.R.
\newblock {Supersymmetry, Cosmology and New TeV Physics}.
\newblock {\em Phys. Rev. Lett.} {\bf 1982}, {\em 48},~223.
\newblock {\url{https://doi.org/10.1103/PhysRevLett.48.223}}.

\bibitem[Jungman et~al.(1996)Jungman, Kamionkowski, and Griest]{Jungman:1995df}
Jungman, G.; Kamionkowski, M.; Griest, K.
\newblock {Supersymmetric dark matter}.
\newblock {\em Phys. Rept.} {\bf 1996}, {\em 267},~195--373.
  %\href{http://arxiv.org/abs/hep-ph/9506380}{{\normalfont [hep-ph/9506380]}}.
\newblock {\url{https://doi.org/10.1016/0370-1573(95)00058-5}}.

\bibitem[Bertone et~al.(2005)Bertone, Hooper, and Silk]{Bertone:2004pz}
Bertone, G.; Hooper, D.; Silk, J.
\newblock {Particle dark matter: Evidence, candidates and constraints}.
\newblock {\em Phys. Rept.} {\bf 2005}, {\em 405},~279--390.
  %\href{http://arxiv.org/abs/hep-ph/0404175}{{\normalfont [hep-ph/0404175]}}.
\newblock {\url{https://doi.org/10.1016/j.physrep.2004.08.031}}.

\bibitem[Weinberg(1972)]{Weinberg:1972kfs}
Weinberg, S.
\newblock {\em {Gravitation and Cosmology: Principles and Applications of the
  General Theory of Relativity}}; John Wiley and Sons: New York, NY, USA, 1972.

\bibitem[Peccei and Quinn(1977)]{Peccei:1977hh}
Peccei, R.D.; Quinn, H.R.
\newblock {CP Conservation in the Presence of Instantons}.
\newblock {\em Phys. Rev. Lett.} {\bf 1977}, {\em 38},~1440--1443.
\newblock {\url{https://doi.org/10.1103/PhysRevLett.38.1440}}.

\bibitem[Cirelli et~al.(2024)Cirelli, Strumia, and Zupan]{Cirelli:2024ssz}
Cirelli, M.; Strumia, A.; Zupan, J.
\newblock {Dark Matter}. {\bf 2024}.
\newblock {\url{https://doi.org/10.48550/arXiv.2406.01705}}.

\bibitem[Hu et~al.(2000)Hu, Barkana, and Gruzinov]{Hu:2000ke}
Hu, W.; Barkana, R.; Gruzinov, A.
\newblock {Cold and fuzzy dark matter}.
\newblock {\em Phys. Rev. Lett.} {\bf 2000}, {\em 85},~1158--1161.
  %\href{http://arxiv.org/abs/astro-ph/0003365}{{\normalfont
  %[astro-ph/0003365]}}.
\newblock {\url{https://doi.org/10.1103/PhysRevLett.85.1158}}.

\bibitem[Hui et~al.(2017)Hui, Ostriker, Tremaine, and Witten]{Hui:2016ltb}
Hui, L.; Ostriker, J.P.; Tremaine, S.; Witten, E.
\newblock {Ultralight scalars as cosmological dark matter}.
\newblock {\em Phys. Rev. D} {\bf 2017}, {\em 95},~043541.
  %\href{http://arxiv.org/abs/1610.08297}{{\normalfont
  %[arXiv:astro-ph.CO/1610.08297]}}.
\newblock {\url{https://doi.org/10.1103/PhysRevD.95.043541}}.

\bibitem[Ferreira(2021)]{Ferreira:2020fam}
Ferreira, E.G.M.
\newblock {Ultra-light dark matter}.
\newblock {\em Astron. Astrophys. Rev.} {\bf 2021}, {\em 29},~7.
  %\href{http://arxiv.org/abs/2005.03254}{{\normalfont
  %[arXiv:astro-ph.CO/2005.03254]}}.
\newblock {\url{https://doi.org/10.1007/s00159-021-00135-6}}.

\bibitem[Carr et~al.(2021)Carr, Kohri, Sendouda, and Yokoyama]{Carr:2020gox}
Carr, B.; Kohri, K.; Sendouda, Y.; Yokoyama, J.
\newblock {Constraints on primordial black holes}.
\newblock {\em Rept. Prog. Phys.} {\bf 2021}, {\em 84},~116902.
  %\href{http://arxiv.org/abs/2002.12778}{{\normalfont
  %[arXiv:astro-ph.CO/2002.12778]}}.
\newblock {\url{https://doi.org/10.1088/1361-6633/ac1e31}}.

\bibitem[Carr and Kuhnel(2020)]{Carr:2020xqk}
Carr, B.; Kuhnel, F.
\newblock {Primordial Black Holes as Dark Matter: Recent Developments}.
\newblock {\em Ann. Rev. Nucl. Part. Sci.} {\bf 2020}, {\em 70},~355--394.
  %\href{http://arxiv.org/abs/2006.02838}{{\normalfont
  %[arXiv:astro-ph.CO/2006.02838]}}.
\newblock {\url{https://doi.org/10.1146/annurev-nucl-050520-125911}}.

\bibitem[Baryakhtar et~al.(2022)]{Baryakhtar:2022hbu}
Baryakhtar, M.; Caputo, R.; Croon, D.; Perez, K.; Berti, E.; Bramante, J.; Buschmann, M.; Brito, R.; Chen, T.Y.; Cole, P.S.;  et~al.
\newblock {Dark Matter In Extreme Astrophysical Environments}.
\newblock  \emph{Proc. Snowmass 2021}
 \textbf{2022}, \emph{3}.

\bibitem[Aprile et~al.(2018)]{XENON:2018voc}
Aprile, E.; Aalbers, J.; Agostini, F.; Alfonsi, M.; Althueser, L.; Amaro, F.D.; Anthony, M.; Arneodo, F.; Baudis, L.; Bauermeister, B.;  et~al.
\newblock {Dark Matter Search Results from a One Ton-Year Exposure of XENON1T}.
\newblock {\em Phys. Rev. Lett.} {\bf 2018}, {\em 121},~111302.
  %\href{http://arxiv.org/abs/1805.12562}{{\normalfont
  %[arXiv:astro-ph.CO/1805.12562]}}.
\newblock {\url{https://doi.org/10.1103/PhysRevLett.121.111302}}.

\bibitem[Aprile et~al.(2019{\natexlab{a}})]{XENON:2019rxp}
Aprile, E.; Aalbers, J.; Agostini, F.; Alfonsi, M.; Althueser, L.; Amaro, F.D.; Anthony, M.; Antochi, V.C.; Arneodo, F.; Baudis, L.;  et~al.
\newblock {Constraining the spin-dependent WIMP-nucleon cross sections with
  XENON1T}.
\newblock {\em Phys. Rev. Lett.} {\bf 2019}, {\em 122},~141301.
  %\href{http://arxiv.org/abs/1902.03234}{{\normalfont
  %[arXiv:astro-ph.CO/1902.03234]}}.
\newblock {\url{https://doi.org/10.1103/PhysRevLett.122.141301}}.

\bibitem[Aprile et~al.(2019{\natexlab{b}})]{XENON:2019gfn}
Aprile, E.; Aalbers, J.; Agostini, F.; Alfonsi, M.; Althueser, L.; Amaro, F.D.; Antochi, V.C.; Angelino, E.; Arneodo, F.; Barge, D.;  et~al.
\newblock {Light Dark Matter Search with Ionization Signals in XENON1T}.
\newblock {\em Phys. Rev. Lett.} {\bf 2019}, {\em 123},~251801.
  %\href{http://arxiv.org/abs/1907.11485}{{\normalfont
  %[arXiv:hep-ex/1907.11485]}}.
\newblock {\url{https://doi.org/10.1103/PhysRevLett.123.251801}}.

\bibitem[Aprile et~al.(2019{\natexlab{c}})]{XENON:2019zpr}
Aprile, E.; Aalbers, J.; Agostini, F.; Alfonsi, M.; Althueser, L.; Amaro, F.D.; Antochi, V.C.; Angelino, E.; Arneodo, F.; Barge, D.;  et~al.
\newblock {Search for Light Dark Matter Interactions Enhanced by the Migdal
  Effect or Bremsstrahlung in XENON1T}.
\newblock {\em Phys. Rev. Lett.} {\bf 2019}, {\em 123},~241803.
  %\href{http://arxiv.org/abs/1907.12771}{{\normalfont
  %[arXiv:hep-ex/1907.12771]}}.
\newblock {\url{https://doi.org/10.1103/PhysRevLett.123.241803}}.

\bibitem[Aprile et~al.(2021)]{XENON:2020gfr}
Aprile, E.; Aalbers, J.; Agostini, F.; Ahmed, Maouloud, S.; Alfonsi, M.; Althueser, L.; Amaro, F.D.; Andaloro, S.; Antochi, V.C.; Angelino, E.;  et~al.
\newblock {Search for Coherent Elastic Scattering of Solar $^8$B Neutrinos in
  the XENON1T Dark Matter Experiment}.
\newblock {\em Phys. Rev. Lett.} {\bf 2021}, {\em 126},~091301.
  %\href{http://arxiv.org/abs/2012.02846}{{\normalfont
  %[arXiv:hep-ex/2012.02846]}}.
\newblock {\url{https://doi.org/10.1103/PhysRevLett.126.091301}}.

\bibitem[Aprile et~al.(2020)]{XENON:2020kmp}
Aprile, E.; Aalbers, J.; Agostini, F.; Alfonsi, M.; Althueser, L.; Amaro, F.D.; Antochi, V.C.; Angelino, E.; Angevaare, J.R.; Arneodo, F.;  et~al.
\newblock {Projected WIMP sensitivity of the XENONnT dark matter experiment}.
\newblock {\em JCAP} {\bf 2020}, {\em 11},~031.
  %\href{http://arxiv.org/abs/2007.08796}{{\normalfont
  %[arXiv:physics.ins-det/2007.08796]}}.
\newblock {\url{https://doi.org/10.1088/1475-7516/2020/11/031}}.

\bibitem[Bernabei et~al.(2010)]{DAMA:2010gpn}
Bernabei, R.; Belli, P.; Cappella, F.; Cerulli, R.; Dai, C.J.; d’Angelo, A.; He, H.L.; Incicchitti, A.; Kuang, H.H.; Ma, X.H.;  et~al.
\newblock {New results from DAMA/LIBRA}.
\newblock {\em Eur. Phys. J. C} {\bf 2010}, {\em 67},~39--49.
  %\href{http://arxiv.org/abs/1002.1028}{{\normalfont
  %[arXiv:astro-ph.GA/1002.1028]}}.
\newblock {\url{https://doi.org/10.1140/epjc/s10052-010-1303-9}}.

\bibitem[Akerib et~al.(2014)]{LUX:2013afz}
Akerib, D.S.; Araújo, H.M.; Bai, X.; Bailey, A.J.; Balajthy, J.; Bedikian, S.; Bernard, E.; Bernstein, A.; Bolozdynya, A.; Bradley, A.;  et~al.
\newblock {First results from the LUX dark matter experiment at the Sanford
  Underground Research Facility}.
\newblock {\em Phys. Rev. Lett.} {\bf 2014}, {\em 112},~091303.
  %\href{http://arxiv.org/abs/1310.8214}{{\normalfont
  %[arXiv:astro-ph.CO/1310.8214]}}.
\newblock {\url{https://doi.org/10.1103/PhysRevLett.112.091303}}.

\bibitem[Angloher et~al.(2016)]{CRESST:2015txj}
Angloher, G.; Bento, A.; Bucci, C.; Canonica, L.; Defay, X.; Erb, A.; von Feilitzsch, F.; Iachellini, N.F.; Gorla, P.; Gütlein, A.;  et~al.
\newblock {Results on light dark matter particles with a low-threshold
  CRESST-II detector}.
\newblock {\em Eur. Phys. J. C} {\bf 2016}, {\em 76},~25.
  %\href{http://arxiv.org/abs/1509.01515}{{\normalfont
  %[arXiv:astro-ph.CO/1509.01515]}}.
\newblock {\url{https://doi.org/10.1140/epjc/s10052-016-3877-3}}.

\bibitem[Abdelhameed et~al.(2019)]{CRESST:2019jnq}
Abdelhameed, A.H.; Angloher, G.; Bauer, P.; Bento, A.; Bertoldo, E.; Bucci, C.; Canonica, L.; D’Addabbo, A.; Defay, X.; Di Lorenzo, S.;  et~al.
\newblock {First results from the CRESST-III low-mass dark matter program}.
\newblock {\em Phys. Rev. D} {\bf 2019}, {\em 100},~102002.
  %\href{http://arxiv.org/abs/1904.00498}{{\normalfont
  %[arXiv:astro-ph.CO/1904.00498]}}.
\newblock {\url{https://doi.org/10.1103/PhysRevD.100.102002}}.

\bibitem[Agnese et~al.(2014)]{SuperCDMS:2014cds}
Agnese, R.; Anderson, A.J.; Asai, M.; Balakishiyeva, D.; Basu, Thakur, R.; Bauer, D.A.; Beaty, J.; Billard, J.; Borgl, ; A.; Bowles, M.A.;  et~al.
\newblock {Search for Low-Mass Weakly Interacting Massive Particles with
  SuperCDMS}.
\newblock {\em Phys. Rev. Lett.} {\bf 2014}, {\em 112},~241302.
  %\href{http://arxiv.org/abs/1402.7137}{{\normalfont
  %[arXiv:hep-ex/1402.7137]}}.
\newblock {\url{https://doi.org/10.1103/PhysRevLett.112.241302}}.

\bibitem[Agnese et~al.(2018)]{SuperCDMS:2017nns}
Agnese, R.; Anderson, A.J.; Aralis, T.; Aramaki, T.; Arnquist, I.J.; Baker, W.; Balakishiyeva, D.; Barker, D.; Basu, Thakur, R.; Bauer, D.A.;  et~al.
\newblock {Low-mass dark matter search with CDMSlite}.
\newblock {\em Phys. Rev. D} {\bf 2018}, {\em 97},~022002.
  %\href{http://arxiv.org/abs/1707.01632}{{\normalfont
  %[arXiv:astro-ph.CO/1707.01632]}}.
\newblock {\url{https://doi.org/10.1103/PhysRevD.97.022002}}.

\bibitem[Vahsen et~al.(2020)]{Vahsen:2020pzb}
Vahsen, S.E.; O'Hare, C.A.J.; Lynch, W.A.; Spooner, N.J.C.; Baracchini, E.; Barbeau, P.; Battat, J.B.R.; Crow, B.; Deaconu, C.; Eldridge, C.;  et~al.
\newblock {CYGNUS: Feasibility of a nuclear recoil observatory with directional
  sensitivity to dark matter and neutrinos}. \emph{arXiv} {\bf 2020}, arXiv:2008.12587.
\newblock  %\href{http://arxiv.org/abs/2008.12587}{{\normalfont
  %[arXiv:physics.ins-det/2008.12587]}}.

\bibitem[Atwood et~al.(2009)]{Fermi-LAT:2009ihh}
Atwood, W.B.; Abdo, A.A.; Ackermann, M.; Althouse, W.; Anderson, B.; Axelsson, M.; Baldini, L.; Ballet, J.; B.; ; D.L.; Barbiellini, G.;  et~al.
\newblock {The Large Area Telescope on the Fermi Gamma-ray Space Telescope
  Mission}.
\newblock {\em Astrophys. J.} {\bf 2009}, {\em 697},~1071--1102.
  %\href{http://arxiv.org/abs/0902.1089}{{\normalfont
  %[arXiv:astro-ph.IM/0902.1089]}}.
\newblock {\url{https://doi.org/10.1088/0004-637X/697/2/1071}}.

\bibitem[Massey et~al.(2007)]{Massey:2007wb}
Massey, R.; Rhodes, J.; Ellis, R.; Scoville, N.; Leauthaud, A.; Finoguenov, A.; Capak, P.; Bacon, D.; Aussel, H.; Kneib, J.P.;  et~al.
\newblock {Dark matter maps reveal cosmic scaffolding}.
\newblock {\em Nature} {\bf 2007}, {\em 445},~286.
  %\href{http://arxiv.org/abs/astro-ph/0701594}{{\normalfont
  %[astro-ph/0701594]}}.
\newblock {\url{https://doi.org/10.1038/nature05497}}.

\bibitem[Hooper(2019)]{Hooper:2018kfv}
Hooper, D.
\newblock {TASI Lectures on Indirect Searches For Dark Matter}.
\newblock {\em PoS} {\bf 2019}, {\em TASI2018},~010.
  %\href{http://arxiv.org/abs/1812.02029}{{\normalfont
  %[arXiv:hep-ph/1812.02029]}}.

\bibitem[Battaglieri et~al.(2017)]{Battaglieri:2017aum}
Battaglieri, M.; Belloni, A.; Chou, A.; Cushman, P.; Echenard, B.; Essig, R.; Estrada, J.; Feng, J.L.; Flaugher, B.; Fox, P.J.;  et~al.
\newblock {US Cosmic Visions: New Ideas in Dark Matter 2017: Community Report}.
\newblock  \emph{Proc. U.S. Cosm. Visions New Ideas Dark Matter} \textbf{2017}, \emph{7}.  %\href{http://arxiv.org/abs/1707.04591}{{\normalfont
  %[arXiv:hep-ph/1707.04591]}}.

\bibitem[Bertone and Hooper(2018)]{Bertone:2016nfn}
Bertone, G.; Hooper, D.
\newblock {History of dark matter}.
\newblock {\em Rev. Mod. Phys.} {\bf 2018}, {\em 90},~045002.
  %\href{http://arxiv.org/abs/1605.04909}{{\normalfont
  %[arXiv:astro-ph.CO/1605.04909]}}.
\newblock {\url{https://doi.org/10.1103/RevModPhys.90.045002}}.

\bibitem[Weinberg(1978)]{Weinberg:1977ma}
Weinberg, S.
\newblock {A New Light Boson?}
\newblock {\em Phys. Rev. Lett.} {\bf 1978}, {\em 40},~223--226.
\newblock {\url{https://doi.org/10.1103/PhysRevLett.40.223}}.

\bibitem[Marsh(2016)]{Marsh:2015xka}
Marsh, D.J.E.
\newblock {Axion Cosmology}.
\newblock {\em Phys. Rept.} {\bf 2016}, {\em 643},~1--79.
  %\href{http://arxiv.org/abs/1510.07633}{{\normalfont
  %[arXiv:astro-ph.CO/1510.07633]}}.
\newblock {\url{https://doi.org/10.1016/j.physrep.2016.06.005}}.

\bibitem[Eilers et~al.(2019)Eilers, Hogg, Rix, and Ness]{Eilers:2019gqs}
Eilers, A.C.; Hogg, D.W.; Rix, H.W.; Ness, M.K.
\newblock {The Circular Velocity Curve of the Milky Way from 5 to 25 kpc}.
\newblock {\em Astrophys. J.} {\bf 2019}, {\em 871},~120.
\newblock {\url{https://doi.org/10.3847/1538-4357/aaf648}}.

\bibitem[Bell et~al.(2020)Bell, Busoni, Robles, and Virgato]{Bell:2020jou}
Bell, N.F.; Busoni, G.; Robles, S.; Virgato, M.
\newblock {Improved Treatment of Dark Matter Capture in Neutron Stars}.
\newblock {\em JCAP} {\bf 2020}, {\em 9},~028.
  %\href{http://arxiv.org/abs/2004.14888}{{\normalfont
  %[arXiv:hep-ph/2004.14888]}}.
\newblock {\url{https://doi.org/10.1088/1475-7516/2020/09/028}}.

\bibitem[Busoni(2022)]{Busoni:2021zoe}
Busoni, G.
\newblock {Capture of Dark Matter in Neutron Stars}.
\newblock {\em Moscow Univ. Phys. Bull.} {\bf 2022}, {\em 77},~301--305.
  %\href{http://arxiv.org/abs/2201.00048}{{\normalfont
  %[arXiv:hep-ph/2201.00048]}}.
\newblock {\url{https://doi.org/10.3103/S0027134922020205}}.

\bibitem[Kouvaris(2008)]{Kouvaris:2007ay}
Kouvaris, C.
\newblock {WIMP Annihilation and Cooling of Neutron Stars}.
\newblock {\em Phys. Rev. D} {\bf 2008}, {\em 77},~023006.
  %\href{http://arxiv.org/abs/0708.2362}{{\normalfont
  %[arXiv:astro-ph/0708.2362]}}.
\newblock {\url{https://doi.org/10.1103/PhysRevD.77.023006}}.

\bibitem[Feng(2010)]{Feng:2010gw}
Feng, J.L.
\newblock {Dark Matter Candidates from Particle Physics and Methods of
  Detection}.
\newblock {\em Ann. Rev. Astron. Astrophys.} {\bf 2010}, {\em 48},~495--545.
  %\href{http://arxiv.org/abs/1003.0904}{{\normalfont
  %[arXiv:astro-ph.CO/1003.0904]}}.
\newblock {\url{https://doi.org/10.1146/annurev-astro-082708-101659}}.

\bibitem[Arcadi et~al.(2018)Arcadi, Dutra, Ghosh, Lindner, Mambrini, Pierre,
  Profumo, and Queiroz]{Arcadi:2017kky}
Arcadi, G.; Dutra, M.; Ghosh, P.; Lindner, M.; Mambrini, Y.; Pierre, M.;
  Profumo, S.; Queiroz, F.S.
\newblock {The waning of the WIMP? A review of models, searches, and
  constraints}.
\newblock {\em Eur. Phys. J. C} {\bf 2018}, {\em 78},~203.
  %\href{http://arxiv.org/abs/1703.07364}{{\normalfont
  %[arXiv:hep-ph/1703.07364]}}.
\newblock {\url{https://doi.org/10.1140/epjc/s10052-018-5662-y}}.

\bibitem[Bertoni et~al.(2013)Bertoni, Nelson, and Reddy]{Bertoni:2013bsa}
Bertoni, B.; Nelson, A.E.; Reddy, S.
\newblock {Dark Matter Thermalization in Neutron Stars}.
\newblock {\em Phys. Rev. D} {\bf 2013}, {\em 88},~123505.
  %\href{http://arxiv.org/abs/1309.1721}{{\normalfont
  %[arXiv:hep-ph/1309.1721]}}.
\newblock {\url{https://doi.org/10.1103/PhysRevD.88.123505}}.

\bibitem[Bell et~al.(2024)Bell, Busoni, Robles, and Virgato]{Bell:2023ysh}
Bell, N.F.; Busoni, G.; Robles, S.; Virgato, M.
\newblock {Thermalization and annihilation of dark matter in neutron stars}.
\newblock {\em JCAP} {\bf 2024}, {\em 4},~006.
  %\href{http://arxiv.org/abs/2312.11892}{{\normalfont
  %[arXiv:hep-ph/2312.11892]}}.
\newblock {\url{https://doi.org/10.1088/1475-7516/2024/04/006}}.

\bibitem[Giannotti et~al.(2016)Giannotti, Irastorza, Redondo, and
  Ringwald]{Giannotti:2015kwo}
Giannotti, M.; Irastorza, I.; Redondo, J.; Ringwald, A.
\newblock {Cool WISPs for stellar cooling excesses}.
\newblock {\em JCAP} {\bf 2016}, {\em 5},~057.
  %\href{http://arxiv.org/abs/1512.08108}{{\normalfont
  %[arXiv:astro-ph.HE/1512.08108]}}.
\newblock {\url{https://doi.org/10.1088/1475-7516/2016/05/057}}.

\bibitem[Heinke and Ho(2010)]{Heinke:2010cr}
Heinke, C.O.; Ho, W.C.G.
\newblock {Direct Observation of the Cooling of the Cassiopeia A Neutron Star}.
\newblock {\em Astrophys. J. Lett.} {\bf 2010}, {\em 719},~L167--L171.
  %\href{http://arxiv.org/abs/1007.4719}{{\normalfont
  %[arXiv:astro-ph.HE/1007.4719]}}.
\newblock {\url{https://doi.org/10.1088/2041-8205/719/2/L167}}.

\bibitem[Shternin et~al.(2011)Shternin, Yakovlev, Heinke, Ho, and
  Patnaude]{Shternin:2010qi}
Shternin, P.S.; Yakovlev, D.G.; Heinke, C.O.; Ho, W.C.G.; Patnaude, D.J.
\newblock {Cooling neutron star in the Cassiopeia\textasciitilde{}A supernova
  remnant: Evidence for superfluidity in the core}.
\newblock {\em Mon. Not. Roy. Astron. Soc.} {\bf 2011}, {\em 412},~L108--L112.
  %\href{http://arxiv.org/abs/1012.0045}{{\normalfont
  %[arXiv:astro-ph.SR/1012.0045]}}.
\newblock {\url{https://doi.org/10.1111/j.1745-3933.2011.01015.x}}.

\bibitem[Giangrandi et~al.(2024)Giangrandi, \'Avila, Sagun, Ivanytskyi, and
  Provid\^encia]{Giangrandi:2024qdb}
Giangrandi, E.; \'Avila, A.; Sagun, V.; Ivanytskyi, O.; Provid\^encia, C.
\newblock {The Impact of Asymmetric Dark Matter on the Thermal Evolution of
  Nucleonic and Hyperonic Compact Stars}.
\newblock {\em Particles} {\bf 2024}, {\em 7},~179--200.
  %\href{http://arxiv.org/abs/2401.03295}{{\normalfont
  %[arXiv:astro-ph.HE/2401.03295]}}.
\newblock {\url{https://doi.org/10.3390/particles7010010}}.

\bibitem[Hook and Huang(2018)]{Hook:2017psm}
Hook, A.; Huang, J.
\newblock {Probing axions with neutron star inspirals and other stellar
  processes}.
\newblock {\em JHEP} {\bf 2018}, {\em 6},~036.
  %\href{http://arxiv.org/abs/1708.08464}{{\normalfont
  %[arXiv:hep-ph/1708.08464]}}.
\newblock {\url{https://doi.org/10.1007/JHEP06(2018)036}}.

\bibitem[Kumar~Poddar et~al.(2019)Kumar~Poddar, Mohanty, and
  Jana]{KumarPoddar:2019ceq}
Kumar~Poddar, T.; Mohanty, S.; Jana, S.
\newblock {Vector gauge boson radiation from compact binary systems in a gauged
  $L_\mu-L_\tau$ scenario}.
\newblock {\em Phys. Rev. D} {\bf 2019}, {\em 100},~123023.
  %\href{http://arxiv.org/abs/1908.09732}{{\normalfont
  %[arXiv:hep-ph/1908.09732]}}.
\newblock {\url{https://doi.org/10.1103/PhysRevD.100.123023}}.

\bibitem[Kumar~Poddar et~al.(2020)Kumar~Poddar, Mohanty, and
  Jana]{KumarPoddar:2019jxe}
Kumar~Poddar, T.; Mohanty, S.; Jana, S.
\newblock {Constraints on ultralight axions from compact binary systems}.
\newblock {\em Phys. Rev. D} {\bf 2020}, {\em 101},~083007.
  %\href{http://arxiv.org/abs/1906.00666}{{\normalfont
  %[arXiv:hep-ph/1906.00666]}}.
\newblock {\url{https://doi.org/10.1103/PhysRevD.101.083007}}.

\bibitem[Seymour and Yagi(2020)]{Seymour:2019tir}
Seymour, B.C.; Yagi, K.
\newblock {Probing Massive Scalar Fields from a Pulsar in a Stellar Triple
  System}.
\newblock {\em Class. Quant. Grav.} {\bf 2020}, {\em 37},~145008.
  %\href{http://arxiv.org/abs/1908.03353}{{\normalfont
  %[arXiv:gr-qc/1908.03353]}}.
\newblock {\url{https://doi.org/10.1088/1361-6382/ab9933}}.

\bibitem[Dror et~al.(2021)Dror, Lehmann, Patel, and Profumo]{Dror:2021wrl}
Dror, J.A.; Lehmann, B.V.; Patel, H.H.; Profumo, S.
\newblock {Discovering new forces with gravitational waves from supermassive
  black holes}.
\newblock {\em Phys. Rev. D} {\bf 2021}, {\em 104},~083021.
  %\href{http://arxiv.org/abs/2105.04559}{{\normalfont
  %[arXiv:astro-ph.CO/2105.04559]}}.
\newblock {\url{https://doi.org/10.1103/PhysRevD.104.083021}}.

\bibitem[Lambiase and Poddar(2024)]{Lambiase:2024dqe}
Lambiase, G.; Poddar, T.K.
\newblock {Electrophilic scalar hair from rotating magnetized stars and effects
  of cosmic neutrino background}. \emph{arXiv} {\bf 2024}, arXiv:2404.18309.
\newblock  %\href{http://arxiv.org/abs/2404.18309}{{\normalfont
  %[arXiv:hep-ph/2404.18309]}}.

\bibitem[Cline and Cornell(2018)]{Cline:2018ami}
Cline, J.M.; Cornell, J.M.
\newblock {Dark decay of the neutron}.
\newblock {\em JHEP} {\bf 2018}, {\em 7},~081.
  %\href{http://arxiv.org/abs/1803.04961}{{\normalfont
  %[arXiv:hep-ph/1803.04961]}}.
\newblock {\url{https://doi.org/10.1007/JHEP07(2018)081}}.

\bibitem[Husain and Thomas(2023)]{Husain:2022brl}
Husain, W.; Thomas, A.W.
\newblock {Novel neutron decay mode inside neutron stars}.
\newblock {\em J. Phys. G} {\bf 2023}, {\em 50},~015202.
  %\href{http://arxiv.org/abs/2206.11262}{{\normalfont
  %[arXiv:hep-ph/2206.11262]}}.
\newblock {\url{https://doi.org/10.1088/1361-6471/aca1d5}}.

\bibitem[Husain et~al.(2022)Husain, Motta, and Thomas]{Husain:2022bxl}
Husain, W.; Motta, T.F.; Thomas, A.W.
\newblock {Consequences of neutron decay inside neutron stars}.
\newblock {\em JCAP} {\bf 2022}, {\em 10},~028.
  %\href{http://arxiv.org/abs/2203.02758}{{\normalfont
  %[arXiv:hep-ph/2203.02758]}}.
\newblock {\url{https://doi.org/10.1088/1475-7516/2022/10/028}}.

\bibitem[Ellis et~al.(2018)Ellis, Hektor, H\"utsi, Kannike, Marzola, Raidal,
  and Vaskonen]{Ellis:2017jgp}
Ellis, J.; Hektor, A.; H\"utsi, G.; Kannike, K.; Marzola, L.; Raidal, M.;
  Vaskonen, V.
\newblock {Search for Dark Matter Effects on Gravitational Signals from Neutron
  Star Mergers}.
\newblock {\em Phys. Lett. B} {\bf 2018}, {\em 781},~607--610.
  %\href{http://arxiv.org/abs/1710.05540}{{\normalfont
  %[arXiv:astro-ph.CO/1710.05540]}}.
\newblock {\url{https://doi.org/10.1016/j.physletb.2018.04.048}}.

\bibitem[Kopp et~al.(2018)Kopp, Laha, Opferkuch, and Shepherd]{Kopp:2018jom}
Kopp, J.; Laha, R.; Opferkuch, T.; Shepherd, W.
\newblock {Cuckoo\textquoteright{}s eggs in neutron stars: can LIGO hear chirps
  from the dark sector?}
\newblock {\em JHEP} {\bf 2018}, {\em 11},~096.
  %\href{http://arxiv.org/abs/1807.02527}{{\normalfont
  %[arXiv:hep-ph/1807.02527]}}.
\newblock {\url{https://doi.org/10.1007/JHEP11(2018)096}}.

\bibitem[Bhattacharya et~al.(2023)Bhattacharya, Dasgupta, Laha, and
  Ray]{Bhattacharya:2023stq}
Bhattacharya, S.; Dasgupta, B.; Laha, R.; Ray, A.
\newblock {Can LIGO Detect Nonannihilating Dark Matter?}
\newblock {\em Phys. Rev. Lett.} {\bf 2023}, {\em 131},~091401.
  %\href{http://arxiv.org/abs/2302.07898}{{\normalfont
  %[arXiv:hep-ph/2302.07898]}}.
\newblock {\url{https://doi.org/10.1103/PhysRevLett.131.091401}}.

\bibitem[Abbott et~al.(2020)]{LIGOScientific:2020zkf}
Abbott, R.; Abbott, T.D.; Abraham, S.; Acernese, F.; Ackley, K.; Adams, C.; Adhikari, R.X.; Adya, V.B.; Affeldt, C.; Agathos, M.;  et~al.
\newblock {GW190814: Gravitational Waves from the Coalescence of a 23 Solar
  Mass Black Hole with a 2.6 Solar Mass Compact Object}.
\newblock {\em Astrophys. J. Lett.} {\bf 2020}, {\em 896},~L44.
  %\href{http://arxiv.org/abs/2006.12611}{{\normalfont
  %[arXiv:astro-ph.HE/2006.12611]}}.
\newblock {\url{https://doi.org/10.3847/2041-8213/ab960f}}.

\bibitem[Most et~al.(2020)Most, Papenfort, Weih, and Rezzolla]{Most:2020bba}
Most, E.R.; Papenfort, L.J.; Weih, L.R.; Rezzolla, L.
\newblock {A lower bound on the maximum mass if the secondary in GW190814 was
  once a rapidly spinning neutron star}.
\newblock {\em Mon. Not. Roy. Astron. Soc.} {\bf 2020}, {\em 499},~L82--L86.
  %\href{http://arxiv.org/abs/2006.14601}{{\normalfont
  %[arXiv:astro-ph.HE/2006.14601]}}.
\newblock {\url{https://doi.org/10.1093/mnrasl/slaa168}}.

\bibitem[Das et~al.(2021)Das, Kumar, and Patra]{Das:2021yny}
Das, H.C.; Kumar, A.; Patra, S.K.
\newblock {Dark matter admixed neutron star as a possible compact component in
  the GW190814 merger event}.
\newblock {\em Phys. Rev. D} {\bf 2021}, {\em 104},~063028.
  %\href{http://arxiv.org/abs/2109.01853}{{\normalfont
  %[arXiv:astro-ph.HE/2109.01853]}}.
\newblock {\url{https://doi.org/10.1103/PhysRevD.104.063028}}.

\bibitem[Hild et~al.(2011)]{Hild:2010id}
Hild, S.; Abernathy, M.; Acernese, F.E.; Amaro-Seoane, P.; Andersson, N.; Arun, K.; Barone, F.; Barr, B.; Barsuglia, M.; Beker, M.;  et~al.
\newblock {Sensitivity Studies for Third-Generation Gravitational Wave
  Observatories}.
\newblock {\em Class. Quant. Grav.} {\bf 2011}, {\em 28},~094013.
  %\href{http://arxiv.org/abs/1012.0908}{{\normalfont %[arXiv:gr-qc/1012.0908]}}.
\newblock {\url{https://doi.org/10.1088/0264-9381/28/9/094013}}.

\bibitem[Branchesi et~al.(2023)]{Branchesi:2023mws}
Branchesi, M.; Maggiore, M.; Alonso, D.; Badger, C.; Banerjee, B.; Beirnaert, F.; Belgacem, E.; Bhagwat, S.; Boileau, G.; Borhanian, S.;  et~al.
\newblock {Science with the Einstein Telescope: A comparison of different
  designs}.
\newblock {\em JCAP} {\bf 2023}, {\em 7},~068.
  %\href{http://arxiv.org/abs/2303.15923}{{\normalfont
  %[arXiv:gr-qc/2303.15923]}}.
\newblock {\url{https://doi.org/10.1088/1475-7516/2023/07/068}}.

\bibitem[Abbott et~al.(2017)]{LIGOScientific:2016wof}
Abbott, B.P.; Abbott, R.; Abbott, T.D.; Abernathy, M.R.; Ackley, K.; Adams, C.; Addesso, P.; Adhikari, R.X.; Adya, V.B.; Affeldt, C.; et~al.
\newblock {Exploring the Sensitivity of Next Generation Gravitational Wave
  Detectors}.
\newblock {\em Class. Quant. Grav.} {\bf 2017}, {\em 34},~044001.
  %\href{http://arxiv.org/abs/1607.08697}{{\normalfont
  %[arXiv:astro-ph.IM/1607.08697]}}.
\newblock {\url{https://doi.org/10.1088/1361-6382/aa51f4}}.

\bibitem[Evans et~al.(2023)]{Evans:2023euw}
Evans, M.; Corsi, A.; Afle, C.; Ananyeva, A.; Arun, K.G.; Ballmer, S.; B.; Bandopadhyay, A.; Barsotti, L.; Baryakhtar, M.; Berger, E.;  et~al.
\newblock {Cosmic Explorer: A Submission to the NSF MPSAC ngGW Subcommittee}. \emph{arXiv}
  {\bf 2023}, arXiv:2306.13745.
\newblock  %\href{http://arxiv.org/abs/2306.13745}{{\normalfont
  %[arXiv:astro-ph.IM/2306.13745]}}.

\bibitem[Gendreau et~al.(2012)Gendreau, Arzoumanian, and
  Okajima]{Gendreau:NICER}
Gendreau, K.; Arzoumanian, Z.; Okajima, T.
\newblock The Neutron star Interior Composition ExploreR (NICER): An Explorer
  mission of opportunity for soft x-ray timing spectroscopy.
\newblock {\em Proc.  SPIE -Int. Soc. Opt. Eng.} {\bf 2012}, {\em 8443},~322--329.
\newblock {\url{https://doi.org/10.1117/12.926396}}.

\bibitem[Bezares et~al.(2019)Bezares, Vigan\`o, and
  Palenzuela]{Bezares:2019jcb}
Bezares, M.; Vigan\`o, D.; Palenzuela, C.
\newblock {Gravitational wave signatures of dark matter cores in binary neutron
  star mergers by using numerical simulations}.
\newblock {\em Phys. Rev. D} {\bf 2019}, {\em 100},~044049.
  %\href{http://arxiv.org/abs/1905.08551}{{\normalfont
  %[arXiv:gr-qc/1905.08551]}}.
\newblock {\url{https://doi.org/10.1103/PhysRevD.100.044049}}.

\bibitem[Paschalidis et~al.(2015)Paschalidis, East, Pretorius, and
  Shapiro]{Paschalidis:2015mla}
Paschalidis, V.; East, W.E.; Pretorius, F.; Shapiro, S.L.
\newblock {One-arm Spiral Instability in Hypermassive Neutron Stars Formed by
  Dynamical-Capture Binary Neutron Star Mergers}.
\newblock {\em Phys. Rev. D} {\bf 2015}, {\em 92},~121502.
  %\href{http://arxiv.org/abs/1510.03432}{{\normalfont
  %[arXiv:astro-ph.HE/1510.03432]}}.
\newblock {\url{https://doi.org/10.1103/PhysRevD.92.121502}}.

\bibitem[Sun and Wen(2024)]{Sun:2023cqr}
Sun, H.; Wen, D.
\newblock {New criterion for the existence of dark matter in neutron stars}.
\newblock {\em Phys. Rev. D} {\bf 2024}, {\em 109},~123037.
  %\href{http://arxiv.org/abs/2312.17288}{{\normalfont
  %[arXiv:astro-ph.HE/2312.17288]}}.
\newblock {\url{https://doi.org/10.1103/PhysRevD.109.123037}}.

\bibitem[Goldman and Nussinov(1989)]{Goldman:1989nd}
Goldman, I.; Nussinov, S.
\newblock {Weakly Interacting Massive Particles and Neutron Stars}.
\newblock {\em Phys. Rev. D} {\bf 1989}, {\em 40},~3221--3230.
\newblock {\url{https://doi.org/10.1103/PhysRevD.40.3221}}.

\bibitem[Bramante and Raj(2024)]{Bramante:2023djs}
Bramante, J.; Raj, N.
\newblock {Dark matter in compact stars}.
\newblock {\em Phys. Rept.} {\bf 2024}, {\em 1052},~1--48.
  %\href{http://arxiv.org/abs/2307.14435}{{\normalfont
  %[arXiv:hep-ph/2307.14435]}}.
\newblock {\url{https://doi.org/10.1016/j.physrep.2023.12.001}}.

\bibitem[Garani et~al.(2019)Garani, Genolini, and Hambye]{Garani:2018kkd}
Garani, R.; Genolini, Y.; Hambye, T.
\newblock {New Analysis of Neutron Star Constraints on Asymmetric Dark Matter}.
\newblock {\em JCAP} {\bf 2019}, {\em 5},~035.
  %\href{http://arxiv.org/abs/1812.08773}{{\normalfont
  %[arXiv:hep-ph/1812.08773]}}.
\newblock {\url{https://doi.org/10.1088/1475-7516/2019/05/035}}.

\bibitem[Garani and Palomares-Ruiz(2022)]{Garani:2021feo}
Garani, R.; Palomares-Ruiz, S.
\newblock {Evaporation of dark matter from celestial bodies}.
\newblock {\em JCAP} {\bf 2022}, {\em 5},~042.
  %\href{http://arxiv.org/abs/2104.12757}{{\normalfont
  %[arXiv:hep-ph/2104.12757]}}.
\newblock {\url{https://doi.org/10.1088/1475-7516/2022/05/042}}.

\bibitem[Bell et~al.(2018)Bell, Busoni, and Robles]{Bell:2018pkk}
Bell, N.F.; Busoni, G.; Robles, S.
\newblock {Heating up Neutron Stars with Inelastic Dark Matter}.
\newblock {\em JCAP} {\bf 2018}, {\em 9},~018.
  %\href{http://arxiv.org/abs/1807.02840}{{\normalfont
  %[arXiv:hep-ph/1807.02840]}}.
\newblock {\url{https://doi.org/10.1088/1475-7516/2018/09/018}}.

\bibitem[Hamaguchi et~al.(2019)Hamaguchi, Nagata, and
  Yanagi]{Hamaguchi:2019oev}
Hamaguchi, K.; Nagata, N.; Yanagi, K.
\newblock {Dark Matter Heating vs. Rotochemical Heating in Old Neutron Stars}.
\newblock {\em Phys. Lett. B} {\bf 2019}, {\em 795},~484--489.
  %\href{http://arxiv.org/abs/1905.02991}{{\normalfont
  %[arXiv:hep-ph/1905.02991]}}.
\newblock {\url{https://doi.org/10.1016/j.physletb.2019.06.060}}.

\bibitem[Gonzalez and Reisenegger(2010)]{Gonzalez:2010ta}
Gonzalez, D.; Reisenegger, A.
\newblock {Internal Heating of Old Neutron Stars: Contrasting Different
  Mechanisms}.
\newblock {\em Astron. Astrophys.} {\bf 2010}, {\em 522},~A16.
  %\href{http://arxiv.org/abs/1005.5699}{{\normalfont
  %[arXiv:astro-ph.HE/1005.5699]}}.
\newblock {\url{https://doi.org/10.1051/0004-6361/201015084}}.

\bibitem[Gusakov(2002)]{Gusakov:2002hh}
Gusakov, M.E.
\newblock {Neutrino emission from superfluid neutron star cores: Various types
  of neutron pairing}.
\newblock {\em Astron. Astrophys.} {\bf 2002}, {\em 389},~702.
  %\href{http://arxiv.org/abs/astro-ph/0204334}{{\normalfont
  %[astro-ph/0204334]}}.
\newblock {\url{https://doi.org/10.1051/0004-6361:20020602}}.

\bibitem[Page et~al.(2004)Page, Lattimer, Prakash, and Steiner]{Page:2004fy}
Page, D.; Lattimer, J.M.; Prakash, M.; Steiner, A.W.
\newblock {Minimal cooling of neutron stars: A New paradigm}.
\newblock {\em Astrophys. J. Suppl.} {\bf 2004}, {\em 155},~623--650.
  %\href{http://arxiv.org/abs/astro-ph/0403657}{{\normalfont
  %[astro-ph/0403657]}}.
\newblock {\url{https://doi.org/10.1086/424844}}.

\bibitem[Flowers et~al.(1976)Flowers, Ruderman, and Sutherland]{Flowers:1976ux}
Flowers, E.; Ruderman, M.; Sutherland, P.
\newblock {Neutrino pair emission from finite-temperature neutron superfluid
  and the cooling of young neutron stars}.
\newblock {\em Astrophys. J.} {\bf 1976}, {\em 205},~541.
\newblock {\url{https://doi.org/10.1086/154308}}.

\bibitem[Yakovlev et~al.(1999)Yakovlev, Kaminker, and
  Levenfish]{Yakovlev:1998wr}
Yakovlev, D.G.; Kaminker, A.D.; Levenfish, K.P.
\newblock {Neutrino emission due to Cooper pairing of nucleons in cooling
  neutron stars}.
\newblock {\em Astron. Astrophys.} {\bf 1999}, {\em 343},~650.
  %\href{http://arxiv.org/abs/astro-ph/9812366}{{\normalfont
  %[astro-ph/9812366]}}.

\bibitem[Kaminker et~al.(1999)Kaminker, Haensel, and Yakovlev]{Kaminker:1999ez}
Kaminker, A.D.; Haensel, P.; Yakovlev, D.G.
\newblock {Neutrino emission due to proton pairing in neutron stars}.
\newblock {\em Astron. Astrophys.} {\bf 1999}, {\em 345},~L14--L16.
  %\href{http://arxiv.org/abs/astro-ph/9904166}{{\normalfont
  %[astro-ph/9904166]}}.

\bibitem[Yakovlev et~al.(2001)Yakovlev, Kaminker, Gnedin, and
  Haensel]{Yakovlev:2000jp}
Yakovlev, D.G.; Kaminker, A.D.; Gnedin, O.Y.; Haensel, P.
\newblock {Neutrino emission from neutron stars}.
\newblock {\em Phys. Rept.} {\bf 2001}, {\em 354},~1.
  %\href{http://arxiv.org/abs/astro-ph/0012122}{{\normalfont
  %[astro-ph/0012122]}}.
\newblock {\url{https://doi.org/10.1016/S0370-1573(00)00131-9}}.

\bibitem[Ikhsanov(2003)]{Ikhsanov:2003dt}
Ikhsanov, N.R.
\newblock {On the accretion luminosity of isolated neutron stars}.
\newblock {\em Astron. Astrophys.} {\bf 2003}, {\em 399},~1147--1150.
  %\href{http://arxiv.org/abs/astro-ph/0301076}{{\normalfont
  %[astro-ph/0301076]}}.
\newblock {\url{https://doi.org/10.1051/0004-6361:20021889}}.

\bibitem[Yin et~al.(2011)Yin, van Heugten, Diederix, Kater, Vink, and
  Stoof]{Yin:2011aa}
Yin, S.; van Heugten, J.J.R.M.; Diederix, J.; Kater, M.; Vink, J.; Stoof,
  H.T.C.
\newblock {Cooling curves for neutron stars with hadronic matter and quark
  matter}. \emph{arXiv} {\bf 2011}, arXiv:1112.1880.
\newblock  %\href{http://arxiv.org/abs/1112.1880}{{\normalfont
  %[arXiv:astro-ph.HE/1112.1880]}}.

\bibitem[Carroll and Ostlie(2017)]{book:Carroll_Ostlie_2017}
Carroll, B.W.; Ostlie, D.A.
\newblock {\em {An Introduction to Modern Astrophysics}}; Cambridge University
  Press: Cambridge, UK, 2017.

\bibitem[Chatterjee et~al.(2023)Chatterjee, Garani, Jain, Kanodia, Kumar, and
  Vempati]{Chatterjee:2022dhp}
Chatterjee, S.; Garani, R.; Jain, R.K.; Kanodia, B.; Kumar, M.S.N.; Vempati,
  S.K.
\newblock {Faint light of old neutron stars and detectability at the James Webb
  Space Telescope}.
\newblock {\em Phys. Rev. D} {\bf 2023}, {\em 108},~L021301.
  %\href{http://arxiv.org/abs/2205.05048}{{\normalfont
  %[arXiv:astro-ph.HE/2205.05048]}}.
\newblock {\url{https://doi.org/10.1103/PhysRevD.108.L021301}}.

\bibitem[Gardner et~al.(2006)]{Gardner:2006ky}
Gardner, J.P.; Mather, J.C.; Clampin, M.; Doyon, R.; Greenhouse, M.A.; Hammel, H.B.; Hutchings, J.B.; Jakobsen, P.; Lilly, S.J.; Long, K.S.;  et~al.
\newblock {The James Webb Space Telescope}.
\newblock {\em Space Sci. Rev.} {\bf 2006}, {\em 123},~485.
  %\href{http://arxiv.org/abs/astro-ph/0606175}{{\normalfont
  %[astro-ph/0606175]}}.
\newblock {\url{https://doi.org/10.1007/s11214-006-8315-7}}.

\bibitem[Kalirai(2018)]{Kalirai:2018qfg}
Kalirai, J.
\newblock {Scientific Discovery with the James Webb Space Telescope}.
\newblock {\em Contemp. Phys.} {\bf 2018}, {\em 59},~251--290.
  %\href{http://arxiv.org/abs/1805.06941}{{\normalfont
  %[arXiv:astro-ph.IM/1805.06941]}}.
\newblock {\url{https://doi.org/10.1080/00107514.2018.1467648}}.

\bibitem[Rigby et~al.(2023)Rigby, Perrin, et~al.]{2023PASP..135d8001R}
Rigby, J.; Perrin, M.; McElwain, M.; Kimble, R.; Friedman, S.; Lallo, M.; Doyon, R.; Feinberg, L.; Ferruit, P.; Glasse, A.;  et~al.
\newblock {The Science Performance of JWST as Characterized in Commissioning}.
\newblock {\em Publ. Astron. Soc. Pac.} {\bf
  2023}, {\em 135}, 048001.
\newblock {\url{https://doi.org/10.1088/1538-3873/acb293}}.

\bibitem[Yakovlev et~al.(1999)Yakovlev, Levenfish, and
  Shibanov]{Yakovlev:1999sk}
Yakovlev, D.G.; Levenfish, K.P.; Shibanov, Y.A.
\newblock {Cooling neutron stars and superfluidity in their interiors}.
\newblock {\em Phys. Usp.} {\bf 1999}, {\em 42},~737--778.
  %\href{http://arxiv.org/abs/astro-ph/9906456}{{\normalfont
  %[astro-ph/9906456]}}.
\newblock {\url{https://doi.org/10.1070/PU1999v042n08ABEH000556}}.

\bibitem[Yanagi(2019)]{Yanagi:2019zne}
Yanagi, K.
\newblock {Thermal Evolution of Neutron Stars as a Probe of Physics Beyond the
  Standard Model}.
\newblock PhD Thesis. The University of Tokyo, Tokyo, Japan, 2019.
  %\href{http://arxiv.org/abs/2003.08199}{{\normalfont
  %[arXiv:hep-ph/2003.08199]}}.

\bibitem[Fernandez and Reisenegger(2005)]{Fernandez:2005cg}
Fernandez, R.; Reisenegger, A.
\newblock {Rotochemical heating in millisecond pulsars. Formalism and
  non-superfluid case}.
\newblock {\em Astrophys. J.} {\bf 2005}, {\em 625},~291--306.
  %\href{http://arxiv.org/abs/astro-ph/0502116}{{\normalfont
  %[astro-ph/0502116]}}.
\newblock {\url{https://doi.org/10.1086/429551}}.

\bibitem[Villain and Haensel(2005)]{Villain:2005ns}
Villain, L.; Haensel, P.
\newblock {Non-equilibrium beta processes in superfluid neutron star cores}.
\newblock {\em Astron. Astrophys.} {\bf 2005}, {\em 444},~539.
  %\href{http://arxiv.org/abs/astro-ph/0504572}{{\normalfont
  %[astro-ph/0504572]}}.
\newblock {\url{https://doi.org/10.1051/0004-6361:20053313}}.

\bibitem[Petrovich and Reisenegger(2010)]{Petrovich:2009yh}
Petrovich, C.; Reisenegger, A.
\newblock {Rotochemical heating in millisecond pulsars: modified Urca reactions
  with uniform Cooper pairing gaps}.
\newblock {\em Astron. Astrophys.} {\bf 2010}, {\em 521},~A77.
  %\href{http://arxiv.org/abs/0912.2564}{{\normalfont
  %[arXiv:astro-ph.HE/0912.2564]}}.
\newblock {\url{https://doi.org/10.1051/0004-6361/200913861}}.

\bibitem[Gonz\'alez-Jim\'enez et~al.(2015)Gonz\'alez-Jim\'enez, Petrovich, and
  Reisenegger]{Gonzalez-Jimenez:2014iia}
Gonz\'alez-Jim\'enez, N.; Petrovich, C.; Reisenegger, A.
\newblock {Rotochemical heating of millisecond and classical pulsars with
  anisotropic and density-dependent superfluid gap models}.
\newblock {\em Mon. Not. Roy. Astron. Soc.} {\bf 2015}, {\em 447},~2073.
  %\href{http://arxiv.org/abs/1411.6500}{{\normalfont
  %[arXiv:astro-ph.SR/1411.6500]}}.
\newblock {\url{https://doi.org/10.1093/mnras/stu2558}}.

\bibitem[Yanagi et~al.(2020)Yanagi, Nagata, and Hamaguchi]{Yanagi:2019vrr}
Yanagi, K.; Nagata, N.; Hamaguchi, K.
\newblock {Cooling Theory Faced with Old Warm Neutron Stars: Role of
  Non-Equilibrium Processes with Proton and Neutron Gaps}.
\newblock {\em Mon. Not. Roy. Astron. Soc.} {\bf 2020}, {\em 492},~5508--5523.
  %\href{http://arxiv.org/abs/1904.04667}{{\normalfont
  %[arXiv:astro-ph.HE/1904.04667]}}.
\newblock {\url{https://doi.org/10.1093/mnras/staa076}}.

\bibitem[Kargaltsev et~al.(2004)Kargaltsev, Pavlov, and
  Romani]{Kargaltsev:2003eb}
Kargaltsev, O.; Pavlov, G.G.; Romani, R.W.
\newblock {Ultraviolet emission from the millisecond pulsar j0437-4715}.
\newblock {\em Astrophys. J.} {\bf 2004}, {\em 602},~327--335.
  %\href{http://arxiv.org/abs/astro-ph/0310854}{{\normalfont
  %[astro-ph/0310854]}}.
\newblock {\url{https://doi.org/10.1086/380993}}.

\bibitem[Rangelov et~al.(2017)Rangelov, Pavlov, Kargaltsev, Reisenegger,
  Guillot, Reyes, and van Kerkwijk]{Rangelov:2016syg}
Rangelov, B.; Pavlov, G.G.; Kargaltsev, O.; Reisenegger, A.; Guillot, S.;
  Reyes, C.; van Kerkwijk, M.H.
\newblock {Hubble Space Telescope Detection of the Millisecond Pulsar
  J2124\ensuremath{-}3358 and its Far-ultraviolet Bow Shock Nebula}.
\newblock {\em Astrophys. J.} {\bf 2017}, {\em 835},~264.
  %\href{http://arxiv.org/abs/1701.00002}{{\normalfont
  %[arXiv:astro-ph.HE/1701.00002]}}.
\newblock {\url{https://doi.org/10.3847/1538-4357/835/2/264}}.

\bibitem[Pavlov et~al.(2017)Pavlov, Rangelov, Kargaltsev, Reisenegger, Guillot,
  and Reyes]{Pavlov:2017eeu}
Pavlov, G.G.; Rangelov, B.; Kargaltsev, O.; Reisenegger, A.; Guillot, S.;
  Reyes, C.
\newblock {Old but still warm: Far-UV detection of PSR B0950+08}.
\newblock {\em Astrophys. J.} {\bf 2017}, {\em 850},~79.
  %\href{http://arxiv.org/abs/1710.06448}{{\normalfont
  %[arXiv:astro-ph.HE/1710.06448]}}.
\newblock {\url{https://doi.org/10.3847/1538-4357/aa947c}}.

\bibitem[Gonzalez-Caniulef et~al.(2019)Gonzalez-Caniulef, Guillot, and
  Reisenegger]{Gonzalez-Caniulef:2019wzi}
Gonzalez-Caniulef, D.; Guillot, S.; Reisenegger, A.
\newblock {Neutron star radius measurement from the ultraviolet and soft X-ray
  thermal emission of PSR J0437\ensuremath{-}4715}.
\newblock {\em Mon. Not. Roy. Astron. Soc.} {\bf 2019}, {\em 490},~5848--5859.
  %\href{http://arxiv.org/abs/1904.12114}{{\normalfont
  %[arXiv:astro-ph.HE/1904.12114]}}.
\newblock {\url{https://doi.org/10.1093/mnras/stz2941}}.

\bibitem[Garani et~al.(2021)Garani, Gupta, and Raj]{Garani:2020wge}
Garani, R.; Gupta, A.; Raj, N.
\newblock {Observing the thermalization of dark matter in neutron stars}.
\newblock {\em Phys. Rev. D} {\bf 2021}, {\em 103},~043019.
  %\href{http://arxiv.org/abs/2009.10728}{{\normalfont
  %[arXiv:hep-ph/2009.10728]}}.
\newblock {\url{https://doi.org/10.1103/PhysRevD.103.043019}}.

\bibitem[Gaisser et~al.(1986)Gaisser, Steigman, and Tilav]{Gaisser:1986ha}
Gaisser, T.K.; Steigman, G.; Tilav, S.
\newblock {Limits on Cold Dark Matter Candidates from Deep Underground
  Detectors}.
\newblock {\em Phys. Rev. D} {\bf 1986}, {\em 34},~2206.
\newblock {\url{https://doi.org/10.1103/PhysRevD.34.2206}}.

\bibitem[Zentner(2009)]{Zentner:2009is}
Zentner, A.R.
\newblock {High-Energy Neutrinos From Dark Matter Particle Self-Capture Within
  the Sun}.
\newblock {\em Phys. Rev. D} {\bf 2009}, {\em 80},~063501.
  %\href{http://arxiv.org/abs/0907.3448}{{\normalfont
  %[arXiv:astro-ph.HE/0907.3448]}}.
\newblock {\url{https://doi.org/10.1103/PhysRevD.80.063501}}.

\bibitem[Chen and Lin(2018)]{Chen:2018ohx}
Chen, C.S.; Lin, Y.H.
\newblock {Reheating neutron stars with the annihilation of self-interacting
  dark matter}.
\newblock {\em JHEP} {\bf 2018}, {\em 8},~069.
  %\href{http://arxiv.org/abs/1804.03409}{{\normalfont
  %[arXiv:hep-ph/1804.03409]}}.
\newblock {\url{https://doi.org/10.1007/JHEP08(2018)069}}.

\bibitem[G\"uver et~al.(2014)G\"uver, Erkoca, Hall~Reno, and
  Sarcevic]{Guver:2012ba}
G\"uver, T.; Erkoca, A.E.; Hall~Reno, M.; Sarcevic, I.
\newblock {On the capture of dark matter by neutron stars}.
\newblock {\em JCAP} {\bf 2014}, {\em 5},~013.
  %\href{http://arxiv.org/abs/1201.2400}{{\normalfont
  %[arXiv:hep-ph/1201.2400]}}.
\newblock {\url{https://doi.org/10.1088/1475-7516/2014/05/013}}.

\bibitem[Del~Popolo et~al.(2020)Del~Popolo, Le~Delliou, and
  Deliyergiyev]{DelPopolo:2020hel}
Del~Popolo, A.; Le~Delliou, M.; Deliyergiyev, M.
\newblock {Neutron Stars and Dark Matter}.
\newblock {\em Universe} {\bf 2020}, {\em 6},~222.
  %\href{http://arxiv.org/abs/2410.06078}{{\normalfont
  %[arXiv:astro-ph.CO/2410.06078]}}.
\newblock {\url{https://doi.org/10.3390/universe6120222}}.

\bibitem[Acevedo et~al.(2020)Acevedo, Bramante, Leane, and
  Raj]{Acevedo:2019agu}
Acevedo, J.F.; Bramante, J.; Leane, R.K.; Raj, N.
\newblock {Warming Nuclear Pasta with Dark Matter: Kinetic and Annihilation
  Heating of Neutron Star Crusts}.
\newblock {\em JCAP} {\bf 2020}, {\em 3},~038.
  %\href{http://arxiv.org/abs/1911.06334}{{\normalfont
  %[arXiv:hep-ph/1911.06334]}}.
\newblock {\url{https://doi.org/10.1088/1475-7516/2020/03/038}}.

\bibitem[McDermott et~al.(2012)McDermott, Yu, and Zurek]{McDermott:2011jp}
McDermott, S.D.; Yu, H.B.; Zurek, K.M.
\newblock {Constraints on Scalar Asymmetric Dark Matter from Black Hole
  Formation in Neutron Stars}.
\newblock {\em Phys. Rev. D} {\bf 2012}, {\em 85},~023519.
  %\href{http://arxiv.org/abs/1103.5472}{{\normalfont
  %[arXiv:hep-ph/1103.5472]}}.
\newblock {\url{https://doi.org/10.1103/PhysRevD.85.023519}}.

\bibitem[Berryman et~al.(2022)Berryman, Gardner, and Zakeri]{Berryman:2022zic}
Berryman, J.M.; Gardner, S.; Zakeri, M.
\newblock {Neutron Stars with Baryon Number Violation, Probing Dark Sectors}.
\newblock {\em Symmetry} {\bf 2022}, {\em 14},~518.
  %\href{http://arxiv.org/abs/2201.02637}{{\normalfont
  %[arXiv:hep-ph/2201.02637]}}.
\newblock {\url{https://doi.org/10.3390/sym14030518}}.

\bibitem[Kaplan et~al.(2009)Kaplan, Luty, and Zurek]{Kaplan:2009ag}
Kaplan, D.E.; Luty, M.A.; Zurek, K.M.
\newblock {Asymmetric Dark Matter}.
\newblock {\em Phys. Rev. D} {\bf 2009}, {\em 79},~115016.
  %\href{http://arxiv.org/abs/0901.4117}{{\normalfont
  %[arXiv:hep-ph/0901.4117]}}.
\newblock {\url{https://doi.org/10.1103/PhysRevD.79.115016}}.

\bibitem[Petraki and Volkas(2013)]{Petraki:2013wwa}
Petraki, K.; Volkas, R.R.
\newblock {Review of asymmetric dark matter}.
\newblock {\em Int. J. Mod. Phys. A} {\bf 2013}, {\em 28},~1330028.
  %\href{http://arxiv.org/abs/1305.4939}{{\normalfont
  %[arXiv:hep-ph/1305.4939]}}.
\newblock {\url{https://doi.org/10.1142/S0217751X13300287}}.

\bibitem[Zurek(2014)]{Zurek:2013wia}
Zurek, K.M.
\newblock {Asymmetric Dark Matter: Theories, Signatures, and Constraints}.
\newblock {\em Phys. Rept.} {\bf 2014}, {\em 537},~91--121.
  %\href{http://arxiv.org/abs/1308.0338}{{\normalfont
  %[arXiv:hep-ph/1308.0338]}}.
\newblock {\url{https://doi.org/10.1016/j.physrep.2013.12.001}}.

\bibitem[Bramante and Linden(2014)]{Bramante:2014zca}
Bramante, J.; Linden, T.
\newblock {Detecting Dark Matter with Imploding Pulsars in the Galactic
  Center}.
\newblock {\em Phys. Rev. Lett.} {\bf 2014}, {\em 113},~191301.
  %\href{http://arxiv.org/abs/1405.1031}{{\normalfont
  %[arXiv:astro-ph.HE/1405.1031]}}.
\newblock {\url{https://doi.org/10.1103/PhysRevLett.113.191301}}.

\bibitem[Acevedo and Bramante(2019)]{Acevedo:2019gre}
Acevedo, J.F.; Bramante, J.
\newblock {Supernovae Sparked By Dark Matter in White Dwarfs}.
\newblock {\em Phys. Rev. D} {\bf 2019}, {\em 100},~043020.
  %\href{http://arxiv.org/abs/1904.11993}{{\normalfont
  %[arXiv:hep-ph/1904.11993]}}.
\newblock {\url{https://doi.org/10.1103/PhysRevD.100.043020}}.

\bibitem[Acevedo et~al.(2021)Acevedo, Bramante, Goodman, Kopp, and
  Opferkuch]{Acevedo:2020gro}
Acevedo, J.F.; Bramante, J.; Goodman, A.; Kopp, J.; Opferkuch, T.
\newblock {Dark Matter, Destroyer of Worlds: Neutrino, Thermal, and Existential
  Signatures from Black Holes in the Sun and Earth}.
\newblock {\em JCAP} {\bf 2021}, {\em 4},~026.
  %\href{http://arxiv.org/abs/2012.09176}{{\normalfont
  %[arXiv:hep-ph/2012.09176]}}.
\newblock {\url{https://doi.org/10.1088/1475-7516/2021/04/026}}.

\bibitem[Kouvaris and Tinyakov(2011)]{Kouvaris:2010jy}
Kouvaris, C.; Tinyakov, P.
\newblock {Constraining Asymmetric Dark Matter through observations of compact
  stars}.
\newblock {\em Phys. Rev. D} {\bf 2011}, {\em 83},~083512.
  %\href{http://arxiv.org/abs/1012.2039}{{\normalfont
  %[arXiv:astro-ph.HE/1012.2039]}}.
\newblock {\url{https://doi.org/10.1103/PhysRevD.83.083512}}.

\bibitem[Gould et~al.(1990)Gould, Draine, Romani, and Nussinov]{Gould:1989gw}
Gould, A.; Draine, B.T.; Romani, R.W.; Nussinov, S.
\newblock {Neutron Stars: Graveyard of Charged Dark Matter}.
\newblock {\em Phys. Lett. B} {\bf 1990}, {\em 238},~337--343.
\newblock {\url{https://doi.org/10.1016/0370-2693(90)91745-W}}.

\bibitem[Starkman et~al.(1990)Starkman, Gould, Esmailzadeh, and
  Dimopoulos]{Starkman:1990nj}
Starkman, G.D.; Gould, A.; Esmailzadeh, R.; Dimopoulos, S.
\newblock {Opening the Window on Strongly Interacting Dark Matter}.
\newblock {\em Phys. Rev. D} {\bf 1990}, {\em 41},~3594.
\newblock {\url{https://doi.org/10.1103/PhysRevD.41.3594}}.

\bibitem[Mack et~al.(2007)Mack, Beacom, and Bertone]{Mack:2007xj}
Mack, G.D.; Beacom, J.F.; Bertone, G.
\newblock {Towards Closing the Window on Strongly Interacting Dark Matter:
  Far-Reaching Constraints from Earth's Heat Flow}.
\newblock {\em Phys. Rev. D} {\bf 2007}, {\em 76},~043523.
  %\href{http://arxiv.org/abs/0705.4298}{{\normalfont
  %[arXiv:astro-ph/0705.4298]}}.
\newblock {\url{https://doi.org/10.1103/PhysRevD.76.043523}}.

\bibitem[Bramante et~al.(2020)Bramante, Buchanan, Goodman, and
  Lodhi]{Bramante:2019fhi}
Bramante, J.; Buchanan, A.; Goodman, A.; Lodhi, E.
\newblock {Terrestrial and Martian Heat Flow Limits on Dark Matter}.
\newblock {\em Phys. Rev. D} {\bf 2020}, {\em 101},~043001.
  %\href{http://arxiv.org/abs/1909.11683}{{\normalfont
  %[arXiv:hep-ph/1909.11683]}}.
\newblock {\url{https://doi.org/10.1103/PhysRevD.101.043001}}.

\bibitem[Kouvaris and Tinyakov(2011)]{Kouvaris:2011fi}
Kouvaris, C.; Tinyakov, P.
\newblock {Excluding Light Asymmetric Bosonic Dark Matter}.
\newblock {\em Phys. Rev. Lett.} {\bf 2011}, {\em 107},~091301.
  %\href{http://arxiv.org/abs/1104.0382}{{\normalfont
  %[arXiv:astro-ph.CO/1104.0382]}}.
\newblock {\url{https://doi.org/10.1103/PhysRevLett.107.091301}}.

\bibitem[Kouvaris and Tinyakov(2013)]{Kouvaris:2012dz}
Kouvaris, C.; Tinyakov, P.
\newblock {(Not)-constraining heavy asymmetric bosonic dark matter}.
\newblock {\em Phys. Rev. D} {\bf 2013}, {\em 87},~123537.
  %\href{http://arxiv.org/abs/1212.4075}{{\normalfont
  %[arXiv:astro-ph.HE/1212.4075]}}.
\newblock {\url{https://doi.org/10.1103/PhysRevD.87.123537}}.

\bibitem[Jamison(2013)]{Jamison:2013yya}
Jamison, A.O.
\newblock {Effects of gravitational confinement on bosonic asymmetric dark
  matter in stars}.
\newblock {\em Phys. Rev. D} {\bf 2013}, {\em 88},~035004.
  %\href{http://arxiv.org/abs/1304.3773}{{\normalfont
  %[arXiv:hep-ph/1304.3773]}}.
\newblock {\url{https://doi.org/10.1103/PhysRevD.88.035004}}.

\bibitem[Garani et~al.(2022)Garani, Tytgat, and Vandecasteele]{Garani:2022quc}
Garani, R.; Tytgat, M.H.G.; Vandecasteele, J.
\newblock {Condensed dark matter with a Yukawa interaction}.
\newblock {\em Phys. Rev. D} {\bf 2022}, {\em 106},~116003.
  %\href{http://arxiv.org/abs/2207.06928}{{\normalfont
  %[arXiv:hep-ph/2207.06928]}}.
\newblock {\url{https://doi.org/10.1103/PhysRevD.106.116003}}.

\bibitem[Herrera and Santos(1997)]{Herrera:1997plx}
Herrera, L.; Santos, N.O.
\newblock {Local anisotropy in self-gravitating systems}.
\newblock {\em Phys. Rept.} {\bf 1997}, {\em 286},~53--130.
\newblock {\url{https://doi.org/10.1016/S0370-1573(96)00042-7}}.

\bibitem[Chavanis(2011)]{Chavanis:2011zi}
Chavanis, P.H.
\newblock {Mass-radius relation of Newtonian self-gravitating Bose-Einstein
  condensates with short-range interactions: I. Analytical results}.
\newblock {\em Phys. Rev. D} {\bf 2011}, {\em 84},~043531.
  %\href{http://arxiv.org/abs/1103.2050}{{\normalfont
  %[arXiv:astro-ph.CO/1103.2050]}}.
\newblock {\url{https://doi.org/10.1103/PhysRevD.84.043531}}.

\bibitem[Page(1976)]{Page:1976df}
Page, D.N.
\newblock {Particle Emission Rates from a Black Hole: Massless Particles from
  an Uncharged, Nonrotating Hole}.
\newblock {\em Phys. Rev. D} {\bf 1976}, {\em 13},~198--206.
\newblock {\url{https://doi.org/10.1103/PhysRevD.13.198}}.

\bibitem[MacGibbon and Webber(1990)]{MacGibbon:1990zk}
MacGibbon, J.H.; Webber, B.R.
\newblock {Quark and gluon jet emission from primordial black holes: The
  instantaneous spectra}.
\newblock {\em Phys. Rev. D} {\bf 1990}, {\em 41},~3052--3079.
\newblock {\url{https://doi.org/10.1103/PhysRevD.41.3052}}.

\bibitem[MacGibbon(1991)]{MacGibbon:1991tj}
MacGibbon, J.H.
\newblock {Quark and gluon jet emission from primordial black holes. 2. The
  Lifetime emission}.
\newblock {\em Phys. Rev. D} {\bf 1991}, {\em 44},~376--392.
\newblock {\url{https://doi.org/10.1103/PhysRevD.44.376}}.

\bibitem[Bramante et~al.(2018)Bramante, Linden, and Tsai]{Bramante:2017ulk}
Bramante, J.; Linden, T.; Tsai, Y.D.
\newblock {Searching for dark matter with neutron star mergers and quiet
  kilonovae}.
\newblock {\em Phys. Rev. D} {\bf 2018}, {\em 97},~055016.
  %\href{http://arxiv.org/abs/1706.00001}{{\normalfont
  %[arXiv:hep-ph/1706.00001]}}.
\newblock {\url{https://doi.org/10.1103/PhysRevD.97.055016}}.

\bibitem[Ruppin et~al.(2014)Ruppin, Billard, Figueroa-Feliciano, and
  Strigari]{Ruppin:2014bra}
Ruppin, F.; Billard, J.; Figueroa-Feliciano, E.; Strigari, L.
\newblock {Complementarity of dark matter detectors in light of the neutrino
  background}.
\newblock {\em Phys. Rev. D} {\bf 2014}, {\em 90},~083510.
  %\href{http://arxiv.org/abs/1408.3581}{{\normalfont
  %[arXiv:hep-ph/1408.3581]}}.
\newblock {\url{https://doi.org/10.1103/PhysRevD.90.083510}}.

\bibitem[Aalbers et~al.(2023)]{LZ:2022lsv}
Aalbers, J.; Akerib, D.S.; Akerlof, C.W.; Al Musalhi, A.K.; Alder, F.; Alqahtani, A.; Alsum, S.K.; Amarasinghe, C.S.; Ames, A.; Anderson, T.J.;  et~al.
\newblock {First Dark Matter Search Results from the LUX-ZEPLIN (LZ)
  Experiment}.
\newblock {\em Phys. Rev. Lett.} {\bf 2023}, {\em 131},~041002.
  %\href{http://arxiv.org/abs/2207.03764}{{\normalfont
  %[arXiv:hep-ex/2207.03764]}}.
\newblock {\url{https://doi.org/10.1103/PhysRevLett.131.041002}}.

\bibitem[Aalbers et~al.(2024)]{LZCollaboration:2024lux}
Aalbers, J.; Akerib, D.S.; Al Musalhi, A.K.; Alder, F.; Amarasinghe, C.S.; Ames, A.; Anderson, T.J.; Angelides, N.; Araújo, H.M.; Armstrong, J.E.;  et~al.
\newblock {Dark Matter Search Results from 4.2 Tonne-Years of Exposure of the
  LUX-ZEPLIN (LZ) Experiment}. \emph{arXiv} {\bf 2024}, arXiv:2410.17036.
\newblock  %\href{http://arxiv.org/abs/2410.17036}{{\normalfont
  %[arXiv:hep-ex/2410.17036]}}.

\bibitem[Bondi and Hoyle(1944)]{1944MNRAS.104..273B}
Bondi, H.; Hoyle, F.
\newblock {On the mechanism of accretion by stars}.
\newblock {\em Mon. Not. Roy. Astron. Soc.} {\bf 1944}, {\em 104},~273.
\newblock {\url{https://doi.org/10.1093/mnras/104.5.273}}.

\bibitem[Mestel(1954)]{1954MNRAS.114..437M}
Mestel, L.
\newblock {The influence of stellar radiation on the rate of accretion}.
\newblock {\em Mon. Not. Roy. Astron. Soc.} {\bf 1954}, {\em 114},~437.
\newblock {\url{https://doi.org/10.1093/mnras/114.4.437}}.

\bibitem[Bondi(1952)]{1952MNRAS.112..195B}
Bondi, H.
\newblock {On spherically symmetrical accretion}.
\newblock {\em Mon. Not. Roy. Astron. Soc.} {\bf 1952}, {\em 112},~195.
\newblock {\url{https://doi.org/10.1093/mnras/112.2.195}}.

\bibitem[East and Lehner(2019)]{East:2019dxt}
East, W.E.; Lehner, L.
\newblock {Fate of a neutron star with an endoparasitic black hole and
  implications for dark matter}.
\newblock {\em Phys. Rev. D} {\bf 2019}, {\em 100},~124026.
  %\href{http://arxiv.org/abs/1909.07968}{{\normalfont
  %[arXiv:gr-qc/1909.07968]}}.
\newblock {\url{https://doi.org/10.1103/PhysRevD.100.124026}}.

\bibitem[Kouvaris and Tinyakov(2014)]{Kouvaris:2013kra}
Kouvaris, C.; Tinyakov, P.
\newblock {Growth of Black Holes in the interior of Rotating Neutron Stars}.
\newblock {\em Phys. Rev. D} {\bf 2014}, {\em 90},~043512.
  %\href{http://arxiv.org/abs/1312.3764}{{\normalfont
  %[arXiv:astro-ph.SR/1312.3764]}}.
\newblock {\url{https://doi.org/10.1103/PhysRevD.90.043512}}.

\bibitem[Giffin et~al.(2022)Giffin, Lloyd, McDermott, and
  Profumo]{Giffin:2021kgb}
Giffin, P.; Lloyd, J.; McDermott, S.D.; Profumo, S.
\newblock {Neutron star quantum death by small black holes}.
\newblock {\em Phys. Rev. D} {\bf 2022}, {\em 105},~123030.
  %\href{http://arxiv.org/abs/2105.06504}{{\normalfont
  %[arXiv:hep-ph/2105.06504]}}.
\newblock {\url{https://doi.org/10.1103/PhysRevD.105.123030}}.

\bibitem[Doran et~al.(2005)Doran, Lasenby, Dolan, and Hinder]{Doran:2005vm}
Doran, C.; Lasenby, A.; Dolan, S.; Hinder, I.
\newblock {Fermion absorption cross section of a Schwarzschild black hole}.
\newblock {\em Phys. Rev. D} {\bf 2005}, {\em 71},~124020.
  %\href{http://arxiv.org/abs/gr-qc/0503019}{{\normalfont [gr-qc/0503019]}}.
\newblock {\url{https://doi.org/10.1103/PhysRevD.71.124020}}.

\bibitem[Bramante et~al.(2014)Bramante, Fukushima, Kumar, and
  Stopnitzky]{Bramante:2013nma}
Bramante, J.; Fukushima, K.; Kumar, J.; Stopnitzky, E.
\newblock {Bounds on self-interacting fermion dark matter from observations of
  old neutron stars}.
\newblock {\em Phys. Rev. D} {\bf 2014}, {\em 89},~015010.
  %\href{http://arxiv.org/abs/1310.3509}{{\normalfont
  %[arXiv:hep-ph/1310.3509]}}.
\newblock {\url{https://doi.org/10.1103/PhysRevD.89.015010}}.

\bibitem[Hawking(1974)]{Hawking:1974rv}
Hawking, S.W.
\newblock {Black hole explosions}.
\newblock {\em Nature} {\bf 1974}, {\em 248},~30--31.
\newblock {\url{https://doi.org/10.1038/248030a0}}.

\bibitem[Ray(2023)]{Ray:2023auh}
Ray, A.
\newblock {Celestial objects as strongly-interacting nonannihilating dark
  matter detectors}.
\newblock {\em Phys. Rev. D} {\bf 2023}, {\em 107},~083012.
  %\href{http://arxiv.org/abs/2301.03625}{{\normalfont
  %[arXiv:hep-ph/2301.03625]}}.
\newblock {\url{https://doi.org/10.1103/PhysRevD.107.083012}}.

\bibitem[Basumatary et~al.(2024)Basumatary, Raj, and Ray]{Basumatary:2024uwo}
Basumatary, U.; Raj, N.; Ray, A.
\newblock {Beyond Hawking evaporation of black holes formed by dark matter in
  compact stars}. \emph{Phys. Rev. D} {\bf 2024}, \emph{111}, L041306.
\newblock  %\href{http://arxiv.org/abs/2410.22702}{{\normalfont
  %[arXiv:hep-ph/2410.22702]}}.

\bibitem[Smith et~al.(2023)]{Fermi-LAT:2023zzt}
Smith, D.A.; Abdollahi, S.; Ajello, M.; Bailes, M.; Baldini, L.; Ballet, J.; Baring, M.G.; Bassa, C.; Gonzalez, J.B.; Bellazzini, R.;  et~al.
\newblock {The Third Fermi Large Area Telescope Catalog of Gamma-Ray Pulsars}.
\newblock {\em Astrophys. J.} {\bf 2023}, {\em 958},~191.
  %\href{http://arxiv.org/abs/2307.11132}{{\normalfont
  %[arXiv:astro-ph.HE/2307.11132]}}.
\newblock {\url{https://doi.org/10.3847/1538-4357/acee67}}.

\bibitem[Bramante and Elahi(2015)]{Bramante:2015dfa}
Bramante, J.; Elahi, F.
\newblock {Higgs portals to pulsar collapse}.
\newblock {\em Phys. Rev. D} {\bf 2015}, {\em 91},~115001.
  %\href{http://arxiv.org/abs/1504.04019}{{\normalfont
  %[arXiv:hep-ph/1504.04019]}}.
\newblock {\url{https://doi.org/10.1103/PhysRevD.91.115001}}.

\bibitem[Perna et~al.(2003)Perna, Narayan, Rybicki, Stella, and
  Treves]{Perna:2003ck}
Perna, R.; Narayan, R.; Rybicki, G.; Stella, L.; Treves, A.
\newblock {Bondi accretion and the problem of the missing isolated neutron
  stars}.
\newblock {\em Astrophys. J.} {\bf 2003}, {\em 594},~936--942.
  %\href{http://arxiv.org/abs/astro-ph/0305421}{{\normalfont
  %[astro-ph/0305421]}}.
\newblock {\url{https://doi.org/10.1086/377091}}.

\bibitem[Dexter and O'Leary(2014)]{Dexter:2013xga}
Dexter, J.; O'Leary, R.M.
\newblock {The Peculiar Pulsar Population of the Central Parsec}.
\newblock {\em Astrophys. J. Lett.} {\bf 2014}, {\em 783},~L7.
  %\href{http://arxiv.org/abs/1310.7022}{{\normalfont
  %[arXiv:astro-ph.GA/1310.7022]}}.
\newblock {\url{https://doi.org/10.1088/2041-8205/783/1/L7}}.

\bibitem[Suresh et~al.(2022)Suresh, Cordes, Chatterjee, Gajjar, Perez, Siemion,
  Lebofsky, MacMahon, and Ng]{Suresh:2022vmf}
Suresh, A.; Cordes, J.M.; Chatterjee, S.; Gajjar, V.; Perez, K.I.; Siemion,
  A.P.V.; Lebofsky, M.; MacMahon, D.H.E.; Ng, C.
\newblock {4\textendash{}8 GHz Fourier-domain Searches for Galactic Center
  Pulsars}.
\newblock {\em Astrophys. J.} {\bf 2022}, {\em 933},~121.
  %\href{http://arxiv.org/abs/2203.00036}{{\normalfont
  %[arXiv:astro-ph.HE/2203.00036]}}.
\newblock {\url{https://doi.org/10.3847/1538-4357/ac74c0}}.

\bibitem[Fuller and Ott(2015)]{Fuller:2014rza}
Fuller, J.; Ott, C.
\newblock {Dark Matter-induced Collapse of Neutron Stars: A Possible Link
  Between Fast Radio Bursts and the Missing Pulsar Problem}.
\newblock {\em Mon. Not. Roy. Astron. Soc.} {\bf 2015}, {\em 450},~L71--L75.
  %\href{http://arxiv.org/abs/1412.6119}{{\normalfont
  %[arXiv:astro-ph.HE/1412.6119]}}.
\newblock {\url{https://doi.org/10.1093/mnrasl/slv049}}.

\bibitem[Bramante(2015)]{Bramante:2015cua}
Bramante, J.
\newblock {Dark matter ignition of type Ia supernovae}.
\newblock {\em Phys. Rev. Lett.} {\bf 2015}, {\em 115},~141301.
  %\href{http://arxiv.org/abs/1505.07464}{{\normalfont
  %[arXiv:hep-ph/1505.07464]}}.
\newblock {\url{https://doi.org/10.1103/PhysRevLett.115.141301}}.

\bibitem[Bramante and Linden(2016)]{Bramante:2016mzo}
Bramante, J.; Linden, T.
\newblock {On the $r$-Process Enrichment of Dwarf Spheroidal Galaxies}.
\newblock {\em Astrophys. J.} {\bf 2016}, {\em 826},~57.
  %\href{http://arxiv.org/abs/1601.06784}{{\normalfont
  %[arXiv:astro-ph.HE/1601.06784]}}.
\newblock {\url{https://doi.org/10.3847/0004-637X/826/1/57}}.

\bibitem[Randall et~al.(2008)Randall, Markevitch, Clowe, Gonzalez, and
  Bradac]{Randall:2008ppe}
Randall, S.W.; Markevitch, M.; Clowe, D.; Gonzalez, A.H.; Bradac, M.
\newblock {Constraints on the Self-Interaction Cross-Section of Dark Matter
  from Numerical Simulations of the Merging Galaxy Cluster 1E 0657-56}.
\newblock {\em Astrophys. J.} {\bf 2008}, {\em 679},~1173--1180.
  %\href{http://arxiv.org/abs/0704.0261}{{\normalfont
  %[arXiv:astro-ph/0704.0261]}}.
\newblock {\url{https://doi.org/10.1086/587859}}.

\bibitem[Marrod\'an~Undagoitia and Rauch(2016)]{MarrodanUndagoitia:2015veg}
Marrod\'an~Undagoitia, T.; Rauch, L.
\newblock {Dark matter direct-detection experiments}.
\newblock {\em J. Phys. G} {\bf 2016}, {\em 43},~013001.
  %\href{http://arxiv.org/abs/1509.08767}{{\normalfont
  %[arXiv:physics.ins-det/1509.08767]}}.
\newblock {\url{https://doi.org/10.1088/0954-3899/43/1/013001}}.

\bibitem[Akerib et~al.(2017)]{LUX:2016ggv}
Akerib, D.S.; Alsum, S.; Araújo, H.M.; Bai, X.; Bailey, A.J.; Balajthy, J.; Beltrame, P.; Bernard, E.P.; Bernstein, A.; Biesiadzinski, T.P.;  et~al.
\newblock {Results from a search for dark matter in the complete LUX exposure}.
\newblock {\em Phys. Rev. Lett.} {\bf 2017}, {\em 118},~021303.
  %\href{http://arxiv.org/abs/1608.07648}{{\normalfont
  %[arXiv:astro-ph.CO/1608.07648]}}.
\newblock {\url{https://doi.org/10.1103/PhysRevLett.118.021303}}.

\bibitem[Aprile et~al.(2023)]{XENON:2023cxc}
Aprile, E.; Abe, K.; Agostini, F.; Ahmed, Maouloud, S.; Althueser, L.; Andrieu, B.; Angelino, E.; Angevaare, J.R.; Antochi, V.C.; Antón, Martin, D.;  et~al.
\newblock {First Dark Matter Search with Nuclear Recoils from the XENONnT
  Experiment}.
\newblock {\em Phys. Rev. Lett.} {\bf 2023}, {\em 131},~041003.
  %\href{http://arxiv.org/abs/2303.14729}{{\normalfont
  %[arXiv:hep-ex/2303.14729]}}.
\newblock {\url{https://doi.org/10.1103/PhysRevLett.131.041003}}.

\bibitem[Ajaj et~al.(2019)]{DEAP:2019yzn}
Ajaj, R.; Amaudruz P. A.; Araujo G. R.; Baldwin M.; Batygov M.; Beltran B.; Bina C. E.; Bonatt J.; Boulay M. G.; Broerman B. et~al.
\newblock {Search for dark matter with a 231-day exposure of liquid argon using
  DEAP-3600 at SNOLAB}.
\newblock {\em Phys. Rev. D} {\bf 2019}, {\em 100},~022004.
  %\href{http://arxiv.org/abs/1902.04048}{{\normalfont
  %[arXiv:astro-ph.CO/1902.04048]}}.
\newblock {\url{https://doi.org/10.1103/PhysRevD.100.022004}}.

\bibitem[Amole et~al.(2019)]{PICO:2019vsc}
Amole, C.;  et~al.
\newblock {Dark Matter Search Results from the Complete Exposure of the PICO-60
  C3F8 Bubble Chamber}.
\newblock {\em Phys. Rev. D} {\bf 2019}, {\em 100},~022001.
  %\href{http://arxiv.org/abs/1902.04031}{{\normalfont
  %[arXiv:astro-ph.CO/1902.04031]}}.
\newblock {\url{https://doi.org/10.1103/PhysRevD.100.022001}}.

\bibitem[Nelson et~al.(2019)Nelson, Reddy, and Zhou]{Nelson:2018xtr}
Nelson, A.; Reddy, S.; Zhou, D.
\newblock {Dark halos around neutron stars and gravitational waves}.
\newblock {\em JCAP} {\bf 2019}, {\em 7},~012.
  %\href{http://arxiv.org/abs/1803.03266}{{\normalfont
  %[arXiv:hep-ph/1803.03266]}}.
\newblock {\url{https://doi.org/10.1088/1475-7516/2019/07/012}}.

\bibitem[Deliyergiyev et~al.(2019)Deliyergiyev, Del~Popolo, Tolos, Le~Delliou,
  Lee, and Burgio]{Deliyergiyev:2019vti}
Deliyergiyev, M.; Del~Popolo, A.; Tolos, L.; Le~Delliou, M.; Lee, X.; Burgio,
  F.
\newblock {Dark compact objects: an extensive overview}.
\newblock {\em Phys. Rev. D} {\bf 2019}, {\em 99},~063015.
  %\href{http://arxiv.org/abs/1903.01183}{{\normalfont
  %[arXiv:gr-qc/1903.01183]}}.
\newblock {\url{https://doi.org/10.1103/PhysRevD.99.063015}}.

\bibitem[Oppenheimer and Volkoff(1939)]{PhysRev.55.374}
Oppenheimer, J.R.; Volkoff, G.M.
\newblock On Massive Neutron Cores.
\newblock {\em Phys. Rev.} {\bf 1939}, {\em 55},~374--381.
\newblock {\url{https://doi.org/10.1103/PhysRev.55.374}}.

\bibitem[Tolman(1939)]{Tolman:1939jz}
Tolman, R.C.
\newblock {Static solutions of Einstein's field equations for spheres of
  fluid}.
\newblock {\em Phys. Rev.} {\bf 1939}, {\em 55},~364--373.
\newblock {\url{https://doi.org/10.1103/PhysRev.55.364}}.

\bibitem[Burgio et~al.(2021)Burgio, Schulze, Vidana, and Wei]{Burgio:2021vgk}
Burgio, G.F.; Schulze, H.J.; Vidana, I.; Wei, J.B.
\newblock {Neutron stars and the nuclear equation of state}.
\newblock {\em Prog. Part. Nucl. Phys.} {\bf 2021}, {\em 120},~103879.
  %\href{http://arxiv.org/abs/2105.03747}{{\normalfont
  %[arXiv:nucl-th/2105.03747]}}.
\newblock {\url{https://doi.org/10.1016/j.ppnp.2021.103879}}.

\bibitem[Chavanis and Harko(2012)]{Chavanis:2011cz}
Chavanis, P.H.; Harko, T.
\newblock {Bose-Einstein Condensate general relativistic stars}.
\newblock {\em Phys. Rev. D} {\bf 2012}, {\em 86},~064011.
  %\href{http://arxiv.org/abs/1108.3986}{{\normalfont
  %[arXiv:astro-ph.SR/1108.3986]}}.
\newblock {\url{https://doi.org/10.1103/PhysRevD.86.064011}}.

\bibitem[Giangrandi et~al.(2023)Giangrandi, Sagun, Ivanytskyi, Provid\^encia,
  and Dietrich]{Giangrandi:2022wht}
Giangrandi, E.; Sagun, V.; Ivanytskyi, O.; Provid\^encia, C.; Dietrich, T.
\newblock {The Effects of Self-interacting Bosonic Dark Matter on Neutron Star
  Properties}.
\newblock {\em Astrophys. J.} {\bf 2023}, {\em 953},~115.
  %\href{http://arxiv.org/abs/2209.10905}{{\normalfont
  %[arXiv:astro-ph.HE/2209.10905]}}.
\newblock {\url{https://doi.org/10.3847/1538-4357/ace104}}.

\bibitem[Rafiei~Karkevandi et~al.(2021)Rafiei~Karkevandi, Shakeri, Sagun, and
  Ivanytskyi]{RafieiKarkevandi:2021hcc}
Rafiei~Karkevandi, D.; Shakeri, S.; Sagun, V.; Ivanytskyi, O.
\newblock {Tidal deformability as a probe of dark matter in neutron stars}.
\newblock In Proceedings of the 16th Marcel Grossmann Meeting on~Recent
  Developments in Theoretical and Experimental General Relativity, Astrophysics
  and Relativistic Field Theories, Virtual Meeting, 05/07/2021 to 10/07/2021; Volume {12}.
  %\href{http://arxiv.org/abs/2112.14231}{{\normalfont
  %[arXiv:astro-ph.HE/2112.14231]}}.
\newblock {\url{https://doi.org/10.1142/9789811269776_0307}}.

\bibitem[Karkevandi et~al.(2022)Karkevandi, Shakeri, Sagun, and
  Ivanytskyi]{Karkevandi:2021ygv}
Karkevandi, D.R.; Shakeri, S.; Sagun, V.; Ivanytskyi, O.
\newblock {Bosonic dark matter in neutron stars and its effect on gravitational
  wave signal}.
\newblock {\em Phys. Rev. D} {\bf 2022}, {\em 105},~023001.
  %\href{http://arxiv.org/abs/2109.03801}{{\normalfont
  %[arXiv:astro-ph.HE/2109.03801]}}.
\newblock {\url{https://doi.org/10.1103/PhysRevD.105.023001}}.

\bibitem[Narain et~al.(2006)Narain, Schaffner-Bielich, and
  Mishustin]{Narain:2006kx}
Narain, G.; Schaffner-Bielich, J.; Mishustin, I.N.
\newblock {Compact stars made of fermionic dark matter}.
\newblock {\em Phys. Rev. D} {\bf 2006}, {\em 74},~063003.
  %\href{http://arxiv.org/abs/astro-ph/0605724}{{\normalfont
  %[astro-ph/0605724]}}.
\newblock {\url{https://doi.org/10.1103/PhysRevD.74.063003}}.

\bibitem[Goldman et~al.(2013)Goldman, Mohapatra, Nussinov, Rosenbaum, and
  Teplitz]{Goldman:2013qla}
Goldman, I.; Mohapatra, R.N.; Nussinov, S.; Rosenbaum, D.; Teplitz, V.
\newblock {Possible Implications of Asymmetric Fermionic Dark Matter for
  Neutron Stars}.
\newblock {\em Phys. Lett. B} {\bf 2013}, {\em 725},~200--207.
  %\href{http://arxiv.org/abs/1305.6908}{{\normalfont
  %[arXiv:astro-ph.CO/1305.6908]}}.
\newblock {\url{https://doi.org/10.1016/j.physletb.2013.07.017}}.

\bibitem[Leung et~al.(2011)Leung, Chu, and Lin]{Leung:2011zz}
Leung, S.C.; Chu, M.C.; Lin, L.M.
\newblock {Dark-matter admixed neutron stars}.
\newblock {\em Phys. Rev. D} {\bf 2011}, {\em 84},~107301.
  %\href{http://arxiv.org/abs/1111.1787}{{\normalfont
  %[arXiv:astro-ph.CO/1111.1787]}}.
\newblock {\url{https://doi.org/10.1103/PhysRevD.84.107301}}.

\bibitem[Ivanytskyi et~al.(2020)Ivanytskyi, Sagun, and
  Lopes]{Ivanytskyi:2019wxd}
Ivanytskyi, O.; Sagun, V.; Lopes, I.
\newblock {Neutron stars: New constraints on asymmetric dark matter}.
\newblock {\em Phys. Rev. D} {\bf 2020}, {\em 102},~063028.
  %\href{http://arxiv.org/abs/1910.09925}{{\normalfont
  %[arXiv:astro-ph.HE/1910.09925]}}.
\newblock {\url{https://doi.org/10.1103/PhysRevD.102.063028}}.

\bibitem[Karkevandi et~al.(2024)Karkevandi, Shahrbaf, Shakeri, and
  Typel]{Karkevandi:2024vov}
Karkevandi, D.R.; Shahrbaf, M.; Shakeri, S.; Typel, S.
\newblock {Exploring the Distribution and Impact of Bosonic Dark Matter in
  Neutron Stars}.
\newblock {\em Particles} {\bf 2024}, {\em 7},~201--213.
  %\href{http://arxiv.org/abs/2402.18696}{{\normalfont
  %[arXiv:astro-ph.HE/2402.18696]}}.
\newblock {\url{https://doi.org/10.3390/particles7010011}}.

\bibitem[Shakeri and Karkevandi(2024)]{Shakeri:2022dwg}
Shakeri, S.; Karkevandi, D.R.
\newblock {Bosonic dark matter in light of the NICER precise mass-radius
  measurements}.
\newblock {\em Phys. Rev. D} {\bf 2024}, {\em 109},~043029.
  %\href{http://arxiv.org/abs/2210.17308}{{\normalfont
  %[arXiv:astro-ph.HE/2210.17308]}}.
\newblock {\url{https://doi.org/10.1103/PhysRevD.109.043029}}.

\bibitem[Liu et~al.(2023)Liu, Wei, Li, Burgio, and Schulze]{Liu:2023ecz}
Liu, H.M.; Wei, J.B.; Li, Z.H.; Burgio, G.F.; Schulze, H.J.
\newblock {Dark matter effects on the properties of neutron stars: Optical
  radii}.
\newblock {\em Phys. Dark Univ.} {\bf 2023}, {\em 42},~101338.
  %\href{http://arxiv.org/abs/2307.11313}{{\normalfont
  %[arXiv:nucl-th/2307.11313]}}.
\newblock {\url{https://doi.org/10.1016/j.dark.2023.101338}}.

\bibitem[Liu et~al.(2024)Liu, Wei, Li, Burgio, Das, and Schulze]{Liu:2024rix}
Liu, H.M.; Wei, J.B.; Li, Z.H.; Burgio, G.F.; Das, H.C.; Schulze, H.J.
\newblock {Dark matter effects on the properties of neutron stars: Compactness
  and tidal deformability}. \emph{Phys. Rev. D} {\bf 2024}, \emph{110}, 023024.
\newblock  %\href{http://arxiv.org/abs/2403.17024}{{\normalfont
  %[arXiv:nucl-th/2403.17024]}}.

\bibitem[Jockel and Sagunski(2024)]{Jockel:2023rrm}
Jockel, C.; Sagunski, L.
\newblock {Fermion Proca Stars: Vector Dark Matter Admixed Neutron Stars}.
\newblock {\em Particles} {\bf 2024}, {\em 7},~52--79.
  %\href{http://arxiv.org/abs/2310.17291}{{\normalfont
  %[arXiv:gr-qc/2310.17291]}}.
\newblock {\url{https://doi.org/10.3390/particles7010004}}.

\bibitem[Diedrichs et~al.(2023)Diedrichs, Becker, Jockel, Christian, Sagunski,
  and Schaffner-Bielich]{Diedrichs:2023trk}
Diedrichs, R.F.; Becker, N.; Jockel, C.; Christian, J.E.; Sagunski, L.;
  Schaffner-Bielich, J.
\newblock {Tidal deformability of fermion-boson stars: Neutron stars admixed
  with ultralight dark matter}.
\newblock {\em Phys. Rev. D} {\bf 2023}, {\em 108},~064009.
  %\href{http://arxiv.org/abs/2303.04089}{{\normalfont
  %[arXiv:gr-qc/2303.04089]}}.
\newblock {\url{https://doi.org/10.1103/PhysRevD.108.064009}}.

\bibitem[Ciancarella et~al.(2021)Ciancarella, Pannarale, Addazi, and
  Marciano]{Ciancarella:2020msu}
Ciancarella, R.; Pannarale, F.; Addazi, A.; Marciano, A.
\newblock {Constraining mirror dark matter inside neutron stars}.
\newblock {\em Phys. Dark Univ.} {\bf 2021}, {\em 32},~100796.
  %\href{http://arxiv.org/abs/2010.12904}{{\normalfont
  %[arXiv:astro-ph.HE/2010.12904]}}.
\newblock {\url{https://doi.org/10.1016/j.dark.2021.100796}}.

\bibitem[Berezhiani et~al.(2021)Berezhiani, Biondi, Mannarelli, and
  Tonelli]{Berezhiani:2020zck}
Berezhiani, Z.; Biondi, R.; Mannarelli, M.; Tonelli, F.
\newblock {Neutron-mirror neutron mixing and neutron stars}.
\newblock {\em Eur. Phys. J. C} {\bf 2021}, {\em 81},~1036.
  %\href{http://arxiv.org/abs/2012.15233}{{\normalfont
  %[arXiv:astro-ph.HE/2012.15233]}}.
\newblock {\url{https://doi.org/10.1140/epjc/s10052-021-09806-1}}.

\bibitem[Kain(2021)]{Kain:2021hpk}
Kain, B.
\newblock {Dark matter admixed neutron stars}.
\newblock {\em Phys. Rev. D} {\bf 2021}, {\em 103},~043009.
  %\href{http://arxiv.org/abs/2102.08257}{{\normalfont
  %[arXiv:gr-qc/2102.08257]}}.
\newblock {\url{https://doi.org/10.1103/PhysRevD.103.043009}}.

\bibitem[Berezhiani(2022)]{Berezhiani:2021src}
Berezhiani, Z.
\newblock {Antistars or Antimatter Cores in Mirror Neutron Stars?}
\newblock {\em Universe} {\bf 2022}, {\em 8},~313.
  %\href{http://arxiv.org/abs/2106.11203}{{\normalfont
  %[arXiv:astro-ph.HE/2106.11203]}}.
\newblock {\url{https://doi.org/10.3390/universe8060313}}.

\bibitem[Yang et~al.(2021)Yang, Pi, and Zheng]{Yang:2021bpe}
Yang, S.H.; Pi, C.M.; Zheng, X.P.
\newblock {Strange stars with a mirror-dark-matter core confronting with the
  observations of compact stars}.
\newblock {\em Phys. Rev. D} {\bf 2021}, {\em 104},~083016.
  %\href{http://arxiv.org/abs/2103.05159}{{\normalfont
  %[arXiv:astro-ph.HE/2103.05159]}}.
\newblock {\url{https://doi.org/10.1103/PhysRevD.104.083016}}.

\bibitem[Yang and Pi(2024)]{Yang:2024sxi}
Yang, S.H.; Pi, C.M.
\newblock {Color-flavor locked strange stars admixed with mirror dark matter
  and the observations of compact stars}.
\newblock {\em JCAP} {\bf 2024}, {\em 9},~052.
  %\href{http://arxiv.org/abs/2402.14262}{{\normalfont
  %[arXiv:astro-ph.HE/2402.14262]}}.
\newblock {\url{https://doi.org/10.1088/1475-7516/2024/09/052}}.

\bibitem[Yang et~al.(2024)Yang, Pi, and Weber]{Yang:2024ycl}
Yang, S.H.; Pi, C.M.; Weber, F.
\newblock {Strange stars admixed with mirror dark matter: confronting
  observations of XTE J1814-338}. \emph{arXiv} {\bf 2024}, arXiv:2409.15969.
\newblock  %\href{http://arxiv.org/abs/2409.15969}{{\normalfont
  %[arXiv:astro-ph.HE/2409.15969]}}.

\bibitem[Rezaei(2023)]{Rezaei:2023iif}
Rezaei, Z.
\newblock {Fuzzy dark matter in relativistic stars}.
\newblock {\em Mon. Not. Roy. Astron. Soc.} {\bf 2023}, {\em 524},~2015--2024.
  %\href{http://arxiv.org/abs/2306.17665}{{\normalfont
  %[arXiv:astro-ph.HE/2306.17665]}}.
\newblock {\url{https://doi.org/10.1093/mnras/stad1975}}.

\bibitem[Shahrbaf et~al.(2022)Shahrbaf, Blaschke, Typel, Farrar, and
  Alvarez-Castillo]{Shahrbaf:2022upc}
Shahrbaf, M.; Blaschke, D.; Typel, S.; Farrar, G.R.; Alvarez-Castillo, D.E.
\newblock {Sexaquark dilemma in neutron stars and its solution by quark
  deconfinement}.
\newblock {\em Phys. Rev. D} {\bf 2022}, {\em 105},~103005.
  %\href{http://arxiv.org/abs/2202.00652}{{\normalfont
  %[arXiv:nucl-th/2202.00652]}}.
\newblock {\url{https://doi.org/10.1103/PhysRevD.105.103005}}.

\bibitem[Shahrbaf et~al.(2024)Shahrbaf, Karkevandi, and
  Typel]{Shahrbaf:2024gdm}
Shahrbaf, M.; Karkevandi, D.R.; Typel, S.
\newblock {Constraints on the mass of a bosonic dark matter candidate within
  the DD2Y-T model}. \emph{arXiv} {\bf 2024}, arXiv:2402.18686.
\newblock  %\href{http://arxiv.org/abs/2402.18686}{{\normalfont
  %[arXiv:nucl-th/2402.18686]}}.

\bibitem[Chen et~al.(2015)Chen, Suyama, and Yokoyama]{Chen:2015zmx}
Chen, P.; Suyama, T.; Yokoyama, J.
\newblock {Spontaneous scalarization: asymmetron as dark matter}.
\newblock {\em Phys. Rev. D} {\bf 2015}, {\em 92},~124016.
  %\href{http://arxiv.org/abs/1508.01384}{{\normalfont
  %[arXiv:gr-qc/1508.01384]}}.
\newblock {\url{https://doi.org/10.1103/PhysRevD.92.124016}}.

\bibitem[Morisaki and Suyama(2017)]{Morisaki:2017nit}
Morisaki, S.; Suyama, T.
\newblock {Spontaneous scalarization with an extremely massive field and heavy
  neutron stars}.
\newblock {\em Phys. Rev. D} {\bf 2017}, {\em 96},~084026.
  %\href{http://arxiv.org/abs/1707.02809}{{\normalfont
  %[arXiv:gr-qc/1707.02809]}}.
\newblock {\url{https://doi.org/10.1103/PhysRevD.96.084026}}.

\bibitem[Sotani and Kokkotas(2017)]{Sotani:2017pfj}
Sotani, H.; Kokkotas, K.D.
\newblock {Maximum mass limit of neutron stars in scalar-tensor gravity}.
\newblock {\em Phys. Rev. D} {\bf 2017}, {\em 95},~044032.
  %\href{http://arxiv.org/abs/1702.00874}{{\normalfont
  %[arXiv:gr-qc/1702.00874]}}.
\newblock {\url{https://doi.org/10.1103/PhysRevD.95.044032}}.

\bibitem[Davoudiasl(2024)]{Davoudiasl:2024grq}
Davoudiasl, H.
\newblock {Gravitationally misaligned ultralight dark matter and implications
  for neutron stars}.
\newblock {\em Phys. Rev. D} {\bf 2024}, {\em 110},~095020.
  %\href{http://arxiv.org/abs/2408.12667}{{\normalfont
  %[arXiv:hep-ph/2408.12667]}}.
\newblock {\url{https://doi.org/10.1103/PhysRevD.110.095020}}.

\bibitem[Reddy and Zhou(2022)]{Reddy:2021rln}
Reddy, S.; Zhou, D.
\newblock {Dark lepton superfluid in protoneutron stars}.
\newblock {\em Phys. Rev. D} {\bf 2022}, {\em 105},~023026.
  %\href{http://arxiv.org/abs/2107.06279}{{\normalfont
  %[arXiv:hep-ph/2107.06279]}}.
\newblock {\url{https://doi.org/10.1103/PhysRevD.105.023026}}.

\bibitem[Fujiwara et~al.(2024)Fujiwara, Hamaguchi, Nagata, and
  Ramirez-Quezada]{Fujiwara:2023hlj}
Fujiwara, M.; Hamaguchi, K.; Nagata, N.; Ramirez-Quezada, M.E.
\newblock {Vortex creep heating vs. dark matter heating in neutron stars}.
\newblock {\em Phys. Lett. B} {\bf 2024}, {\em 848},~138341.
  %\href{http://arxiv.org/abs/2309.02633}{{\normalfont
  %[arXiv:hep-ph/2309.02633]}}.
\newblock {\url{https://doi.org/10.1016/j.physletb.2023.138341}}.

\bibitem[Panotopoulos and Lopes(2017)]{Panotopoulos:2017idn}
Panotopoulos, G.; Lopes, I.
\newblock {Dark matter effect on realistic equation of state in neutron stars}.
\newblock {\em Phys. Rev. D} {\bf 2017}, {\em 96},~083004.
  %\href{http://arxiv.org/abs/1709.06312}{{\normalfont
  %[arXiv:hep-ph/1709.06312]}}.
\newblock {\url{https://doi.org/10.1103/PhysRevD.96.083004}}.

\bibitem[Das et~al.(2019)Das, Malik, and Nayak]{Das:2018frc}
Das, A.; Malik, T.; Nayak, A.C.
\newblock {Confronting nuclear equation of state in the presence of dark matter
  using GW170817 observation in relativistic mean field theory approach}.
\newblock {\em Phys. Rev. D} {\bf 2019}, {\em 99},~043016.
  %\href{http://arxiv.org/abs/1807.10013}{{\normalfont
  %[arXiv:hep-ph/1807.10013]}}.
\newblock {\url{https://doi.org/10.1103/PhysRevD.99.043016}}.

\bibitem[Das et~al.(2021{\natexlab{a}})Das, Kumar, Kumar, Biswal, and
  Patra]{Das:2020ptd}
Das, H.C.; Kumar, A.; Kumar, B.; Biswal, S.K.; Patra, S.K.
\newblock {Impacts of dark matter on the curvature of the neutron star}.
\newblock {\em JCAP} {\bf 2021}, {\em 1},~007.
  %\href{http://arxiv.org/abs/2007.05382}{{\normalfont
  %[arXiv:nucl-th/2007.05382]}}.
\newblock {\url{https://doi.org/10.1088/1475-7516/2021/01/007}}.

\bibitem[Das et~al.(2021{\natexlab{b}})Das, Kumar, and Patra]{Das:2021wku}
Das, H.C.; Kumar, A.; Patra, S.K.
\newblock {Effects of dark matter on the in-spiral properties of the binary
  neutron stars}.
\newblock {\em Mon. Not. Roy. Astron. Soc.} {\bf 2021}, {\em 507},~4053--4060.
  %\href{http://arxiv.org/abs/2104.01815}{{\normalfont
  %[arXiv:astro-ph.HE/2104.01815]}}.
\newblock {\url{https://doi.org/10.1093/mnras/stab2387}}.

\bibitem[Sen and Guha(2021)]{Sen:2021wev}
Sen, D.; Guha, A.
\newblock {Implications of feebly interacting dark sector on neutron star
  properties and constraints from GW170817}.
\newblock {\em Mon. Not. Roy. Astron. Soc.} {\bf 2021}, {\em 504},~3354--3363.
  %\href{http://arxiv.org/abs/2104.06141}{{\normalfont
  %[arXiv:hep-ph/2104.06141]}}.
\newblock {\url{https://doi.org/10.1093/mnras/stab1056}}.

\bibitem[Lenzi et~al.(2023)Lenzi, Dutra, Louren\c{c}o, Lopes, and
  Menezes]{Lenzi:2022ypb}
Lenzi, C.H.; Dutra, M.; Louren\c{c}o, O.; Lopes, L.L.; Menezes, D.P.
\newblock {Dark matter effects on hybrid star properties}.
\newblock {\em Eur. Phys. J. C} {\bf 2023}, {\em 83},~266.
  %\href{http://arxiv.org/abs/2212.12615}{{\normalfont
  %[arXiv:hep-ph/2212.12615]}}.
\newblock {\url{https://doi.org/10.1140/epjc/s10052-023-11416-y}}.

\bibitem[Dutra et~al.(2022)Dutra, Lenzi, and Louren\c{c}o]{Dutra:2022mxl}
Dutra, M.; Lenzi, C.H.; Louren\c{c}o, O.
\newblock {Dark particle mass effects on neutron star properties from a
  short-range correlated hadronic model}.
\newblock {\em Mon. Not. Roy. Astron. Soc.} {\bf 2022}, {\em 517},~4265--4274.
  %\href{http://arxiv.org/abs/2211.10263}{{\normalfont
  %[arXiv:nucl-th/2211.10263]}}.
\newblock {\url{https://doi.org/10.1093/mnras/stac2986}}.

\bibitem[Flores et~al.(2024)Flores, Lenzi, Dutra, Louren\c{c}o, and
  Arba\~nil]{Flores:2024hts}
Flores, C.V.; Lenzi, C.H.; Dutra, M.; Louren\c{c}o, O.; Arba\~nil, J.D.V.
\newblock {Gravitational wave asteroseismology of dark matter hadronic stars}.
\newblock {\em Phys. Rev. D} {\bf 2024}, {\em 109},~083021.
  %\href{http://arxiv.org/abs/2402.12600}{{\normalfont
  %[arXiv:hep-ph/2402.12600]}}.
\newblock {\url{https://doi.org/10.1103/PhysRevD.109.083021}}.

\bibitem[Hong and Ren(2024)]{Hong:2024sey}
Hong, B.; Ren, Z.
\newblock {Mixed dark matter models for the peculiar compact object in remnant
  HESS J1731-347 and their implications for gravitational wave properties}.
\newblock {\em Phys. Rev. D} {\bf 2024}, {\em 109},~023002.
\newblock {\url{https://doi.org/10.1103/PhysRevD.109.023002}}.

\bibitem[Thakur et~al.(2024)Thakur, Malik, Das, Jha, Sharma, and
  Provid\^encia]{Thakur:2024btu}
Thakur, P.; Malik, T.; Das, A.; Jha, T.K.; Sharma, B.K.; Provid\^encia, C.
\newblock {Feasibility of dark matter admixed neutron star based on recent
  observational constraints}. \emph{arXiv} {\bf 2024}, arXiv:2408.03780.
\newblock  %\href{http://arxiv.org/abs/2408.03780}{{\normalfont
  %[arXiv:nucl-th/2408.03780]}}.

\bibitem[Guha and Sen(2021)]{Guha:2021njn}
Guha, A.; Sen, D.
\newblock {Feeble DM-SM interaction via new scalar and vector mediators in
  rotating neutron stars}.
\newblock {\em JCAP} {\bf 2021}, {\em 9},~027.
  %\href{http://arxiv.org/abs/2106.10353}{{\normalfont
  %[arXiv:hep-ph/2106.10353]}}.
\newblock {\url{https://doi.org/10.1088/1475-7516/2021/09/027}}.

\bibitem[Guha and Sen(2024)]{Guha:2024pnn}
Guha, A.; Sen, D.
\newblock {Constraining the mass of fermionic dark matter from its feeble
  interaction with hadronic matter via dark mediators in neutron stars}.
\newblock {\em Phys. Rev. D} {\bf 2024}, {\em 109},~043038.
  %\href{http://arxiv.org/abs/2401.14419}{{\normalfont
  %[arXiv:astro-ph.HE/2401.14419]}}.
\newblock {\url{https://doi.org/10.1103/PhysRevD.109.043038}}.

\bibitem[Motta et~al.(2018)Motta, Guichon, and Thomas]{Motta:2018bil}
Motta, T.F.; Guichon, P.A.M.; Thomas, A.W.
\newblock {Neutron to Dark Matter Decay in Neutron Stars}.
\newblock {\em Int. J. Mod. Phys. A} {\bf 2018}, {\em 33},~1844020.
  %\href{http://arxiv.org/abs/1806.00903}{{\normalfont
  %[arXiv:nucl-th/1806.00903]}}.
\newblock {\url{https://doi.org/10.1142/S0217751X18440207}}.

\bibitem[Shirke et~al.(2023)Shirke, Ghosh, Chatterjee, Sagunski, and
  Schaffner-Bielich]{Shirke:2023ktu}
Shirke, S.; Ghosh, S.; Chatterjee, D.; Sagunski, L.; Schaffner-Bielich, J.
\newblock {R-modes as a new probe of dark matter in neutron stars}.
\newblock {\em JCAP} {\bf 2023}, {\em 12},~008.
  %\href{http://arxiv.org/abs/2305.05664}{{\normalfont
  %[arXiv:astro-ph.HE/2305.05664]}}.
\newblock {\url{https://doi.org/10.1088/1475-7516/2023/12/008}}.

\bibitem[Shirke et~al.(2024)Shirke, Pradhan, Chatterjee, Sagunski, and
  Schaffner-Bielich]{Shirke:2024ymc}
Shirke, S.; Pradhan, B.K.; Chatterjee, D.; Sagunski, L.; Schaffner-Bielich, J.
\newblock {Effects of dark matter on f-mode oscillations of neutron stars}.
\newblock {\em Phys. Rev. D} {\bf 2024}, {\em 110},~063025.
  %\href{http://arxiv.org/abs/2403.18740}{{\normalfont
  %[arXiv:gr-qc/2403.18740]}}.
\newblock {\url{https://doi.org/10.1103/PhysRevD.110.063025}}.

\bibitem[Bastero-Gil et~al.(2024)Bastero-Gil, Huertas-Roldan, and
  Santos]{Bastero-Gil:2024kjo}
Bastero-Gil, M.; Huertas-Roldan, T.; Santos, D.
\newblock {Neutron decay anomaly, neutron stars, and dark matter}.
\newblock {\em Phys. Rev. D} {\bf 2024}, {\em 110},~083003.
  %\href{http://arxiv.org/abs/2403.08666}{{\normalfont
  %[arXiv:astro-ph.CO/2403.08666]}}.
\newblock {\url{https://doi.org/10.1103/PhysRevD.110.083003}}.

\bibitem[Bauswein et~al.(2012)Bauswein, Janka, Hebeler, and
  Schwenk]{Bauswein:2012ya}
Bauswein, A.; Janka, H.T.; Hebeler, K.; Schwenk, A.
\newblock {Equation-of-state dependence of the gravitational-wave signal from
  the ring-down phase of neutron-star mergers}.
\newblock {\em Phys. Rev. D} {\bf 2012}, {\em 86},~063001.
  %\href{http://arxiv.org/abs/1204.1888}{{\normalfont
  %[arXiv:astro-ph.SR/1204.1888]}}.
\newblock {\url{https://doi.org/10.1103/PhysRevD.86.063001}}.

\bibitem[Das et~al.(2022)Das, Malik, and Nayak]{Das:2020ecp}
Das, A.; Malik, T.; Nayak, A.C.
\newblock {Dark matter admixed neutron star properties in light of
  gravitational wave observations: A two fluid approach}.
\newblock {\em Phys. Rev. D} {\bf 2022}, {\em 105},~123034.
  %\href{http://arxiv.org/abs/2011.01318}{{\normalfont
  %[arXiv:nucl-th/2011.01318]}}.
\newblock {\url{https://doi.org/10.1103/PhysRevD.105.123034}}.

\bibitem[Hippert et~al.(2023)Hippert, Dillingham, Tan, Curtin, Noronha-Hostler,
  and Yunes]{Hippert:2022snq}
Hippert, M.; Dillingham, E.; Tan, H.; Curtin, D.; Noronha-Hostler, J.; Yunes,
  N.
\newblock {Dark matter or regular matter in neutron stars? How to tell the
  difference from the coalescence of compact objects}.
\newblock {\em Phys. Rev. D} {\bf 2023}, {\em 107},~115028.
  %\href{http://arxiv.org/abs/2211.08590}{{\normalfont
  %[arXiv:astro-ph.HE/2211.08590]}}.
\newblock {\url{https://doi.org/10.1103/PhysRevD.107.115028}}.

\bibitem[Raithel et~al.(2018)Raithel, \"Ozel, and Psaltis]{Raithel:2018ncd}
Raithel, C.; \"Ozel, F.; Psaltis, D.
\newblock {Tidal deformability from GW170817 as a direct probe of the neutron
  star radius}.
\newblock {\em Astrophys. J. Lett.} {\bf 2018}, {\em 857},~L23.
  %\href{http://arxiv.org/abs/1803.07687}{{\normalfont
  %[arXiv:astro-ph.HE/1803.07687]}}.
\newblock {\url{https://doi.org/10.3847/2041-8213/aabcbf}}.

\bibitem[Zhao and Lattimer(2018)]{Zhao:2018nyf}
Zhao, T.; Lattimer, J.M.
\newblock {Tidal Deformabilities and Neutron Star Mergers}.
\newblock {\em Phys. Rev. D} {\bf 2018}, {\em 98},~063020.
  %\href{http://arxiv.org/abs/1808.02858}{{\normalfont
  %[arXiv:astro-ph.HE/1808.02858]}}.
\newblock {\url{https://doi.org/10.1103/PhysRevD.98.063020}}.

\bibitem[Hinderer(2008)]{Hinderer:2007mb}
Hinderer, T.
\newblock {Tidal Love numbers of neutron stars}.
\newblock {\em Astrophys. J.} {\bf 2008}, {\em 677},~1216--1220.
  %\href{http://arxiv.org/abs/0711.2420}{{\normalfont
  %[arXiv:astro-ph/0711.2420]}}.
\newblock Erratum in \emph{Astrophys. J.} \textbf{2009}, \emph{697}, 964;
  {\url{https://doi.org/10.1086/533487}}.

\bibitem[Flanagan and Hinderer(2008)]{Flanagan:2007ix}
Flanagan, E.E.; Hinderer, T.
\newblock {Constraining neutron star tidal Love numbers with gravitational wave
  detectors}.
\newblock {\em Phys. Rev. D} {\bf 2008}, {\em 77},~021502.
  %\href{http://arxiv.org/abs/0709.1915}{{\normalfont
  %[arXiv:astro-ph/0709.1915]}}.
\newblock {\url{https://doi.org/10.1103/PhysRevD.77.021502}}.

\bibitem[Damour and Nagar(2009)]{Damour:2009vw}
Damour, T.; Nagar, A.
\newblock {Relativistic tidal properties of neutron stars}.
\newblock {\em Phys. Rev. D} {\bf 2009}, {\em 80},~084035.
  %\href{http://arxiv.org/abs/0906.0096}{{\normalfont %[arXiv:gr-qc/0906.0096]}}.
\newblock {\url{https://doi.org/10.1103/PhysRevD.80.084035}}.

\bibitem[Hinderer et~al.(2010)Hinderer, Lackey, Lang, and
  Read]{Hinderer:2009ca}
Hinderer, T.; Lackey, B.D.; Lang, R.N.; Read, J.S.
\newblock {Tidal deformability of neutron stars with realistic equations of
  state and their gravitational wave signatures in binary inspiral}.
\newblock {\em Phys. Rev. D} {\bf 2010}, {\em 81},~123016.
  %\href{http://arxiv.org/abs/0911.3535}{{\normalfont
  %[arXiv:astro-ph.HE/0911.3535]}}.
\newblock {\url{https://doi.org/10.1103/PhysRevD.81.123016}}.

\bibitem[Postnikov et~al.(2010)Postnikov, Prakash, and
  Lattimer]{Postnikov:2010yn}
Postnikov, S.; Prakash, M.; Lattimer, J.M.
\newblock {Tidal Love Numbers of Neutron and Self-Bound Quark Stars}.
\newblock {\em Phys. Rev. D} {\bf 2010}, {\em 82},~024016.
  %\href{http://arxiv.org/abs/1004.5098}{{\normalfont
  %[arXiv:astro-ph.SR/1004.5098]}}.
\newblock {\url{https://doi.org/10.1103/PhysRevD.82.024016}}.

\bibitem[Li et~al.(2012)Li, Wang, and Cheng]{Li:2012qf}
Li, X.; Wang, F.; Cheng, K.S.
\newblock {Gravitational effects of condensate dark matter on compact stellar
  objects}.
\newblock {\em JCAP} {\bf 2012}, {\em 10},~031.
  %\href{http://arxiv.org/abs/1210.1748}{{\normalfont
  %[arXiv:astro-ph.CO/1210.1748]}}.
\newblock {\url{https://doi.org/10.1088/1475-7516/2012/10/031}}.

\bibitem[Arbey et~al.(2003)Arbey, Lesgourgues, and Salati]{Arbey:2003sj}
Arbey, A.; Lesgourgues, J.; Salati, P.
\newblock {Galactic halos of fluid dark matter}.
\newblock {\em Phys. Rev. D} {\bf 2003}, {\em 68},~023511.
  %\href{http://arxiv.org/abs/astro-ph/0301533}{{\normalfont
  %[astro-ph/0301533]}}.
\newblock {\url{https://doi.org/10.1103/PhysRevD.68.023511}}.

\bibitem[Su\'arez et~al.(2014)Su\'arez, Robles, and Matos]{Suarez:2013iw}
Su\'arez, A.; Robles, V.H.; Matos, T.
\newblock {A Review on the Scalar Field/Bose-Einstein Condensate Dark Matter
  Model}.
\newblock {\em Astrophys. Space Sci. Proc.} {\bf 2014}, {\em 38},~107--142.
  %\href{http://arxiv.org/abs/1302.0903}{{\normalfont
  %[arXiv:astro-ph.CO/1302.0903]}}.
\newblock {\url{https://doi.org/10.1007/978-3-319-02063-1_9}}.

\bibitem[Rutherford et~al.(2023)Rutherford, Raaijmakers, Prescod-Weinstein, and
  Watts]{Rutherford:2022xeb}
Rutherford, N.; Raaijmakers, G.; Prescod-Weinstein, C.; Watts, A.
\newblock {Constraining bosonic asymmetric dark matter with neutron star
  mass-radius measurements}.
\newblock {\em Phys. Rev. D} {\bf 2023}, {\em 107},~103051.
  %\href{http://arxiv.org/abs/2208.03282}{{\normalfont
  %[arXiv:astro-ph.HE/2208.03282]}}.
\newblock {\url{https://doi.org/10.1103/PhysRevD.107.103051}}.

\bibitem[Shen et~al.(1998)Shen, Toki, Oyamatsu, and Sumiyoshi]{Shen:1998gq}
Shen, H.; Toki, H.; Oyamatsu, K.; Sumiyoshi, K.
\newblock {Relativistic equation of state of nuclear matter for supernova and
  neutron star}.
\newblock {\em Nucl. Phys. A} {\bf 1998}, {\em 637},~435--450.
  %\href{http://arxiv.org/abs/nucl-th/9805035}{{\normalfont [nucl-th/9805035]}}.
\newblock {\url{https://doi.org/10.1016/S0375-9474(98)00236-X}}.

\bibitem[Diener(2008)]{Diener:2008bj}
Diener, J.P.W.
\newblock {Relativistic mean-field theory applied to the study of neutron star
  properties}.
\newblock Master's Thesis,  2008,
\newblock{\url{https://arxiv.org/abs/0806.0747}}

\bibitem[Miao et~al.(2022)Miao, Zhu, Li, and Huang]{Miao:2022rqj}
Miao, Z.; Zhu, Y.; Li, A.; Huang, F.
\newblock {Dark Matter Admixed Neutron Star Properties in the Light of X-Ray
  Pulse Profile Observations}.
\newblock {\em Astrophys. J.} {\bf 2022}, {\em 936},~69.
  %\href{http://arxiv.org/abs/2204.05560}{{\normalfont
  %[arXiv:astro-ph.HE/2204.05560]}}.
\newblock {\url{https://doi.org/10.3847/1538-4357/ac8544}}.

\bibitem[Collier et~al.(2022)Collier, Croon, and Leane]{Collier:2022cpr}
Collier, M.; Croon, D.; Leane, R.K.
\newblock {Tidal Love numbers of novel and admixed celestial objects}.
\newblock {\em Phys. Rev. D} {\bf 2022}, {\em 106},~123027.
  %\href{http://arxiv.org/abs/2205.15337}{{\normalfont
  %[arXiv:gr-qc/2205.15337]}}.
\newblock {\url{https://doi.org/10.1103/PhysRevD.106.123027}}.

\bibitem[Routaray et~al.(2023)Routaray, Mohanty, Das, Ghosh, Kalita, Parmar,
  and Kumar]{Routaray:2023spb}
Routaray, P.; Mohanty, S.R.; Das, H.C.; Ghosh, S.; Kalita, P.J.; Parmar, V.;
  Kumar, B.
\newblock {Investigating dark matter-admixed neutron stars with NITR equation
  of state in light of PSR J0952-0607}.
\newblock {\em JCAP} {\bf 2023}, {\em 10},~073.
  %\href{http://arxiv.org/abs/2304.05100}{{\normalfont
  %[arXiv:nucl-th/2304.05100]}}.
\newblock {\url{https://doi.org/10.1088/1475-7516/2023/10/073}}.

\bibitem[Grippa et~al.(2024)Grippa, Lambiase, and Poddar]{Grippa:2024sfu}
Grippa, F.; Lambiase, G.; Poddar, T.K.
\newblock {Constraints on scalar and vector dark matter admixed neutron stars
  with linear and quadratic couplings}. \emph{arXiv} {\bf 2024}, arXiv:2407.16386.
\newblock  %\href{http://arxiv.org/abs/2407.16386}{{\normalfont
  %[arXiv:hep-ph/2407.16386]}}.

\bibitem[Xiang et~al.(2014)Xiang, Jiang, Zhang, and Yang]{Xiang:2013xwa}
Xiang, Q.F.; Jiang, W.Z.; Zhang, D.R.; Yang, R.Y.
\newblock {Effects of fermionic dark matter on properties of neutron stars}.
\newblock {\em Phys. Rev. C} {\bf 2014}, {\em 89},~025803.
  %\href{http://arxiv.org/abs/1305.7354}{{\normalfont
  %[arXiv:astro-ph.SR/1305.7354]}}.
\newblock {\url{https://doi.org/10.1103/PhysRevC.89.025803}}.


%%%%%%%%%%%%%%%%%%%%%%%%%%%%%%%%%%%%%%%%%%%%%%%%%%%%%%%%%%%%%%%%%%
% NEW CITATIONS
%%%%%%%%%%%%%%%%%%%%%%%%%%%%%%%%%%%%%%%%%%%%%%%%%%%%%%%%%%%%%%%%%%
%NSs
\bibitem[Huang et~al.(2024)]{Huang:2023grj}
Huang, C.; Raaijmakers, G.; Watts, A.; Tolos, L.; Providencia, C. 
\newblock {Constraining a relativistic mean field model using neutron star mass-radius measurements I: nucleonic models};
\newblock {\em Mon. Not. Roy. Astron. Soc.} {\bf 2024}, {\em 529},~4650-4665.
\newblock {\url{https://doi.org/10.1093/mnras/stae844}}.

\bibitem[Grundler et~al.(2025)]{Grundler:2025mcz}
Grundler, X.; Li, B.A. 
\newblock {Bayesian quantification of observability and equation of state of twin stars};
\newblock {\em Phys. Rev. D 112} {\bf 2025}, {\em 10},~103012.
\newblock {\url{https://doi.org/10.1103/hsd4-j54y}}. 

\bibitem[Imam et~al.(2025)]{Imam:2025lut}
Imam, S.M.A.; Patra, N.K. 
\newblock {Bayesian analysis of the neutron star equation of state and model comparison: Insights from PSR J0437+4715, PSR J0614+3329, and other multiphysics data};
\newblock {\em Phys. Rev. D 112} {\bf 2025}, {\em 10},~103018.
\newblock {\url{https://doi.org/10.1103/2fz9-xlv1}}.

\bibitem[Cartaxo et~al.(2025)]{Cartaxo:2025jpi}
Cartaxo, J.; Huang, C.; Malik, T.; Sourav, S.; Yuan, W.L.; Zhou, T.; Liu, X.; Providencia, C. 
\newblock {Covariant Energy Density Functionals for Modeling the Equation of State of Neutron Star Matter: Cross-comparison Analysis Using CompactObject};
\newblock {\em Astrophys. J. Suppl.} {\bf 2026}, {\em 282},~2.
\newblock {\url{https://doi.org/10.3847/1538-4365/ae2310}}. 

%DANSs
\bibitem[Liu et~al.(2025)]{Liu:2025cwy}
Liu, X.Z.; Mahapatra, P.; Huang, C.; Hazarika, A.; Singha, C.; Das, P.K. 
\newblock {Revealing dark matter's role in neutron stars anisotropy: A Bayesian approach using multimessenger observations};
\newblock {\em Phys. Rev. D 112} {\bf 2025}, {\em 8},~083032.
\newblock {\url{https://doi.org/10.1103/zhs6-487x}}.

\bibitem[Arvikar et~al.(2025)]{Arvikar:2025dwl}
Arvikar, P.; Gautam, S.; Venneti, A., Banik, S. 
\newblock {Fermionic versus Bosonic Dark Matter in Neutron Stars: A Bayesian Study with Multi-Density Constraints};
\emph{arXiv} {\bf 2025}, arXiv:2512.13574.
\newblock  %\href{http://arxiv.org/abs/2512.13574}{{\normalfont
  %[arXiv:astro-ph.CO/2512.13574]}}.

\bibitem[Santos et~al.(2025)]{Santos:2025xep}
Santos, R.M.; Nunes, R.C.; Coelho, J.G.; de Araujo, J.C.N. 
\newblock {Observational bounds on dark matter-admixed neutron stars from gravitational wave data};
\newblock {\em Phys. Rev. D 112} {\bf 2025}, {\em 10},~104067.
\newblock {\url{https://doi.org/10.1103/fnqt-sgyc}}.  
%%%%%%%%%%%%%%%%%%%%%%%%%%%%%%%%%%%%%%%%%%%%%%%%%%%%%%%%%%%%%%%%%%
%%%%%%%%%%%%%%%%%%%%%%%%%%%%%%%%%%%%%%%%%%%%%%%%%%%%%%%%%%%%%%%%%%
\bibitem[Khlopov et~al.(1991)Khlopov, Beskin, Bochkarev, Pustylnik, and
  Pustylnik]{Khlopov:1989fj}
Khlopov, M.Y.; Beskin, G.M.; Bochkarev, N.E.; Pustylnik, L.A.; Pustylnik, S.A.
\newblock {Observational Physics of Mirror World}.
\newblock {\em Sov. Astron.} {\bf 1991}, {\em 35},~21.

\bibitem[Bucciantini()]{Bucciantini:Lecture_notes}
Bucciantini, N.
\newblock {Neutron stars: lectures notes}.
\newblock \url{https://www.arcetri.inaf.it/niccolo.bucciantini/files/ns.pdf}.

\bibitem[Hessels et~al.(2006)Hessels, Ransom, Stairs, Freire, Kaspi, and
  Camilo]{Hessels:2006ze}
Hessels, J.W.T.; Ransom, S.M.; Stairs, I.H.; Freire, P.C.C.; Kaspi, V.M.;
  Camilo, F.
\newblock {A radio pulsar spinning at 716-hz}.
\newblock {\em Science} {\bf 2006}, {\em 311},~1901--1904.
  %\href{http://arxiv.org/abs/astro-ph/0601337}{{\normalfont
  %[astro-ph/0601337]}}.
\newblock {\url{https://doi.org/10.1126/science.1123430}}.

\bibitem[Cromartie et~al.(2019)]{NANOGrav:2019jur}
Cromartie, H.T.;  et~al.
\newblock {Relativistic Shapiro delay measurements of an extremely massive
  millisecond pulsar}.
\newblock {\em Nature Astron.} {\bf 2019}, {\em 4},~72--76.
  %\href{http://arxiv.org/abs/1904.06759}{{\normalfont
  %[arXiv:astro-ph.HE/1904.06759]}}.
\newblock {\url{https://doi.org/10.1038/s41550-019-0880-2}}.

\bibitem[Fonseca et~al.(2021)]{Fonseca:2021wxt}
Fonseca, E.;  et~al.
\newblock {Refined Mass and Geometric Measurements of the High-mass PSR
  J0740+6620}.
\newblock {\em Astrophys. J. Lett.} {\bf 2021}, {\em 915},~L12.
  %\href{http://arxiv.org/abs/2104.00880}{{\normalfont
  %[arXiv:astro-ph.HE/2104.00880]}}.
\newblock {\url{https://doi.org/10.3847/2041-8213/ac03b8}}.

\bibitem[Demorest et~al.(2010)Demorest, Pennucci, Ransom, Roberts, and
  Hessels]{Demorest:2010bx}
Demorest, P.; Pennucci, T.; Ransom, S.; Roberts, M.; Hessels, J.
\newblock {Shapiro Delay Measurement of A Two Solar Mass Neutron Star}.
\newblock {\em Nature} {\bf 2010}, {\em 467},~1081--1083.
  %\href{http://arxiv.org/abs/1010.5788}{{\normalfont
  %[arXiv:astro-ph.HE/1010.5788]}}.
\newblock {\url{https://doi.org/10.1038/nature09466}}.

\bibitem[Pastor-Marazuela et~al.(2023)]{Pastor-Marazuela:2022pnp}
Pastor-Marazuela, I.;  et~al.
\newblock {A fast radio burst with submillisecond quasi-periodic structure}.
\newblock {\em Astron. Astrophys.} {\bf 2023}, {\em 678},~A149.
  %\href{http://arxiv.org/abs/2202.08002}{{\normalfont
  %[arXiv:astro-ph.HE/2202.08002]}}.
\newblock {\url{https://doi.org/10.1051/0004-6361/202243339}}.

\bibitem[Miller et~al.(2019)]{Miller:2019cac}
Miller, M.C.;  et~al.
\newblock {PSR J0030+0451 Mass and Radius from $NICER$ Data and Implications
  for the Properties of Neutron Star Matter}.
\newblock {\em Astrophys. J. Lett.} {\bf 2019}, {\em 887},~L24.
  %\href{http://arxiv.org/abs/1912.05705}{{\normalfont
  %[arXiv:astro-ph.HE/1912.05705]}}.
\newblock {\url{https://doi.org/10.3847/2041-8213/ab50c5}}.

\bibitem[Riley et~al.(2019)]{Riley:2019yda}
Riley, T.E.;  et~al.
\newblock {A $NICER$ View of PSR J0030+0451: Millisecond Pulsar Parameter
  Estimation}.
\newblock {\em Astrophys. J. Lett.} {\bf 2019}, {\em 887},~L21.
  %\href{http://arxiv.org/abs/1912.05702}{{\normalfont
  %[arXiv:astro-ph.HE/1912.05702]}}.
\newblock {\url{https://doi.org/10.3847/2041-8213/ab481c}}.

\bibitem[Miller et~al.(2021)]{Miller:2021qha}
Miller, M.C.;  et~al.
\newblock {The Radius of PSR J0740+6620 from NICER and XMM-Newton Data}.
\newblock {\em Astrophys. J. Lett.} {\bf 2021}, {\em 918},~L28.
  %\href{http://arxiv.org/abs/2105.06979}{{\normalfont
  %[arXiv:astro-ph.HE/2105.06979]}}.
\newblock {\url{https://doi.org/10.3847/2041-8213/ac089b}}.

\bibitem[Riley et~al.(2021)]{Riley:2021pdl}
Riley, T.E.;  et~al.
\newblock {A NICER View of the Massive Pulsar PSR J0740+6620 Informed by Radio
  Timing and XMM-Newton Spectroscopy}.
\newblock {\em Astrophys. J. Lett.} {\bf 2021}, {\em 918},~L27.
  %\href{http://arxiv.org/abs/2105.06980}{{\normalfont
  %[arXiv:astro-ph.HE/2105.06980]}}.
\newblock {\url{https://doi.org/10.3847/2041-8213/ac0a81}}.

\bibitem[Gleason et~al.(2022)Gleason, Brown, and Kain]{Gleason:2022eeg}
Gleason, T.; Brown, B.; Kain, B.
\newblock {Dynamical evolution of dark matter admixed neutron stars}.
\newblock {\em Phys. Rev. D} {\bf 2022}, {\em 105},~023010.
  %\href{http://arxiv.org/abs/2201.02274}{{\normalfont
  %[arXiv:gr-qc/2201.02274]}}.
\newblock {\url{https://doi.org/10.1103/PhysRevD.105.023010}}.

\bibitem[Guver et~al.(2010{\natexlab{a}})Guver, Ozel, Cabrera-Lavers, and
  Wroblewski]{Guver:2008gc}
Guver, T.; Ozel, F.; Cabrera-Lavers, A.; Wroblewski, P.
\newblock {The Distance, Mass, and Radius of the Neutron Star in 4U 1608-52}.
\newblock {\em Astrophys. J.} {\bf 2010}, {\em 712},~964--973.
  %\href{http://arxiv.org/abs/0811.3979}{{\normalfont
  %[arXiv:astro-ph/0811.3979]}}.
\newblock {\url{https://doi.org/10.1088/0004-637X/712/2/964}}.

\bibitem[Guver et~al.(2010{\natexlab{b}})Guver, Wroblewski, Camarota, and
  Ozel]{Guver:2010td}
Guver, T.; Wroblewski, P.; Camarota, L.; Ozel, F.
\newblock {The Mass and Radius of the Neutron Star in 4U 1820-30}.
\newblock {\em Astrophys. J.} {\bf 2010}, {\em 719},~1807.
  %\href{http://arxiv.org/abs/1002.3825}{{\normalfont
  %[arXiv:astro-ph.HE/1002.3825]}}.
\newblock {\url{https://doi.org/10.1088/0004-637X/719/2/1807}}.

\bibitem[Ozel et~al.(2010)Ozel, Baym, and Guver]{Ozel:2010fw}
Ozel, F.; Baym, G.; Guver, T.
\newblock {Astrophysical Measurement of the Equation of State of Neutron Star
  Matter}.
\newblock {\em Phys. Rev. D} {\bf 2010}, {\em 82},~101301.
  %\href{http://arxiv.org/abs/1002.3153}{{\normalfont
  %[arXiv:astro-ph.HE/1002.3153]}}.
\newblock {\url{https://doi.org/10.1103/PhysRevD.82.101301}}.

\bibitem[Scordino and Bombaci(2024)]{Scordino:2024ehe}
Scordino, D.; Bombaci, I.
\newblock {Dark matter admixed neutron stars with a realistic nuclear equation
  of state from chiral nuclear interactions}. \emph{J. High Energy Astrophys.} {\bf 2024}, \emph{45}, 371--381.
\newblock  %\href{http://arxiv.org/abs/2405.19251}{{\normalfont
  %[arXiv:astro-ph.HE/2405.19251]}}.

\bibitem[Zdunik et~al.(2004)Zdunik, Haensel, Gourgoulhon, and
  Bejger]{Zdunik:2003vg}
Zdunik, J.L.; Haensel, P.; Gourgoulhon, E.; Bejger, M.
\newblock {Hyperon softening of the EOS of dense matter and the spin evolution
  of isolated neutron stars}.
\newblock {\em Astron. Astrophys.} {\bf 2004}, {\em 416},~1013.
  %\href{http://arxiv.org/abs/astro-ph/0311470}{{\normalfont
  %[astro-ph/0311470]}}.
\newblock {\url{https://doi.org/10.1051/0004-6361:20034387}}.

\bibitem[Weissenborn et~al.(2012)Weissenborn, Chatterjee, and
  Schaffner-Bielich]{Weissenborn:2011kb}
Weissenborn, S.; Chatterjee, D.; Schaffner-Bielich, J.
\newblock {Hyperons and massive neutron stars: the role of hyperon potentials}.
\newblock {\em Nucl. Phys. A} {\bf 2012}, {\em 881},~62--77.
  %\href{http://arxiv.org/abs/1111.6049}{{\normalfont
  %[arXiv:astro-ph.HE/1111.6049]}}.
\newblock {\url{https://doi.org/10.1016/j.nuclphysa.2012.02.012}}.

\bibitem[Vida\~na(2022)]{Vidana:2022tlx}
Vida\~na, I.
\newblock {Neutron stars and the hyperon puzzle}.
\newblock {\em EPJ Web Conf.} {\bf 2022}, {\em 271},~09001.
\newblock {\url{https://doi.org/10.1051/epjconf/202227109001}}.

\bibitem[Schertler et~al.(2000)Schertler, Greiner, Schaffner-Bielich, and
  Thoma]{Schertler:2000xq}
Schertler, K.; Greiner, C.; Schaffner-Bielich, J.; Thoma, M.H.
\newblock {Quark phases in neutron stars and a 'third family' of compact stars
  as a signature for phase transitions}.
\newblock {\em Nucl. Phys. A} {\bf 2000}, {\em 677},~463--490.
  %\href{http://arxiv.org/abs/astro-ph/0001467}{{\normalfont
  %[astro-ph/0001467]}}.
\newblock {\url{https://doi.org/10.1016/S0375-9474(00)00305-5}}.

\bibitem[Abbott et~al.(2019)]{LIGOScientific:2018hze}
Abbott, B.P.;  et~al.
\newblock {Properties of the binary neutron star merger GW170817}.
\newblock {\em Phys. Rev. X} {\bf 2019}, {\em 9},~011001.
  %\href{http://arxiv.org/abs/1805.11579}{{\normalfont
  %[arXiv:gr-qc/1805.11579]}}.
\newblock {\url{https://doi.org/10.1103/PhysRevX.9.011001}}.

\bibitem[Yang et~al.(2022)Yang, Xie, and Liu]{Yang:2022ees}
Yang, R.X.; Xie, F.; Liu, D.J.
\newblock {Tidal Deformability of Neutron Stars in Unimodular Gravity}.
\newblock {\em Universe} {\bf 2022}, {\em 8},~576.
  %\href{http://arxiv.org/abs/2211.00278}{{\normalfont
  %[arXiv:gr-qc/2211.00278]}}.
\newblock {\url{https://doi.org/10.3390/universe8110576}}.

\bibitem[Ecker and Rezzolla(2022)]{Ecker:2022xxj}
Ecker, C.; Rezzolla, L.
\newblock {A General, Scale-independent Description of the Sound Speed in
  Neutron Stars}.
\newblock {\em Astrophys. J. Lett.} {\bf 2022}, {\em 939},~L35.
  %\href{http://arxiv.org/abs/2207.04417}{{\normalfont
  %[arXiv:gr-qc/2207.04417]}}.
\newblock {\url{https://doi.org/10.3847/2041-8213/ac8674}}.

\bibitem[Rezzolla and Zanotti(2013)]{Rezzolla:2013dea}
Rezzolla, L.; Zanotti, O.
\newblock {\em {Relativistic Hydrodynamics}}; Oxford University Press: Oxford, UK, 2013.
\newblock {\url{https://doi.org/10.1093/acprof:oso/9780198528906.001.0001}}.

\bibitem[Bludman and Ruderman(1968)]{Bludman:1968zz}
Bludman, S.A.; Ruderman, M.A.
\newblock {Possibility of the Speed of Sound Exceeding the Speed of Light in
  Ultradense Matter}.
\newblock {\em Phys. Rev.} {\bf 1968}, {\em 170},~1176--1184.
\newblock {\url{https://doi.org/10.1103/PhysRev.170.1176}}.

\bibitem[Hebeler et~al.(2013)Hebeler, Lattimer, Pethick, and
  Schwenk]{Hebeler:2013nza}
Hebeler, K.; Lattimer, J.M.; Pethick, C.J.; Schwenk, A.
\newblock {Equation of state and neutron star properties constrained by nuclear
  physics and observation}.
\newblock {\em Astrophys. J.} {\bf 2013}, {\em 773},~11.
  %\href{http://arxiv.org/abs/1303.4662}{{\normalfont
  %[arXiv:astro-ph.SR/1303.4662]}}.
\newblock {\url{https://doi.org/10.1088/0004-637X/773/1/11}}.

\bibitem[Altiparmak et~al.(2022)Altiparmak, Ecker, and
  Rezzolla]{Altiparmak:2022bke}
Altiparmak, S.; Ecker, C.; Rezzolla, L.
\newblock {On the Sound Speed in Neutron Stars}.
\newblock {\em Astrophys. J. Lett.} {\bf 2022}, {\em 939},~L34.
  %\href{http://arxiv.org/abs/2203.14974}{{\normalfont
  %[arXiv:astro-ph.HE/2203.14974]}}.
\newblock {\url{https://doi.org/10.3847/2041-8213/ac9b2a}}.

\bibitem[Bedaque and Steiner(2015)]{Bedaque:2014sqa}
Bedaque, P.; Steiner, A.W.
\newblock {Sound velocity bound and neutron stars}.
\newblock {\em Phys. Rev. Lett.} {\bf 2015}, {\em 114},~031103.
  %\href{http://arxiv.org/abs/1408.5116}{{\normalfont
  %[arXiv:nucl-th/1408.5116]}}.
\newblock {\url{https://doi.org/10.1103/PhysRevLett.114.031103}}.

\bibitem[Hoyos et~al.(2016)Hoyos, Jokela, Rodr\'\i{}guez~Fern\'andez, and
  Vuorinen]{Hoyos:2016cob}
Hoyos, C.; Jokela, N.; Rodr\'\i{}guez~Fern\'andez, D.; Vuorinen, A.
\newblock {Breaking the sound barrier in AdS/CFT}.
\newblock {\em Phys. Rev. D} {\bf 2016}, {\em 94},~106008.
  %\href{http://arxiv.org/abs/1609.03480}{{\normalfont
  %[arXiv:hep-th/1609.03480]}}.
\newblock {\url{https://doi.org/10.1103/PhysRevD.94.106008}}.

\bibitem[Landry et~al.(2020)Landry, Essick, and Chatziioannou]{Landry:2020vaw}
Landry, P.; Essick, R.; Chatziioannou, K.
\newblock {Nonparametric constraints on neutron star matter with existing and
  upcoming gravitational wave and pulsar observations}.
\newblock {\em Phys. Rev. D} {\bf 2020}, {\em 101},~123007.
  %\href{http://arxiv.org/abs/2003.04880}{{\normalfont
  %[arXiv:astro-ph.HE/2003.04880]}}.
\newblock {\url{https://doi.org/10.1103/PhysRevD.101.123007}}.

\bibitem[Roy and Suyama(2024)]{Roy:2022nwy}
Roy, S.; Suyama, T.
\newblock {On the sound velocity bound in neutron stars}.
\newblock {\em Results Phys.} {\bf 2024}, {\em 61},~107757.
  %\href{http://arxiv.org/abs/2211.07874}{{\normalfont
  %[arXiv:astro-ph.HE/2211.07874]}}.
\newblock {\url{https://doi.org/10.1016/j.rinp.2024.107757}}.

\bibitem[Kurkela et~al.(2010)Kurkela, Romatschke, and Vuorinen]{Kurkela:2009gj}
Kurkela, A.; Romatschke, P.; Vuorinen, A.
\newblock {Cold Quark Matter}.
\newblock {\em Phys. Rev. D} {\bf 2010}, {\em 81},~105021.
  %\href{http://arxiv.org/abs/0912.1856}{{\normalfont
  %[arXiv:hep-ph/0912.1856]}}.
\newblock {\url{https://doi.org/10.1103/PhysRevD.81.105021}}.

\bibitem[Fraga et~al.(2014)Fraga, Kurkela, and Vuorinen]{Fraga:2013qra}
Fraga, E.S.; Kurkela, A.; Vuorinen, A.
\newblock {Interacting quark matter equation of state for compact stars}.
\newblock {\em Astrophys. J. Lett.} {\bf 2014}, {\em 781},~L25.
  %\href{http://arxiv.org/abs/1311.5154}{{\normalfont
  %[arXiv:nucl-th/1311.5154]}}.
\newblock {\url{https://doi.org/10.1088/2041-8205/781/2/L25}}.

\bibitem[Tews et~al.(2018)Tews, Carlson, Gandolfi, and Reddy]{Tews:2018kmu}
Tews, I.; Carlson, J.; Gandolfi, S.; Reddy, S.
\newblock {Constraining the speed of sound inside neutron stars with chiral
  effective field theory interactions and observations}.
\newblock {\em Astrophys. J.} {\bf 2018}, {\em 860},~149.
  %\href{http://arxiv.org/abs/1801.01923}{{\normalfont
  %[arXiv:nucl-th/1801.01923]}}.
\newblock {\url{https://doi.org/10.3847/1538-4357/aac267}}.

\bibitem[Greif et~al.(2019)Greif, Raaijmakers, Hebeler, Schwenk, and
  Watts]{Greif:2018njt}
Greif, S.K.; Raaijmakers, G.; Hebeler, K.; Schwenk, A.; Watts, A.L.
\newblock {Equation of state sensitivities when inferring neutron star and
  dense matter properties}.
\newblock {\em Mon. Not. Roy. Astron. Soc.} {\bf 2019}, {\em 485},~5363--5376.
  %\href{http://arxiv.org/abs/1812.08188}{{\normalfont
  %[arXiv:astro-ph.HE/1812.08188]}}.
\newblock {\url{https://doi.org/10.1093/mnras/stz654}}.

\bibitem[Kojo et~al.(2015)Kojo, Powell, Song, and Baym]{Kojo:2014rca}
Kojo, T.; Powell, P.D.; Song, Y.; Baym, G.
\newblock {Phenomenological QCD equation of state for massive neutron stars}.
\newblock {\em Phys. Rev. D} {\bf 2015}, {\em 91},~045003.
  %\href{http://arxiv.org/abs/1412.1108}{{\normalfont
  %[arXiv:hep-ph/1412.1108]}}.
\newblock {\url{https://doi.org/10.1103/PhysRevD.91.045003}}.

\bibitem[Alsing et~al.(2018)Alsing, Silva, and Berti]{Alsing:2017bbc}
Alsing, J.; Silva, H.O.; Berti, E.
\newblock {Evidence for a maximum mass cut-off in the neutron star mass
  distribution and constraints on the equation of state}.
\newblock {\em Mon. Not. Roy. Astron. Soc.} {\bf 2018}, {\em 478},~1377--1391.
  %\href{http://arxiv.org/abs/1709.07889}{{\normalfont
  %[arXiv:astro-ph.HE/1709.07889]}}.
\newblock {\url{https://doi.org/10.1093/mnras/sty1065}}.

\bibitem[Bitaghsir~Fadafan et~al.(2019)Bitaghsir~Fadafan, Kazemian, and
  Schmitt]{BitaghsirFadafan:2018uzs}
Bitaghsir~Fadafan, K.; Kazemian, F.; Schmitt, A.
\newblock {Towards a holographic quark-hadron continuity}.
\newblock {\em JHEP} {\bf 2019}, {\em 3},~183.
  %\href{http://arxiv.org/abs/1811.08698}{{\normalfont
  %[arXiv:hep-ph/1811.08698]}}.
\newblock {\url{https://doi.org/10.1007/JHEP03(2019)183}}.

\bibitem[McLerran and Reddy(2019)]{McLerran:2018hbz}
McLerran, L.; Reddy, S.
\newblock {Quarkyonic Matter and Neutron Stars}.
\newblock {\em Phys. Rev. Lett.} {\bf 2019}, {\em 122},~122701.
  %\href{http://arxiv.org/abs/1811.12503}{{\normalfont
  %[arXiv:nucl-th/1811.12503]}}.
\newblock {\url{https://doi.org/10.1103/PhysRevLett.122.122701}}.

\bibitem[Bauswein et~al.(2017)Bauswein, Just, Janka, and
  Stergioulas]{Bauswein:2017vtn}
Bauswein, A.; Just, O.; Janka, H.T.; Stergioulas, N.
\newblock {Neutron-star radius constraints from GW170817 and future
  detections}.
\newblock {\em Astrophys. J. Lett.} {\bf 2017}, {\em 850},~L34.
  %\href{http://arxiv.org/abs/1710.06843}{{\normalfont
  %[arXiv:astro-ph.HE/1710.06843]}}.
\newblock {\url{https://doi.org/10.3847/2041-8213/aa9994}}.

\bibitem[Margalit and Metzger(2017)]{Margalit:2017dij}
Margalit, B.; Metzger, B.D.
\newblock {Constraining the Maximum Mass of Neutron Stars From Multi-Messenger
  Observations of GW170817}.
\newblock {\em Astrophys. J. Lett.} {\bf 2017}, {\em 850},~L19.
  %\href{http://arxiv.org/abs/1710.05938}{{\normalfont
  %[arXiv:astro-ph.HE/1710.05938]}}.
\newblock {\url{https://doi.org/10.3847/2041-8213/aa991c}}.

\bibitem[Rezzolla et~al.(2018)Rezzolla, Most, and Weih]{Rezzolla:2017aly}
Rezzolla, L.; Most, E.R.; Weih, L.R.
\newblock {Using gravitational-wave observations and quasi-universal relations
  to constrain the maximum mass of neutron stars}.
\newblock {\em Astrophys. J. Lett.} {\bf 2018}, {\em 852},~L25.
  %\href{http://arxiv.org/abs/1711.00314}{{\normalfont
  %[arXiv:astro-ph.HE/1711.00314]}}.
\newblock {\url{https://doi.org/10.3847/2041-8213/aaa401}}.

\bibitem[Ruiz et~al.(2018)Ruiz, Shapiro, and Tsokaros]{Ruiz:2017due}
Ruiz, M.; Shapiro, S.L.; Tsokaros, A.
\newblock {GW170817, General Relativistic Magnetohydrodynamic Simulations, and
  the Neutron Star Maximum Mass}.
\newblock {\em Phys. Rev. D} {\bf 2018}, {\em 97},~021501.
  %\href{http://arxiv.org/abs/1711.00473}{{\normalfont
  %[arXiv:astro-ph.HE/1711.00473]}}.
\newblock {\url{https://doi.org/10.1103/PhysRevD.97.021501}}.

\bibitem[Shibata et~al.(2017)Shibata, Fujibayashi, Hotokezaka, Kiuchi, Kyutoku,
  Sekiguchi, and Tanaka]{Shibata:2017xdx}
Shibata, M.; Fujibayashi, S.; Hotokezaka, K.; Kiuchi, K.; Kyutoku, K.;
  Sekiguchi, Y.; Tanaka, M.
\newblock {Modeling GW170817 based on numerical relativity and its
  implications}.
\newblock {\em Phys. Rev. D} {\bf 2017}, {\em 96},~123012.
  %\href{http://arxiv.org/abs/1710.07579}{{\normalfont
  %[arXiv:astro-ph.HE/1710.07579]}}.
\newblock {\url{https://doi.org/10.1103/PhysRevD.96.123012}}.

\bibitem[Shibata et~al.(2019)Shibata, Zhou, Kiuchi, and
  Fujibayashi]{Shibata:2019ctb}
Shibata, M.; Zhou, E.; Kiuchi, K.; Fujibayashi, S.
\newblock {Constraint on the maximum mass of neutron stars using GW170817
  event}.
\newblock {\em Phys. Rev. D} {\bf 2019}, {\em 100},~023015.
  %\href{http://arxiv.org/abs/1905.03656}{{\normalfont
  %[arXiv:astro-ph.HE/1905.03656]}}.
\newblock {\url{https://doi.org/10.1103/PhysRevD.100.023015}}.

\bibitem[Nathanail et~al.(2021)Nathanail, Most, and
  Rezzolla]{Nathanail:2021tay}
Nathanail, A.; Most, E.R.; Rezzolla, L.
\newblock {GW170817 and GW190814: tension on the maximum mass}.
\newblock {\em Astrophys. J. Lett.} {\bf 2021}, {\em 908},~L28.
  %\href{http://arxiv.org/abs/2101.01735}{{\normalfont
  %[arXiv:astro-ph.HE/2101.01735]}}.
\newblock {\url{https://doi.org/10.3847/2041-8213/abdfc6}}.

\bibitem[Acernese et~al.(2015)]{VIRGO:2014yos}
Acernese, F.;  et~al.
\newblock {Advanced Virgo: a second-generation interferometric gravitational
  wave detector}.
\newblock {\em Class. Quant. Grav.} {\bf 2015}, {\em 32},~024001.
  %\href{http://arxiv.org/abs/1408.3978}{{\normalfont %[arXiv:gr-qc/1408.3978]}}.
\newblock {\url{https://doi.org/10.1088/0264-9381/32/2/024001}}.

\bibitem[Aasi et~al.(2015)]{LIGOScientific:2014pky}
Aasi, J.;  et~al.
\newblock {Advanced LIGO}.
\newblock {\em Class. Quant. Grav.} {\bf 2015}, {\em 32},~074001.
  %\href{http://arxiv.org/abs/1411.4547}{{\normalfont %[arXiv:gr-qc/1411.4547]}}.
\newblock {\url{https://doi.org/10.1088/0264-9381/32/7/074001}}.

\bibitem[Somiya(2012)]{Somiya:2011np}
Somiya, K.
\newblock {Detector configuration of KAGRA: The Japanese cryogenic
  gravitational-wave detector}.
\newblock {\em Class. Quant. Grav.} {\bf 2012}, {\em 29},~124007.
  %\href{http://arxiv.org/abs/1111.7185}{{\normalfont %[arXiv:gr-qc/1111.7185]}}.
\newblock {\url{https://doi.org/10.1088/0264-9381/29/12/124007}}.

\bibitem[Croon et~al.(2018)Croon, Nelson, Sun, Walker, and
  Xianyu]{Croon:2017zcu}
Croon, D.; Nelson, A.E.; Sun, C.; Walker, D.G.E.; Xianyu, Z.Z.
\newblock {Hidden-Sector Spectroscopy with Gravitational Waves from Binary
  Neutron Stars}.
\newblock {\em Astrophys. J. Lett.} {\bf 2018}, {\em 858},~L2.
  %\href{http://arxiv.org/abs/1711.02096}{{\normalfont
  %[arXiv:hep-ph/1711.02096]}}.
\newblock {\url{https://doi.org/10.3847/2041-8213/aabe76}}.

\bibitem[Choi and Jung(2019)]{Choi:2018axi}
Choi, H.G.; Jung, S.
\newblock {New probe of dark matter-induced fifth force with neutron star
  inspirals}.
\newblock {\em Phys. Rev. D} {\bf 2019}, {\em 99},~015013.
  %\href{http://arxiv.org/abs/1810.01421}{{\normalfont
  %[arXiv:hep-ph/1810.01421]}}.
\newblock {\url{https://doi.org/10.1103/PhysRevD.99.015013}}.

\bibitem[Alexander et~al.(2018)Alexander, McDonough, Sims, and
  Yunes]{Alexander:2018qzg}
Alexander, S.; McDonough, E.; Sims, R.; Yunes, N.
\newblock {Hidden-Sector Modifications to Gravitational Waves From Binary
  Inspirals}.
\newblock {\em Class. Quant. Grav.} {\bf 2018}, {\em 35},~235012.
  %\href{http://arxiv.org/abs/1808.05286}{{\normalfont
  %[arXiv:gr-qc/1808.05286]}}.
\newblock {\url{https://doi.org/10.1088/1361-6382/aaeb5c}}.

\bibitem[Sagunski et~al.(2018)Sagunski, Zhang, Johnson, Lehner, Sakellariadou,
  Liebling, Palenzuela, and Neilsen]{Sagunski:2017nzb}
Sagunski, L.; Zhang, J.; Johnson, M.C.; Lehner, L.; Sakellariadou, M.;
  Liebling, S.L.; Palenzuela, C.; Neilsen, D.
\newblock {Neutron star mergers as a probe of modifications of general
  relativity with finite-range scalar forces}.
\newblock {\em Phys. Rev. D} {\bf 2018}, {\em 97},~064016.
  %\href{http://arxiv.org/abs/1709.06634}{{\normalfont
  %[arXiv:gr-qc/1709.06634]}}.
\newblock {\url{https://doi.org/10.1103/PhysRevD.97.064016}}.

\bibitem[Poddar et~al.(2024)Poddar, Ghoshal, and Lambiase]{Poddar:2023pfj}
Poddar, T.K.; Ghoshal, A.; Lambiase, G.
\newblock {Listening to dark sirens from gravitational waves : Combined effects
  of fifth force, ultralight particle radiation, and eccentricity}.
\newblock {\em Phys. Dark Univ.} {\bf 2024}, {\em 46},~101651.
  %\href{http://arxiv.org/abs/2302.14513}{{\normalfont
  %[arXiv:hep-ph/2302.14513]}}.
\newblock {\url{https://doi.org/10.1016/j.dark.2024.101651}}.

\bibitem[Bauswein and Janka(2012)]{Bauswein:2011tp}
Bauswein, A.; Janka, H.T.
\newblock {Measuring neutron-star properties via gravitational waves from
  binary mergers}.
\newblock {\em Phys. Rev. Lett.} {\bf 2012}, {\em 108},~011101.
  %\href{http://arxiv.org/abs/1106.1616}{{\normalfont
  %[arXiv:astro-ph.SR/1106.1616]}}.
\newblock {\url{https://doi.org/10.1103/PhysRevLett.108.011101}}.

\bibitem[Clark et~al.(2016)Clark, Bauswein, Stergioulas, and
  Shoemaker]{Clark:2015zxa}
Clark, J.A.; Bauswein, A.; Stergioulas, N.; Shoemaker, D.
\newblock {Observing Gravitational Waves From The Post-Merger Phase Of Binary
  Neutron Star Coalescence}.
\newblock {\em Class. Quant. Grav.} {\bf 2016}, {\em 33},~085003.
  %\href{http://arxiv.org/abs/1509.08522}{{\normalfont
  %[arXiv:astro-ph.HE/1509.08522]}}.
\newblock {\url{https://doi.org/10.1088/0264-9381/33/8/085003}}.

\bibitem[Prakash et~al.(2021)Prakash, Radice, Logoteta, Perego, Nedora,
  Bombaci, Kashyap, Bernuzzi, and Endrizzi]{Prakash:2021wpz}
Prakash, A.; Radice, D.; Logoteta, D.; Perego, A.; Nedora, V.; Bombaci, I.;
  Kashyap, R.; Bernuzzi, S.; Endrizzi, A.
\newblock {Signatures of deconfined quark phases in binary neutron star
  mergers}.
\newblock {\em Phys. Rev. D} {\bf 2021}, {\em 104},~083029.
  %\href{http://arxiv.org/abs/2106.07885}{{\normalfont
  %[arXiv:astro-ph.HE/2106.07885]}}.
\newblock {\url{https://doi.org/10.1103/PhysRevD.104.083029}}.

\bibitem[Prakash et~al.(2024)Prakash, Gupta, Breschi, Kashyap, Radice,
  Bernuzzi, Logoteta, and Sathyaprakash]{Prakash:2023afe}
Prakash, A.; Gupta, I.; Breschi, M.; Kashyap, R.; Radice, D.; Bernuzzi, S.;
  Logoteta, D.; Sathyaprakash, B.S.
\newblock {Detectability of QCD phase transitions in binary neutron star
  mergers: Bayesian inference with the next generation gravitational wave
  detectors}.
\newblock {\em Phys. Rev. D} {\bf 2024}, {\em 109},~103008.
  %\href{http://arxiv.org/abs/2310.06025}{{\normalfont
  %[arXiv:gr-qc/2310.06025]}}.
\newblock {\url{https://doi.org/10.1103/PhysRevD.109.103008}}.

\bibitem[Stergioulas et~al.(2011)Stergioulas, Bauswein, Zagkouris, and
  Janka]{Stergioulas:2011gd}
Stergioulas, N.; Bauswein, A.; Zagkouris, K.; Janka, H.T.
\newblock {Gravitational waves and nonaxisymmetric oscillation modes in mergers
  of compact object binaries}.
\newblock {\em Mon. Not. Roy. Astron. Soc.} {\bf 2011}, {\em 418},~427.
  %\href{http://arxiv.org/abs/1105.0368}{{\normalfont %[arXiv:gr-qc/1105.0368]}}.
\newblock {\url{https://doi.org/10.1111/j.1365-2966.2011.19493.x}}.

\bibitem[Takami et~al.(2014)Takami, Rezzolla, and Baiotti]{Takami:2014zpa}
Takami, K.; Rezzolla, L.; Baiotti, L.
\newblock {Constraining the Equation of State of Neutron Stars from Binary
  Mergers}.
\newblock {\em Phys. Rev. Lett.} {\bf 2014}, {\em 113},~091104.
  %\href{http://arxiv.org/abs/1403.5672}{{\normalfont %[arXiv:gr-qc/1403.5672]}}.
\newblock {\url{https://doi.org/10.1103/PhysRevLett.113.091104}}.

\bibitem[Bauswein et~al.(2016)Bauswein, Stergioulas, and
  Janka]{Bauswein:2015vxa}
Bauswein, A.; Stergioulas, N.; Janka, H.T.
\newblock {Exploring properties of high-density matter through remnants of
  neutron-star mergers}.
\newblock {\em Eur. Phys. J. A} {\bf 2016}, {\em 52},~56.
  %\href{http://arxiv.org/abs/1508.05493}{{\normalfont
  %[arXiv:astro-ph.HE/1508.05493]}}.
\newblock {\url{https://doi.org/10.1140/epja/i2016-16056-7}}.

\bibitem[East et~al.(2016)East, Paschalidis, Pretorius, and
  Shapiro]{East:2015vix}
East, W.E.; Paschalidis, V.; Pretorius, F.; Shapiro, S.L.
\newblock {Relativistic Simulations of Eccentric Binary Neutron Star Mergers:
  One-arm Spiral Instability and Effects of Neutron Star Spin}.
\newblock {\em Phys. Rev. D} {\bf 2016}, {\em 93},~024011.
  %\href{http://arxiv.org/abs/1511.01093}{{\normalfont
  %[arXiv:astro-ph.HE/1511.01093]}}.
\newblock {\url{https://doi.org/10.1103/PhysRevD.93.024011}}.

\bibitem[Lehner et~al.(2016{\natexlab{a}})Lehner, Liebling, Palenzuela, and
  Motl]{Lehner:2016wjg}
Lehner, L.; Liebling, S.L.; Palenzuela, C.; Motl, P.M.
\newblock {m=1 instability and gravitational wave signal in binary neutron star
  mergers}.
\newblock {\em Phys. Rev. D} {\bf 2016}, {\em 94},~043003.
  %\href{http://arxiv.org/abs/1605.02369}{{\normalfont
  %[arXiv:gr-qc/1605.02369]}}.
\newblock {\url{https://doi.org/10.1103/PhysRevD.94.043003}}.

\bibitem[Lehner et~al.(2016{\natexlab{b}})Lehner, Liebling, Palenzuela,
  Caballero, O'Connor, Anderson, and Neilsen]{Lehner:2016lxy}
Lehner, L.; Liebling, S.L.; Palenzuela, C.; Caballero, O.L.; O'Connor, E.;
  Anderson, M.; Neilsen, D.
\newblock {Unequal mass binary neutron star mergers and multimessenger
  signals}.
\newblock {\em Class. Quant. Grav.} {\bf 2016}, {\em 33},~184002.
  %\href{http://arxiv.org/abs/1603.00501}{{\normalfont
  %[arXiv:gr-qc/1603.00501]}}.
\newblock {\url{https://doi.org/10.1088/0264-9381/33/18/184002}}.

\bibitem[Radice et~al.(2016)Radice, Bernuzzi, and Ott]{Radice:2016gym}
Radice, D.; Bernuzzi, S.; Ott, C.D.
\newblock {One-armed spiral instability in neutron star mergers and its
  detectability in gravitational waves}.
\newblock {\em Phys. Rev. D} {\bf 2016}, {\em 94},~064011.
  %\href{http://arxiv.org/abs/1603.05726}{{\normalfont
  %[arXiv:gr-qc/1603.05726]}}.
\newblock {\url{https://doi.org/10.1103/PhysRevD.94.064011}}.

\bibitem[Hodges(1993)]{Hodges:1993yb}
Hodges, H.M.
\newblock {Mirror baryons as the dark matter}.
\newblock {\em Phys. Rev. D} {\bf 1993}, {\em 47},~456--459.
\newblock {\url{https://doi.org/10.1103/PhysRevD.47.456}}.

\bibitem[Berezhiani et~al.(2001)Berezhiani, Comelli, and
  Villante]{Berezhiani:2000gw}
Berezhiani, Z.; Comelli, D.; Villante, F.L.
\newblock {The Early mirror universe: Inflation, baryogenesis, nucleosynthesis
  and dark matter}.
\newblock {\em Phys. Lett. B} {\bf 2001}, {\em 503},~362--375.
  %\href{http://arxiv.org/abs/hep-ph/0008105}{{\normalfont [hep-ph/0008105]}}.
\newblock {\url{https://doi.org/10.1016/S0370-2693(01)00217-9}}.

\bibitem[Berezhiani(2004)]{Berezhiani:2003xm}
Berezhiani, Z.
\newblock {Mirror world and its cosmological consequences}.
\newblock {\em Int. J. Mod. Phys. A} {\bf 2004}, {\em 19},~3775--3806.
  %\href{http://arxiv.org/abs/hep-ph/0312335}{{\normalfont [hep-ph/0312335]}}.
\newblock {\url{https://doi.org/10.1142/S0217751X04020075}}.

\bibitem[Ignatiev and Volkas(2003)]{Ignatiev:2003js}
Ignatiev, A.Y.; Volkas, R.R.
\newblock {Mirror dark matter and large scale structure}.
\newblock {\em Phys. Rev. D} {\bf 2003}, {\em 68},~023518.
  %\href{http://arxiv.org/abs/hep-ph/0304260}{{\normalfont [hep-ph/0304260]}}.
\newblock {\url{https://doi.org/10.1103/PhysRevD.68.023518}}.

\bibitem[Foot(2014)]{Foot:2014mia}
Foot, R.
\newblock {Mirror dark matter: Cosmology, galaxy structure and direct
  detection}.
\newblock {\em Int. J. Mod. Phys. A} {\bf 2014}, {\em 29},~1430013.
  %\href{http://arxiv.org/abs/1401.3965}{{\normalfont
  %[arXiv:astro-ph.CO/1401.3965]}}.
\newblock {\url{https://doi.org/10.1142/S0217751X14300130}}.

\bibitem[Emma et~al.(2022)Emma, Schianchi, Pannarale, Sagun, and
  Dietrich]{Emma:2022xjs}
Emma, M.; Schianchi, F.; Pannarale, F.; Sagun, V.; Dietrich, T.
\newblock {Numerical Simulations of Dark Matter Admixed Neutron Star Binaries}.
\newblock {\em Particles} {\bf 2022}, {\em 5},~273--286.
  %\href{http://arxiv.org/abs/2206.10887}{{\normalfont
  %[arXiv:gr-qc/2206.10887]}}.
\newblock {\url{https://doi.org/10.3390/particles5030024}}.

\bibitem[Bauswein et~al.(2023)Bauswein, Guo, Lien, Lin, and
  Wu]{Bauswein:2020kor}
Bauswein, A.; Guo, G.; Lien, J.H.; Lin, Y.H.; Wu, M.R.
\newblock {Compact dark objects in neutron star mergers}.
\newblock {\em Phys. Rev. D} {\bf 2023}, {\em 107},~083002.
  %\href{http://arxiv.org/abs/2012.11908}{{\normalfont
  %[arXiv:astro-ph.HE/2012.11908]}}.
\newblock {\url{https://doi.org/10.1103/PhysRevD.107.083002}}.

\bibitem[R\"uter et~al.(2023)R\"uter, Sagun, Tichy, and
  Dietrich]{Ruter:2023uzc}
R\"uter, H.R.; Sagun, V.; Tichy, W.; Dietrich, T.
\newblock {Quasiequilibrium configurations of binary systems of dark matter
  admixed neutron stars}.
\newblock {\em Phys. Rev. D} {\bf 2023}, {\em 108},~124080.
  %\href{http://arxiv.org/abs/2301.03568}{{\normalfont
  %[arXiv:gr-qc/2301.03568]}}.
\newblock {\url{https://doi.org/10.1103/PhysRevD.108.124080}}.

\bibitem[Diamond and Marques-Tavares(2022)]{Diamond:2021ekg}
Diamond, M.D.; Marques-Tavares, G.
\newblock {\ensuremath{\gamma}-Ray Flashes from Dark Photons in Neutron Star
  Mergers}.
\newblock {\em Phys. Rev. Lett.} {\bf 2022}, {\em 128},~211101.
  %\href{http://arxiv.org/abs/2106.03879}{{\normalfont
  %[arXiv:hep-ph/2106.03879]}}.
\newblock {\url{https://doi.org/10.1103/PhysRevLett.128.211101}}.

\bibitem[Diamond et~al.(2024)Diamond, Fiorillo, Marques-Tavares, Tamborra, and
  Vitagliano]{Diamond:2023cto}
Diamond, M.; Fiorillo, D.F.G.; Marques-Tavares, G.; Tamborra, I.; Vitagliano,
  E.
\newblock {Multimessenger Constraints on Radiatively Decaying Axions from
  GW170817}.
\newblock {\em Phys. Rev. Lett.} {\bf 2024}, {\em 132},~101004.
  %\href{http://arxiv.org/abs/2305.10327}{{\normalfont
  %[arXiv:hep-ph/2305.10327]}}.
\newblock {\url{https://doi.org/10.1103/PhysRevLett.132.101004}}.

\bibitem[Amaro-Seoane et~al.(2017)]{LISA:2017pwj}
Amaro-Seoane, P.;  et~al.
\newblock {Laser Interferometer Space Antenna}. \emph{arXiv} {\bf 2017}, arXiv:1702.00786.
\newblock  %\href{http://arxiv.org/abs/1702.00786}{{\normalfont
  %[arXiv:astro-ph.IM/1702.00786]}}.

\bibitem[Arun et~al.(2022)]{LISA:2022kgy}
Arun, K.G.;  et~al.
\newblock {New horizons for fundamental physics with LISA}.
\newblock {\em Living Rev. Rel.} {\bf 2022}, {\em 25},~4.
  %\href{http://arxiv.org/abs/2205.01597}{{\normalfont
  %[arXiv:gr-qc/2205.01597]}}.
\newblock {\url{https://doi.org/10.1007/s41114-022-00036-9}}.

\bibitem[Seoane et~al.(2023)]{LISA:2022yao}
Seoane, P.A.;  et~al.
\newblock {Astrophysics with the Laser Interferometer Space Antenna}.
\newblock {\em Living Rev. Rel.} {\bf 2023}, {\em 26},~2.
  %\href{http://arxiv.org/abs/2203.06016}{{\normalfont
  %[arXiv:gr-qc/2203.06016]}}.
\newblock {\url{https://doi.org/10.1007/s41114-022-00041-y}}.

\bibitem[Dengler et~al.(2022)Dengler, Schaffner-Bielich, and
  Tolos]{Dengler:2021qcq}
Dengler, Y.; Schaffner-Bielich, J.; Tolos, L.
\newblock {Second Love number of dark compact planets and neutron stars with
  dark matter}.
\newblock {\em Phys. Rev. D} {\bf 2022}, {\em 105},~043013.
  %\href{http://arxiv.org/abs/2111.06197}{{\normalfont
  %[arXiv:astro-ph.HE/2111.06197]}}.
\newblock {\url{https://doi.org/10.1103/PhysRevD.105.043013}}.

\bibitem[Berezhiani and Bento(2006)]{Berezhiani:2005hv}
Berezhiani, Z.; Bento, L.
\newblock {Neutron - mirror neutron oscillations: How fast might they be?}
\newblock {\em Phys. Rev. Lett.} {\bf 2006}, {\em 96},~081801.
  %\href{http://arxiv.org/abs/hep-ph/0507031}{{\normalfont [hep-ph/0507031]}}.
\newblock {\url{https://doi.org/10.1103/PhysRevLett.96.081801}}.

\bibitem[Serebrov et~al.(2008)]{Serebrov:2007gw}
Serebrov, A.P.;  et~al.
\newblock {Experimental search for neutron: Mirror neutron oscillations using
  storage of ultracold neutrons}.
\newblock {\em Phys. Lett. B} {\bf 2008}, {\em 663},~181--185.
  %\href{http://arxiv.org/abs/0706.3600}{{\normalfont
  %[arXiv:nucl-ex/0706.3600]}}.
\newblock {\url{https://doi.org/10.1016/j.physletb.2008.04.014}}.

\bibitem[Fornal and Grinstein(2018)]{Fornal:2018eol}
Fornal, B.; Grinstein, B.
\newblock {Dark Matter Interpretation of the Neutron Decay Anomaly}.
\newblock {\em Phys. Rev. Lett.} {\bf 2018}, {\em 120},~191801.
  %\href{http://arxiv.org/abs/1801.01124}{{\normalfont
  %[arXiv:hep-ph/1801.01124]}}.
\newblock Erratum in \emph{Phys. Rev. Lett.} \textbf{2020}, \emph{124}, 219901;
  {\url{https://doi.org/10.1103/PhysRevLett.120.191801}}.

\bibitem[Pichlmaier et~al.(2010)Pichlmaier, Varlamov, Schreckenbach, and
  Geltenbort]{Pichlmaier:2010zz}
Pichlmaier, A.; Varlamov, V.; Schreckenbach, K.; Geltenbort, P.
\newblock {Neutron lifetime measurement with the UCN trap-in-trap MAMBO II}.
\newblock {\em Phys. Lett. B} {\bf 2010}, {\em 693},~221--226.
\newblock {\url{https://doi.org/10.1016/j.physletb.2010.08.032}}.

\bibitem[Serebrov et~al.(2005)]{Serebrov:2004zf}
Serebrov, A.;  et~al.
\newblock {Measurement of the neutron lifetime using a gravitational trap and a
  low-temperature Fomblin coating}.
\newblock {\em Phys. Lett. B} {\bf 2005}, {\em 605},~72--78.
  %\href{http://arxiv.org/abs/nucl-ex/0408009}{{\normalfont [nucl-ex/0408009]}}.
\newblock {\url{https://doi.org/10.1016/j.physletb.2004.11.013}}.

\bibitem[Steyerl et~al.(2012)Steyerl, Pendlebury, Kaufman, Malik, and
  Desai]{Steyerl:2012zz}
Steyerl, A.; Pendlebury, J.M.; Kaufman, C.; Malik, S.S.; Desai, A.M.
\newblock {Quasielastic scattering in the interaction of ultracold neutrons
  with a liquid wall and application in a reanalysis of the Mambo I
  neutron-lifetime experiment}.
\newblock {\em Phys. Rev. C} {\bf 2012}, {\em 85},~065503.
\newblock {\url{https://doi.org/10.1103/PhysRevC.85.065503}}.

\bibitem[Arzumanov et~al.(2015)Arzumanov, Bondarenko, Chernyavsky, Geltenbort,
  Morozov, Nesvizhevsky, Panin, and Strepetov]{Arzumanov:2015tea}
Arzumanov, S.; Bondarenko, L.; Chernyavsky, S.; Geltenbort, P.; Morozov, V.;
  Nesvizhevsky, V.V.; Panin, Y.; Strepetov, A.
\newblock {A measurement of the neutron lifetime using the method of storage of
  ultracold neutrons and detection of inelastically up-scattered neutrons}.
\newblock {\em Phys. Lett. B} {\bf 2015}, {\em 745},~79--89.
\newblock {\url{https://doi.org/10.1016/j.physletb.2015.04.021}}.

\bibitem[Byrne et~al.(1990)]{Byrne:1990ig}
Byrne, J.;  et~al.
\newblock {Measurement of the neutron lifetime by counting trapped protons}.
\newblock {\em Phys. Rev. Lett.} {\bf 1990}, {\em 65},~289--292.
\newblock {\url{https://doi.org/10.1103/PhysRevLett.65.289}}.

\bibitem[Yue et~al.(2013)Yue, Dewey, Gilliam, Greene, Laptev, Nico, Snow, and
  Wietfeldt]{Yue:2013qrc}
Yue, A.T.; Dewey, M.S.; Gilliam, D.M.; Greene, G.L.; Laptev, A.B.; Nico, J.S.;
  Snow, W.M.; Wietfeldt, F.E.
\newblock {Improved Determination of the Neutron Lifetime}.
\newblock {\em Phys. Rev. Lett.} {\bf 2013}, {\em 111},~222501.
  %\href{http://arxiv.org/abs/1309.2623}{{\normalfont
  %[arXiv:nucl-ex/1309.2623]}}.
\newblock {\url{https://doi.org/10.1103/PhysRevLett.111.222501}}.

\bibitem[Czarnecki et~al.(2018)Czarnecki, Marciano, and
  Sirlin]{Czarnecki:2018okw}
Czarnecki, A.; Marciano, W.J.; Sirlin, A.
\newblock {Neutron Lifetime and Axial Coupling Connection}.
\newblock {\em Phys. Rev. Lett.} {\bf 2018}, {\em 120},~202002.
  %\href{http://arxiv.org/abs/1802.01804}{{\normalfont
  %[arXiv:hep-ph/1802.01804]}}.
\newblock {\url{https://doi.org/10.1103/PhysRevLett.120.202002}}.

\bibitem[Tang et~al.(2018)]{Tang:2018eln}
Tang, Z.;  et~al.
\newblock {Search for the Neutron Decay n$\rightarrow$ X+$\gamma$ where X is a
  dark matter particle}.
\newblock {\em Phys. Rev. Lett.} {\bf 2018}, {\em 121},~022505.
  %\href{http://arxiv.org/abs/1802.01595}{{\normalfont
  %[arXiv:nucl-ex/1802.01595]}}.
\newblock {\url{https://doi.org/10.1103/PhysRevLett.121.022505}}.

\bibitem[McKeen et~al.(2018)McKeen, Nelson, Reddy, and Zhou]{McKeen:2018xwc}
McKeen, D.; Nelson, A.E.; Reddy, S.; Zhou, D.
\newblock {Neutron stars exclude light dark baryons}.
\newblock {\em Phys. Rev. Lett.} {\bf 2018}, {\em 121},~061802.
  %\href{http://arxiv.org/abs/1802.08244}{{\normalfont.
  %[arXiv:hep-ph/1802.08244]}}.
\newblock {\url{https://doi.org/10.1103/PhysRevLett.121.061802}}.

\bibitem[Berezhiani et~al.(2018)Berezhiani, Biondi, Geltenbort,
  Krasnoshchekova, Varlamov, Vassiljev, and Zherebtsov]{Berezhiani:2017jkn}
Berezhiani, Z.; Biondi, R.; Geltenbort, P.; Krasnoshchekova, I.A.; Varlamov,
  V.E.; Vassiljev, A.V.; Zherebtsov, O.M.
\newblock {New experimental limits on neutron-mirror neutron oscillations in
  the presence of mirror magnetic field}.
\newblock {\em Eur. Phys. J. C} {\bf 2018}, {\em 78},~717.
  %\href{http://arxiv.org/abs/1712.05761}{{\normalfont
  %[arXiv:hep-ex/1712.05761]}}.
\newblock {\url{https://doi.org/10.1140/epjc/s10052-018-6189-y}}.

\bibitem[Berezhiani(2019{\natexlab{a}})]{Berezhiani:2018eds}
Berezhiani, Z.
\newblock {Neutron lifetime puzzle and neutron\textendash{}mirror neutron
  oscillation}.
\newblock {\em Eur. Phys. J. C} {\bf 2019}, {\em 79},~484.
  %\href{http://arxiv.org/abs/1807.07906}{{\normalfont
  %[arXiv:hep-ph/1807.07906]}}.
\newblock {\url{https://doi.org/10.1140/epjc/s10052-019-6995-x}}.

\bibitem[Berezhiani(2019{\natexlab{b}})]{Berezhiani:2018udo}
Berezhiani, Z.
\newblock {Neutron lifetime and dark decay of the neutron and hydrogen}.
\newblock {\em LHEP} {\bf 2019}, {\em 2},~118.
  %\href{http://arxiv.org/abs/1812.11089}{{\normalfont
  %[arXiv:hep-ph/1812.11089]}}.
\newblock {\url{https://doi.org/10.31526/lhep.1.2019.118}}.

\bibitem[Ayres et~al.(2022)]{Ayres:2021zbh}
Ayres, N.J.;  et~al.
\newblock {Improved Search for Neutron to Mirror-Neutron Oscillations in the
  Presence of Mirror Magnetic Fields with a Dedicated Apparatus at the PSI UCN
  Source}.
\newblock {\em Symmetry} {\bf 2022}, {\em 14},~503.
  %\href{http://arxiv.org/abs/2111.02794}{{\normalfont
  %[arXiv:physics.ins-det/2111.02794]}}.
\newblock {\url{https://doi.org/10.3390/sym14030503}}.

\bibitem[Fuwa et~al.(2024)]{Fuwa:2024cdf}
Fuwa, Y.;  et~al.
\newblock {Improved measurements of neutron lifetime with cold neutron beam at
  J-PARC}. \emph{arXiv} {\bf 2024}, arXiv:2412.19519.
\newblock  %\href{http://arxiv.org/abs/2412.19519}{{\normalfont
  %[arXiv:nucl-ex/2412.19519]}}.

\end{thebibliography}

\PublishersNote{}
\end{adjustwidth}

\end{document}